\documentclass[12pt]{article}
\usepackage{latexsym}
\usepackage{epsfig,amssymb,euscript}
\usepackage{amsmath}
\usepackage{mathrsfs}
\numberwithin{equation}{section}  
\newsavebox{\ns}
\newsavebox{\dbrane}

\def\be{\begin{equation}}
\def\ee{\end{equation}}
\def\bea{\begin{eqnarray}}
\def\eea{\end{eqnarray}}

\newcommand{\nn}{\nonumber}

\def\Dslash{\,\,{\raise.15ex\hbox{/}\mkern-12mu D}}
\def\Dbarslash{\,\,{\raise.15ex\hbox{/}\mkern-12mu {\bar D}}}
\def\delslash{\,\,{\raise.15ex\hbox{/}\mkern-9mu \partial}}
\def\delbarslash{\,\,{\raise.15ex\hbox{/}\mkern-9mu {\bar\partial}}}
\def\pslash{\,\,{\raise.15ex\hbox{/}\mkern-9mu p}}
\def\calDslash{\,\,{\raise.15ex\hbox{/}\mkern-12mu {\cal D}}}

\newcommand\R{\mathbb{R}}
\newcommand\Z{\mathbb{Z}}

\newcommand\C{\mathbb{C}}

\newcommand\diff{\mathrm{d}}

\newcommand{\vol}{\mathrm{vol}}

\newcommand{\ii}{\mathrm{i}}

\newcommand{\comment}[1]{}

\begin{document}
\begin{titlepage}
\begin{center}
\today

\vskip .5cm
{\Large \bf  Emerging Non-Anomalous Baryonic Symmetries \\ [3.4mm]

 in the AdS$_5$/CFT$_4$ Correspondence} \\[3.4mm]

\vskip 1.3cm

Nessi Benishti \let\thefootnote\relax\footnotetext{\noindent \tt nessibenishti@gmail.com}\\
\vskip 1cm

{\em Rudolf Peierls Centre for Theoretical Physics, \\ University of Oxford, \\
1 Keble Road, Oxford OX1 3NP, U.K.}\\

\vskip 1cm

\end{center}

\begin{abstract}
\noindent 
We study the breaking of baryonic symmetries in the AdS$_5$/CFT$_4$ correspondence for D3 branes at Calabi-Yau three-fold singularities. This leads, for particular VEVs, to the emergence of non-anomalous baryonic symmetries during the renormalization group flow. We claim that these VEVs correspond to critical values of the B-field moduli in the dual supergravity backgrounds. We study in detail the $\C^3/\Z_3$ orbifold, the cone over $\mathbb{F}_0$ and the $\C^3/\Z_5$ orbifold. For the first two examples, we study the dual supergravity backgrounds that correspond to the breaking of the emerging baryonic symmetries and identify the expected Goldstone bosons and global strings in the
infra-red. In doing so we confirm the claim that the emerging symmetries are indeed non-anomalous baryonic symmetries.

\end{abstract}

\vskip .8cm

\vfill
\hrule width 5cm
\vskip 5mm

\end{titlepage}
\pagestyle{plain}
\setcounter{page}{1}
\newcounter{bean}
\baselineskip18pt

\tableofcontents


\section{Introduction}\label{sec:intro}
The AdS/CFT correspondence \cite{Maldacena:1997re} allows one to understand gauge theory dynamics in terms of string theory on some background spacetime. Properties of strongly coupled gauge theories may then be understood geometrically, leading to non-trivial predictions on both sides of the correspondence. The richest of such examples which are both tractable, using current techniques, and also non-trivial, are given by the $(3+1)$d $\mathcal{N} = 1$ gauge theories that arise on a stack of D3-branes probing a singular Calabi-Yau three-fold. When the Calabi-Yau is also toric, one can use toric geometry, which by now is an extremely well developed subject, to study the gauge theories. 

The AdS/CFT correspondence connects the strong coupling regime of such gauge theories with supergravity in a mildly curved geometry. For the case of D3-branes placed at the tips of Calabi-Yau cones over five-dimensional geometries $Y$, the gravity dual is of the form AdS$_5 \times Y$, where $Y$ is a Sasaki-Einstein manifold \cite{Klebanov:1998hh,Acharya:1998db,Morrison:1998cs}. For example, one may take $Y = T^{1,1}$ \cite{Klebanov:1998hh}, or the more recently discovered infinite families of Sasaki-Einstein manifolds, $Y^{p,q}$ \cite{Gauntlett:2004zh,Gauntlett:2004yd} and $L^{a,b,c}$ \cite{Cvetic:2005vk,Martelli:2005wy}. In all these cases, the dual field theories \cite{Martelli:2004wu,Bertolini:2004xf,Benvenuti:2004dy,Benvenuti:2005ja,Franco:2005sm,Butti:2005sw} are conjectured to be supersymmetric gauge theories, at an infra-red (IR) conformal fixed point of the renormalization group (RG). Over the past few years very powerful techniques have been developed \cite{Hanany:2005ve,Franco:2005rj,Franco:2005sm,Hanany:2005ss,Feng:2005gw} to describe these models in terms of quiver and dimer diagrams that provide the relevant information concerning the spectrum and couplings of the corresponding gauge theory, for a review we refer the reader to \cite{Kennaway:2007tq}.

The moduli space of the field theory should at least contain the branch which is dual to the position of the $N$ D3 branes on the geometry. This corresponds to VEVs of mesonic operators in the field theory. In addition to that, the moduli space captures all the resolutions of the cone that the branes probe. The supergravity backgrounds constructed from these resolved cones allow, in general, turning on flat B-field and RR forms. Those that do not vanish at the boundaries correspond to marginal couplings in the field theory and the modes that do vanish are part of the moduli space of the supergravity background. The two kinds of directions just discussed correspond, in the moduli space of the field theory, to VEVs of baryonic operators. Setting non-vanishing VEVs in the field theory leads to the breaking of baryonic symmetries. This also pick a point in the moduli space of the theory and at the same time introduces a scale and thus a RG flow, whose endpoint will be a different SCFT. The supergravity dual of this RG flow was first discussed in the Type IIB context by Klebanov-Witten \cite{Klebanov:1999tb}, and to some extent in the M-theory AdS$_4 \times Y^7$ context in \cite{Benishti:2009ky,Benishti:2010jn}. The corresponding supergravity background has two boundaries that correspond to the two conformal theories in the IR and UV. 

In \cite{Martelli:2008cm} the full moduli space of such field theories was studied and compared to supergravity. A remarkable match was shown between the two. One aim of this paper is to push this a little further by showing that the directions in the moduli space of the field theory that correspond to the B-field moduli in the supergravity are \emph{periodic}. This is obviously expected as the B-field is periodic in string theory due to large gauge transformations.

Quiver gauge theories have in general both anomalous and non-anomalous baryonic symmetries. Breaking of non-anomalous baryonic symmetries in field theory result in the appearance of massless Goldstone bosons and global strings in the IR \cite{Vilenkin:1982ni}. In the AdS/CFT context, this has been studied in the conifold case in \cite{Klebanov:2007cx} and was generalized in \cite{Martelli:2008cm} to general toric Calabi-Yau backgrounds. It was shown that in the supergravity RR four-form fluctuations that are sourced by D3 branes, wrapping blown-up two-cycles, contain these Goldstone bosons. These branes form global strings in the Minkowski space directions around which the Goldstone boson has a monodromy. 

As was shown in \cite{Martelli:2008cm}, for fully resolved geometries, the number of massless modes coming from RR four-form fluctuations is equal to the number of two-cycles in the resolved cone, thus it exceeds the number of broken non-anomalous symmetries in the field theory which is just the number of three-cycles in the Sasaki-Einstein manifold. Moreover, there are massless modes that originate from fluctuations in the B-field and RR two-form. As explained in \cite{Martelli:2008cm} the number of those is the same number as the number of four-cycles in the resolved geometry. In total the number of such additional massless modes is the same as the number of the broken \emph{anomalous} baryonic symmetries in the field theory. However, the breaking of anomalous baryonic symmetries should not result in Goldstone bosons as these are not true symmetries of the quantum theory. Thus the interpretation of the additional massless modes just described in the field theory side is not clear. In \cite{Martelli:2008cm} it was suggested that such modes should be lifted by non-perturbative corrections. In this paper we want to show that the additional massless modes coming from RR four-form fluctuations can in fact be interpreted as Goldstone Bosons in some special cases. 

As a start we study the moduli space of three field theories living on D3 branes probing the $\C^3/\Z_3$ orbifold, the cone over $\mathbb{F}_0$ and the $\C^3/\Z_5$ orbifold. We examine the appearance of non-conformal theories for specific Higgsings. Such phases, we claim, appear whenever the resolved geometry contains four-cycles. Such Higgsing were observed before in the brane tilling context \cite{Hanany:2005ve} where the non-conformality translates to inconsistent tilling. This was also studied in \cite{Krishnan:2008kv} for the $\C^3/\Z_3$ orbifold. 

We identify systematically which VEVs lead to such non-conformal theories for our three examples and suggest gravity interpretations of these RG flows. More specifically, we show that these VEVs correspond in the gravity to specific values of the background compactly-supported B-field modes. These B-field values allow for D3-branes to wrap two-cycles in the resolved background. We suggest that these D3-branes are dual to global strings that appear in the field theory due to the breaking of the non-anomalous baryonic symmetry that emerge along the RG flow. This allows us to interpret the additional massless modes originating from RR four-form fluctuations, that were discussed in the previous paragraph, as Goldstone bosons.

The organization of this paper is as follows. In Section 2 we review the quiver field theories and the dual AdS$_5 \times Y$ supergravity backgrounds of interest. In Section 3 we discuss in detail the $\C^3/\Z_3$ orbifold. We compare its moduli space with the one expected from supergravity. Then we turn to discuss the non-conformal phase and the VEVs that lead to it. We end this section with a supergravity analysis of the corresponding RG flow. In Section 4 we repeat the same analysis for the cone over $\mathbb{F}_0$ theory. In Section 5 we take the first steps in applying our discussion to the $\C^3/\Z_5$ orbifold; we study the moduli space of the field theory and identify the VEVs that lead to non-conformal phases. We end with some concluding comments in Section 6. Finally, a number of relevant calculations and formulae are collected in the appendices.

\section{AdS$_5$ backgrounds and field theories}
In this section we briefly review the $\mathcal{N}=1$ supersymmetric 
quiver field theories of interest, focusing in particular on their vacuum moduli spaces. 
For further details the reader is referred to \cite{Martelli:2008cm} and references therein. 
We shall make extensive use of toric geometry throughout the paper,  
so include a brief summary for completeness (a standard reference is \cite{Fulton}). 
\subsection{Quiver gauge theories} \label{section2}

Our starting point is an $\mathcal{N}=1$ gauge theory in $(3+1)$ dimensions with 
product gauge group $\mathcal{G}\equiv\prod_{i=1}^G U(N_i)$. The matter content will be specified by a quiver diagram with
$G$ nodes. To each arrow in the quiver going from node $i$ to node $j$ we associate 
a chiral superfield $X_{i,j}$ in the bifundamental representation of 
the corresponding two gauge groups. More precisely, we take the convention that $X_{i,j}$ transforms in the $({N}_i,\bar{N}_j)$ representation of the gauge groups at nodes $i$ and $j$, respectively.
When $i=j$ this is understood to be the adjoint representation. 

We want to show now that the $G$ central $U(1)$s in $\mathcal{G}$ become global symmetries in the IR. It is easy to see that the diagonal $U(1)$ does not couple to any matter field. The rest of the $U(1)$s, due to triangle anomalies in the quiver, are divided into anomalous and non-anomalous symmetries. The gauge coupling of the non-anomalous $U(1)$s vanishes in the IR and the gauge fields associated with the anomalous $U(1)$s become massive during the flow. To see which symmetries are anomalous consider the $U(1)_{\vec{q}} \subset U(1)^G$ generated by $\mathcal{A}=\sum_{i}^G q_i\mathcal{A}_i$ where $\mathcal{A}_i$ are the generators of the central $U(1)_i\subset U(N_i)$. The condition for the cancellation of such triangle anomalies $Tr[U(1)_{\vec{q}}SU(N_k)^2]$ is just
\bea\label{anom}
\sum_{X_{i,j}|i=k}N_jq_j-\sum_{X_{i,j}|j=k}N_iq_i=0~.
\eea

Modding out by the global symmetries we see that in the IR the gauge group becomes
\bea\label{SUgroup}
\mathcal{SG}=\prod_{i=1}^G SU(N_i)~.
\eea

\subsubsection{Classical vacuum moduli space}

The classical VMS $\mathscr{M}$ is determined by the following equations 
\begin{eqnarray}\label{VMSeqns}
\nn \partial_{X_{i,j}} W &=& 0~,\\
\mu_i := -\sum\limits_{j=1}^G X_{j,i}^{\dagger} {X_{j,i}} + 
\sum\limits_{k=1}^G  {X_{i,k}} X_{i,k}^{\dagger} &=& 
0~
\end{eqnarray}
which are the F-term and D-term equations respectively, and $W$ is the superpotential of the field theory. 

In the Abelian case the moduli space $\mathscr{M}$ is straightforward to describe. The first equation 
describes the space of F-term solutions, which is by construction an affine 
algebraic set. For the theories we study in this paper, this is itself a toric variety, of dimension 
$4+(G-2) = G+2$. This is the so-called \emph{master space} $\mathscr{F}_{G+2}$, studied 
in detail in \cite{Forcella:2008bb}.

Finally, the combination of imposing the second equation in \eqref{VMSeqns} and identifying by the gauge symmetries
may be described as a K\"ahler quotient of $\mathscr{F}_{G+2}$ by a subgroup 
$U(1)^{G-1}\subset U(1)^G$. This subgroup does not include the diagonal $U(1)$ that does not couple to the bifundamental fields.
In particular, this K\"ahler quotient precisely sets the $\mu_i$ in \eqref{VMSeqns}
equal to zero. To summarize, 
\begin{equation}\label{symp}
\mathscr{M} \cong \mathscr{F}_{G+2} \, //\, U(1)^{G-2}~,
\end{equation}
where the K\"ahler quotient is taken at level zero, implying that $\mathscr{M}$ is a 
K\"ahler cone. We will denote this space as the Abelian mesonic moduli space. For a stack of $N$ coincident D3-branes transverse to a Calabi-Yau three-fold singularity, one expects the moduli space to be the $N$th symmetric product of the three-fold. 

The moment maps associated to the $G-1$ global $U(1)$s can take any non-vanishing values. In the physics literature this often associated with turning on FI parameters $\eta_i$ that contribute to the D-terms\footnote{Strictly speaking, since the $U(1)$s are not gauged no FI parameters can be turned on. The FI-like contributions to the D-terms come from VEVs to fields.}. We will use this terminology during the paper. The K\"ahler quotient in \eqref{symp} should then be taken with respect to 
\begin{eqnarray}\label{moment}
\mu_i := -\sum\limits_{j=1}^G X_{j,i}^{\dagger} {X_{j,i}} + 
\sum\limits_{k=1}^G  {X_{i,k}} X_{i,k}^{\dagger} &=& 
\eta_i~.
\end{eqnarray}
Setting the values of the FI parameters picks a point in $\mathscr{F}_{G+2}$ that solves \eqref{moment}. On the gravity side, as we explain in Section \ref{section3}, these FI parameters correspond to the K\"ahler class of the resolved CY geometry and the compactly-supported $B$-field periods. 

It is then convenient to present the $\mathscr{F}_{G+2}$ moduli space of a field theory with the aid of a FI space. Since $\sum_i \eta_i=0$, as can be seen by summing over the moment maps in \eqref{moment}, this space is just $\R^{G-1}$. The additional phases, corresponding to the moment maps, that survive the K\"ahler quotient are fibered over the FI space to form $(\C^*)^{G-1}$. The structure of $\mathscr{F}_{G+2}$ corresponds(very loosely) to $\mathscr{M}(\eta)$ fibered over $(\C^*)^{G-1}$, where $\mathscr{M}(\eta)$ obtained from \eqref{symp} with the corresponding moment maps levels. As we will see later on, in the corresponding resolved Calabi-Yau $X$, $b_2(X)$ of the $G-1$\footnote{Recall from \cite{Martelli:2008cm} that $G-1=b_2(X)+b_4(X)$, where $b_i(X)$ is the $i$th Betti number of $X$.} directions in the FI space correspond to K\"ahler moduli while the other $b_4(X)$ are dual to the compactly-supported B-field moduli. The fact that the latter are periodic, as expected from the periodicity of the B-field in string theory, is not obvious and we will show it later on in examples.

The FI space is expected to be divided into \textit{chambers}, where each one of them corresponds to the fully resolved geometry. These chambers are separated by \textit{walls}. Part of the walls correspond to singular $\mathscr{M}(\eta)$ spaces. When crossing such a wall the $\mathscr{M}(\eta)$ moduli space undergoes a form of small birational transformation called a flip \cite{flip}. Inside each such chamber the K\"ahler classes vary linearly with respect to the FI parameters and the $\mathscr{M}(\eta)$ spaces are isomorphic. These chambers however are further divided, as we will show later on in this paper, by additional walls that correspond to critical values of the compactly-supported B-field periods. 

\subsubsection{Toric Calabi-Yau three-folds}\label{sec:toric}

An affine toric three-fold variety $X=X_3$ is specified by a \emph{strictly convex 
rational polyhedral cone} $\mathcal{C}_3\subset\R^3$. More invariantly, 
$\R^3$ here is the Lie algebra of a torus $\mathbb{T}^3\cong U(1)^3$ of rank three.
By definition,
$\mathcal{C}_3$ takes the form
\bea
\mathcal{C}_3= \left\{\sum_{a=1}^D \lambda_a v_a\mid\lambda_a\in\R_{\geq 0}\right\}
\eea
where the set of vectors $v_a\in\R^3$, $a=1,\ldots,D$, are the generating 
rays of the cone. The condition of being rational means that 
$v_a\in\mathbb{Q}^3$, and without loss of generality we normalize these 
to be primitive vectors $v_a\in\Z^3$. 
The condition of 
strict convexity is equivalent to saying that $\mathcal{C}_3$ 
is a cone over a compact convex polytope. 

For an affine toric Calabi-Yau three-fold the $v_a$ all have their endpoints in a single hyperplane, 
where the hyperplane is at unit distance from the origin/apex of the cone. By an appropriate 
choice of basis, we may therefore write $v_a=(1,w_a)$ where the $w_a\in\Z^2$ 
are the vertices of the \emph{toric diagram} $\Delta$. The toric diagram is simply the 
convex hull of these lattice points, and so is a compact convex lattice 
polytope in $\R^2$. Any affine toric Calabi-Yau three-fold is specified 
uniquely by $\Delta$, up to $GL(3,\Z)$ 
transformations of the original torus $\mathbb{T}^3\cong U(1)^3$. 
Much of the geometry of affine toric Calabi-Yau three-folds 
reduces to studying these lattice polytopes.

Given a toric diagram $\Delta$, one can recover the corresponding 
Calabi-Yau three-fold via \emph{Delzant's construction}. In physics terms, this would be 
called a gauged linear sigma model (GLSM) description of the three-fold. A minimal 
presentation of the variety is as follows. 
One takes the external vertices $w_a\in\Z^2$, $a=1,\ldots,D$, of the toric diagram $\Delta$
(the smallest set of points whose convex hull is $\Delta$), 
and constructs the linear map
\bea\label{globe}
A&:&\R^D\rightarrow \R^3\nonumber\\
&;&e_a\mapsto v_a~.
\eea
Here $\{e_a\}$ denotes the standard orthonormal basis of $\R^D$. 
The fact that we started with a strictly convex cone implies that 
the map (\ref{globe}) is surjective.
Since $A$ maps lattice points in $\R^D$ to lattice points in $\R^3$, 
there is an induced map of tori
\bea
\mathbb{T}^D=\R^D/\Z^D \rightarrow \mathbb{T}^3=\R^3/\Z^3~.
\eea
The kernel is $\mathcal{G}\cong U(1)^{D-3}\times \Gamma$, where
$\Gamma\cong \Z^3/\mathrm{span}_{\Z}\{v_a\}$ is a finite Abelian group. 
The toric variety $X_3$ is then the K\"ahler quotient
\bea
X_3 \cong \C^D\, //\, \mathcal{G}
\eea
at moment map level zero, so that it is a K\"ahler cone. In GLSM terms, the 
coordinates $p_1,\ldots,p_D$ on $\C^D$ are identified with vacuum expectation values 
of the chiral fields; we shall thus generally refer to these as \emph{p-fields}. 
The moment map equation then arises as a D-term equation, while quotienting by 
$\mathcal{G}$ identifies gauge-equivalent vacua.
There is an induced action of $\mathbb{T}^3\cong U(1)^3\cong U(1)^D/\mathcal{G}$ 
on the K\"ahler variety $X_3$, and the image of the moment map 
is a polyhedral cone $\mathcal{C}_3^*$ which is the \emph{dual cone} to 
the polyhedral cone $\mathcal{C}_3$ with which we began.

\subsection{Gravity backgrounds}\label{section3}


\subsubsection{AdS$_5$ backgrounds}

In this subsection we want to discuss the Type IIB supergravity solution obtained by placing $N$ D3-branes at the tip of Calabi-Yau three-fold which is a cone over Sasaki-Einstein space $Y$. The corresponding metric and five-form are 
\bea\label{D3metric}
g_{10} &=& H^{-1/2}g_{\R^4} + H^{1/2} g_{C(Y)} \\ \label{metric}
G_5 & = & (1+*_{10})\diff H^{-1}\wedge \vol_4
\eea
where the warp factor $H$ reads
\bea\label{Hfull}
H = 1+ \frac{L^4}{r^4}~.
\eea
Here $g_{\R^4}$ is four-dimensional Minkowski space, with volume form $\vol_4$, and
$L$ is a constant given by
\bea\label{Lrel}
L^4 = \frac{(2\pi)^4g_s(\alpha')^2N}{4\vol(Y)}~.
\eea
The near-horizon limit corresponds to the small $r$ limit, thus the metric (\ref{D3metric}) becomes
\bea\label{AdSmetric}
g_{10}\, =\, \frac{L^2}{r^2}\diff r^2 + \frac{r^2}{L^2} g_{\R^4} + L^2 g_Y~.
\eea
This is just the AdS$_5\times Y$ metric. To fix the background one needs to specify the flat background fields. Turning on such fields corresponds to exactly marginal deformations in the $\mathcal{N}=1$ superconformal
field theories living on the boundary \cite{Martelli:2008cm}. 

In the next subsection we want to review the supergravity backgrounds that are obtained by resolving the Calabi-Yau cones. These backgrounds correspond to non-vanishing VEVs of the scalar operators, which Higgses the field theory.

\subsubsection{Symmetry-breaking backgrounds}\label{defsection}

The crepant (partial) resolutions of toric singularities are well understood, being described by toric geometry and hence fans of polyhedral cones. The extended K\"ahler cone for such resolutions is known as the Gel'fand-Kapranov-Zelevinsky(GKZ) fan, or secondary fan. The fan is a collection of polyhedral cones living in $\R^{b_2(X)}$, glued together along their boundaries, such that each cone corresponds to a particular choice of topology for $X$. Implicit here is the fact that $b_2(X)$ is independent of which topology for $X$ we choose. A point inside the polyhedral cone corresponding to a given $X$ is a K\"ahler class on $X$. The boundaries between cones correspond to partial resolutions, where there are further residual singularities, and there is a topology change as one crosses a boundary from one cone into another. Having chosen such an $X$ we must then find a Calabi-Yau metric on $X$ that approaches the given cone metric asymptotically. Such a metric can always be found \cite{craig1,craig2,craig3}. Note that the GKZ fan should match the $\R^{b_2(X)} \subset \R^{G-1}$ part of the FI space in the field theory that corresponds to the K\"ahler moduli.

The ten-dimensional metric in the resolved background is
\bea
g_{10} &=& H^{-1/2}g_{\R^4} + H^{1/2} g_{X} \label{resolved-metric}
\eea
with $G_5$-flux still given by \eqref{D3metric}. Placing $N$ spacetime-filling D3 branes at a point $y\in X$ leads to the warp factor equation
\bea\label{green}
\Delta_x H[y] = -\frac{(2\pi)^4g_s(\alpha^{'})^2N}{\sqrt{\det g_X}} \delta^6(x-y)~.
\eea
Here $\Delta H = \diff^* \diff H = - \nabla^i\nabla_i H$ is the scalar Laplacian of $X$ acting on $H$. 
Having chosen a particular resolution and K\"ahler class, hence metric, we must then find the warp factor $H$ satisfying (\ref{green}). This amounts to finding the Green's function on $X$, and this always exists 
and is unique. A general discussion in the Type IIB context may be found in \cite{Martelli:2007mk}. 

\begin{figure}[ht]
\begin{center} 
\includegraphics[scale=.7]{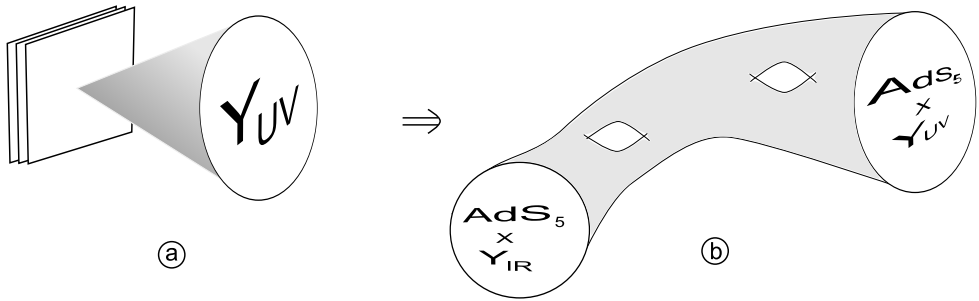} 
\end{center}
\caption{(a) A stack of N D3-branes transverse to the Calabi-Yau cone singularity $\mathcal{C}(Y_{UV})$; (b)
the supergravity geometry describing an RG flow dual to a diagonal Higgsing.}
\label{resolved}
\end{figure}

In the warped metric \eqref{resolved-metric} the point $y\in X$ is effectively sent to infinity, and the geometry has two asymptotically AdS$_5$ regions: one near $r=\infty$ that is asymptotically AdS$_5\times Y$, where $Y=Y_{UV}$ is the base of the unresolved cone, and one near to the point $y$, which is asymptotically AdS$_5\times Y_{IR}$, where $Y_{IR}$ is the base of the tangent cone close to the stack of D3 branes. This is illustrated in Figure~\ref{resolved}. For further discussion, see \cite{Martelli:2007mk, Martelli:2008cm, Benishti:2009ky}.

The $b_2(X)$ K\"ahler moduli are naturally complexified by noting that $H^4(X,Y,\R)\cong H_2(X,\R)\cong \R^{b_2(X)}$ by Poincar\'e duality, and that this group classifies the periods of $C_4$ through four-cycles in $X$, which may either be closed or have a boundary three-cycle on $Y=\partial X$. More precisely, taking into account large gauge transformations leads to the torus $H^2(X,Y,\R)/H^2(X,Y,\Z)\cong U(1)^{b_2(X)}$. 

The supergravity backgrounds may be interpreted as a renormalization group flow from the initial $\mathcal{N}=1$ superconformal field theory to a new SCFT in the IR as was first shown in \cite{Klebanov:1999tb}. There may be additional light particles in the IR, namely Goldstone bosons associated to the spontaneous breaking of non-anomalous baryonic symmetries.

The topology of $X$ in general allows one to turn on various topologically non-trivial flat form-fields. The forms of interest sit in compactly supported cohomology classes and thus correspond to \emph{fixed} marginal coupling in the field theory in the UV. These form field moduli are discussed in \cite{Martelli:2008cm}. In particular we have the NS B-field, as well as the RR two-form $C_2$, which are harmonic two-forms that are $L^2$ normalizable with respect to the unwarped metric, and four-form $C_4$ which is harmonic $L^2$ normalizable four-form with respect to the warped metric. The B-field, which lives in $H^2(X, Y ;\R)/H^2(X, Y ;\Z) = U(1)^{b_4(X)}$, is identified with $b_4(X)$ FI parameters in the field theory. Recall that the other $b_2(X)$ FI parameters correspond to the K\"ahler class of the metric. 

The RR field moduli form a torus $U(1)^{G-1}$ due to large gauge transformations. Supersymmetry pairs the K\"ahler class with $C_4$, and the B-field with $C_2$. In the field theory moduli space, this is reflected by the complexification $(\C^{\star})^{G-1}$ of the global baryonic symmetry group.


\section{The $\mathbb{C}^3/\mathbb{Z}_3$ orbifold}

In this paper we discuss field theories dual to toric Calabi-Yau three-folds with four-cycles in their resolutions. The simplest example of such a theory is the $\mathbb{C}^3/\mathbb{Z}_3$ orbifold theory. The space $\C^3/\Z_3$ is defined as the three-dimensional complex space $\C^3$ under the identification
\bea
\{x_1,x_2,x_2\} \sim \{w\,x_1,w\,x_2,w\,x_3\} \quad , \, w^3=1~.
\eea
The fixed point under this identification is just in the origin, therefore the near horizon geometry close to the point-like $N$ D3 branes is smooth. More specifically, from \eqref{AdSmetric}, one sees that this is just the AdS$_5 \times S^5/\Z_3$ space. In the following we discuss the matching between the moduli space in the field theory and supergravity and the emergence of a non-anomalous global baryonic symmetry after giving specific VEVs.

\subsection{Field theory description}
 
\begin{figure}[ht]
\begin{center} 
\includegraphics[scale=.5]{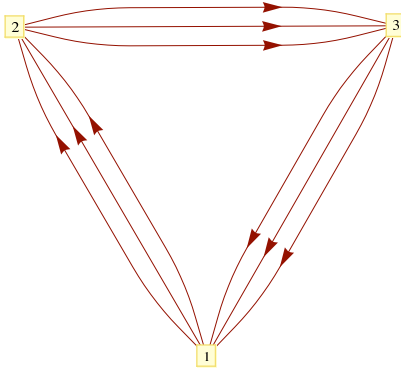} 
\end{center}
\caption{The quiver diagram for a dual of $\mathbb{C}^3/\mathbb{Z}_3$.}
\label{Z3-quiver}
\end{figure}

A stack of $N$ D3 branes propagating on the $\mathbb{C}^3/\mathbb{Z}_3$ orbifold is dual to a $(3+1)$d SCFT described by the quiver in Figure~\ref{Z3-quiver}.
\\

The field theory has a superpotential that reads
\begin{equation}
W={\rm Tr}\Big(\, X_{1,2}^i\, X_{2,3}^j\, X_{3,1}^k\,\Big)\,\epsilon_{ijk}~.
\end{equation}
This theory was studied in detail in \cite{Gukov:1998kn}. In the UV the gauge group is just $U(N)^3$ containing a central $U(1)^3$ factor which become global symmetries in the IR. The diagonal $U(1)$ decouples as no matter field carries charge under it. In addition any other combination of $U(1)$s is anomalous as can be seen from \eqref{anom}. One can choose the following orthogonal generators for these $U(1)$s 
\bea
\mathcal{A}_{B_1}=\mathcal{A}_1-\mathcal{A}_2 \ , \quad \mathcal{A}_{B_2}=\mathcal{A}_1+\mathcal{A}_2-\mathcal{A}_3
\eea
where $\mathcal{A}_i$ are the generators of the corresponding $U(1)$s in the quiver.

\subsubsection{The GLSM description}

We can compute the moduli space in the usual way. By making use of gauge rotations, we can set all the fields to be diagonal. Then, the effective theory reduces to $N$ copies of the $U(1)$ theory. To compute the moduli space of the abelian theory one can use the \textit{forward alogrithm} of \cite{Hanany:2008gx}. The key point is that the algorithm takes the data of the matter content and the superpotential, and produces the GLSM charge matrix $Q_t$. The kernel of this, $G_t$, is a matrix that encodes the toric diagram of the Calabi-Yau three-fold.

\begin{figure}[ht]
\begin{center} 
\includegraphics[scale=1]{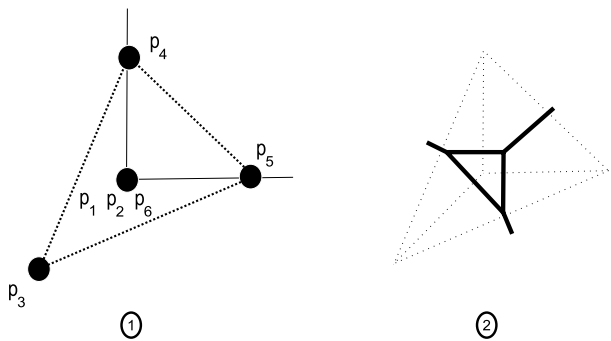} 
\end{center}
\caption{The $\C^3/\Z_3$ geometry. 1. The toric diagram of $\C^3/\Z_3$. 2. The pq-web is presented on top of the triangulation of the toric diagram that corresponds to the fully resolved $\C^3/\Z_3$ orbifold.}
\label{Z3-toric}
\end{figure}

Working in the GLSM description we derive the relations between the $X_{i,j}^k$ fields and the $p$-fields 
\bea
\nn \label{Z3-pfields}
&&X_{1,2}^1=p_1\,p_3 \ , \quad X_{1,2}^2=p_1\,p_4 \ , \quad X_{1,2}^3=p_1\,p_5 \ ,\\  \nn
&&X_{2,3}^1=p_2\,p_3 \ , \quad X_{2,3}^2=p_2\,p_4 \ , \quad X_{2,3}^3=p_2\,p_5 \ ,\\ 
&&X_{3,1}^1=p_6\,p_3 \ , \quad X_{3,1}^2=p_6\,p_4 \ , \quad X_{3,1}^3=p_6\,p_5~.
\eea
Using the forward algorithm one can show that 
\bea \label{Z3-Qt}
Q_t=\left(
\begin{array}{c c c c c c | c} 
 p_1 & p_2 & p_3 & p_4 & p_5 & p_6 & FI \\ \hline
 0 & 0 & 1 & 1 & 1 & -3 & a\\ 
 0 & -1 & 0 & 0 & 0 & 1 & b \\
 1 & 1 & -1 & -1 & -1 & 1 & 
\end{array}\right)~.
\eea
Now it is clear that, using the 2nd and 3rd rows in \eqref{Z3-Qt}, one can derive an expression for $p_1$ and $p_2$, which correspond to internal points, in terms of the other $p$-fields. Explicitly one can write
\bea
|p_1|^2=|p_6|^2+a+b \ , \quad \ |p_2|^2=|p_6|^2-b \ .
\eea
In the singular cone, for which $a=b=0$, $|p_1|$ and $|p_2|$ are fixed by $|p_6|$. Moreover, each row $\ell$ in \eqref{Z3-Qt} encodes the charges of the p-fields under the GLSM $U(1)_{\ell}$ gauge transformation. One can use $U(1)_{2}$ and $U(1)_{3}$ to fix the phases of $p_1$ and $p_2$. So we are left with four degrees of freedom, $p_3$, $p_4$, $p_5$ and $p_6$, and the constraint coming from the first row in the charge matrix is
\bea
|p_3|^2+|p_4|^2+|p_5|^2=3\,|p_6|^2 \ .
\eea
which after the corresponding gauge identification gives $\C^3/\Z_3$. Indeed, taking the FI parameters to zero one can compute the $G_t$ matrix by taking the null-space of $Q_t$ 
\bea
G_t=
\left(
\begin{array}{cccccc}
p_1 & p_2 & p_3 & p_4 & p_5 & p_6 \\ \hline
 1 & 1 & 1 & 1 & 1 & 1 \\
 0 & 0 & -1 & 0 & 1 & 0 \\
 0 & 0 & -1 & 1 & 0 & 0
\end{array}
\right)~,
\eea
the columns of this matrix are the coordinates in the toric diagram as reviewed in Section~\ref{sec:toric}. The toric diagram describes the moduli space and is presented in Figure~\ref{Z3-toric}.
Note that each spacetime field is the product of an inner and an outer p-field.

\subsubsection{Non-conformal phase in $\C^3/\Z_3$ theory} \label{sec:noncon}

In \cite{Krishnan:2008kv} the Higgsing of the $\C^3/\Z_3$ field theory was studied. There the authors noticed an interesting non-conformal phase that appears after giving a VEV to one of the fields and RG flowing to the VEV scale. This corresponds in the GLSM language to giving a VEV to one external and one internal p-field as can be seen from \eqref{Z3-pfields}. 

\begin{figure}[ht]
\begin{center} 
\includegraphics[scale=.8]{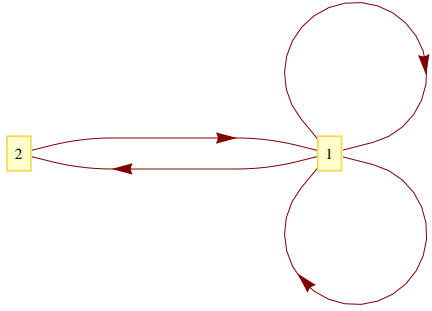} 
\end{center}
\caption{The quiver diagram for the non-conformal phase obtained by Higgsing the $\C^3/\Z_3$ theory. Both nodes are $SU(N)$.}
\label{HVZ-quiver}
\end{figure}

We want now to review how the discussed phase appears. For simplicity we will discuss just diagonal VEVs that correspond to keeping the D3 branes in one stack. Let us concentrate for the time being on the VEV $\| X_{1,2}^1 \|=v\, \mathbb{I}_{N \times N}$. This VEV breaks the anomalous global symmetry $\mathcal{A}_{B_1}$ and leaves $\mathcal{A}_{B_2}$ untouched. After giving such a VEV we see from the following part in the superpotential
\begin{equation}
W=\,v\,{\rm Tr}\Big(\, X_{2,3}^2\, X_{3,1}^3\,-\, X_{2,3}^3\, X_{3,1}^2\,\Big)\,+\,.\,.\,.
\end{equation}
that $X_{2,3}^2$, $X_{3,1}^3$, $X_{2,3}^3$ and $X_{3,1}^2$ become massive. At scales below $v$, these fields should be integrated out. The resulting quiver of the effective theory is presented in Figure~\ref{HVZ-quiver}.

\noindent
The new effective superpotential reads
\begin{equation} \label{HVZ-W}
W={\rm Tr}\Big(\, [X_{1,1}^1,X_{1,1}^2]\, X_{1,2}\, X_{2,1}\,\Big)~.
\end{equation}

We can look for example at node 2. Since it has $N_f=N_c$, this theory cannot be in a conformal fixed point. 
In \cite{Krishnan:2008kv} it was argued that node 2 in this theory confines. This is indeed expected in case that the dynamical scale of node 2 is dominating. This will be consistent with the supergravity analysis that will be presented later on in this paper.
In the IR one can treat the field theory effectively as a copy of $N_f=N_c$ SQCD coupled to a pair of singlets $X_{1,1}^i$ with the superpotential of the non-conformal theory. The IR of the SQCD theory is described in terms of mesons and baryons
\bea
\nn
&&\mathcal{M}^i_j=(X_{1,2})^j_{\alpha}\,(X_{2,1})^{\alpha}_i\, ,\qquad \mathcal{B}^{i_1\cdots i_N}=\epsilon^{\alpha_1\cdots \alpha_N}\, (X_{1,2})^{i_1}_{\alpha_1}\cdots (X_{1,2})_{\alpha_N}^{i_N} \\ &&\tilde{\mathcal{B}}_{i_1\cdots i_N}=\epsilon_{\alpha_1\cdots \alpha_N}\, (X_{2,1})_{i_1}^{\alpha_1}\cdots (X_{2,1})^{\alpha_N}_{i_N}
\eea
where latin indices stand for $SU(N)_1$ while greek ones stand for $SU(N)_2$. Note that the superpotential is then written as
\begin{equation}
W={\rm Tr}\Big(\, \mathcal{M}\,[X_{1,1}^1,\, X_{1,1}^2]\,\Big)~.
\end{equation}
The baryons carry charge under the baryonic symmetry that transform the fields as follows
\begin{equation}
X_{1,2}\rightarrow e^{i\theta}\, X_{1,2}\qquad X_{2,1}\rightarrow e^{-i\theta}\, X_{2,1} ~.
\end{equation}
This, we claim, is a non-anomalous symmetry that was evolved from $\mathcal{A}_{B_2}$ during the RG flow.

As argued by Seiberg \cite{Seiberg:1994bz}, the theory has a quantum mechanically modified moduli space given by
\begin{equation}
{\rm det}\,\mathcal{M}-\mathcal{B}\, \tilde{\mathcal{B}}=\Lambda^{2N}
\end{equation}
where $\Lambda$ is the dynamically generated scale of the SQCD theory. In addition, we have to impose the F-terms coming from the superpotential. Thus, finally, the moduli space is given by the solutions to
\begin{equation}
{\rm det}\,\mathcal{M}-\mathcal{B}\, \tilde{\mathcal{B}}=\Lambda^{2N},\qquad [X_{1,1}^i,\, \mathcal{M}]=0,\qquad [X_{1,1}^1,\,X_{1,1}^2]=0~.
\end{equation}
Note that, since node 2 is confined, in the IR the relevant degrees of freedom are mesons and baryons. As such, the superpotential appears written in terms of the meson $\mathcal{M}$, and thus automatically generates an F-term ensuring that the $X_{1,1}^i$ fields commute. Let us introduce a Lagrange multiplier chiral field $\lambda$, so that we can write
\begin{equation}
W={\rm Tr}\Big(\, \mathcal{M}\,[X_{1,1}^1,\,X_{1,1}^2]\,\Big)+\lambda\, \Big({\rm det}\,\mathcal{M}-\mathcal{B}\, \tilde{\mathcal{B}}-\Lambda^{2N}\Big)~.
\end{equation}

Vanishing VEVs to the entire set of fields is not a point on the moduli space of this theory. As we explain now, a conformal fixed point can be reached by considering non-vanishing VEVs that induce continuation of the RG flow. In \cite{Krishnan:2008kv} the authors discussed the RG flow resulting from a non-vanishing VEV to the mesonic operator $\mathcal{M}$ only. This RG flow ends with the $\mathcal{N}=4$ $SU(N)$ SYM theory in the IR. In this paper however, we will be interested in the flow that results from giving VEVs to the baryons. After setting the only non-vanishing VEV as $\mathcal{B}\tilde{\mathcal{B}}=-\Lambda^{2N}$ and flowing to the new IR the superpotential reduce to
\begin{equation}
W={\rm Tr}\Big(\, \mathcal{M}\,[X_{1,1}^1,\,X_{1,1}^2]\,\Big)~,
\end{equation}
and $SU(N)_1$ flows to strong coupling. So we see that the theory in the IR is just $\mathcal{N}=4$ $SU(N)$ SYM. We expect to see a Goldstone boson and global string due to the fact that the non-anomalous baryonic symmetry was broken. 

It is interesting to notice, as was argued in \cite{Krishnan:2008kv}, that in the gravity there is just one scale, the size of the four-cycle. Thus as expected, in the large $N$ limit in the field theory the scale in which the massive fields are integrated out is also the scale in which node 2 in the non-conformal phase confines. The strong coupling scale for the confining is related to the energy scale $E$ set by the VEV $v$ as:
\bea
\Lambda = E\,e^{-\frac{8\,\pi^2}{2\,N\,g^2_{YM}(E)}}
\eea
where $E = \sqrt{v}$. For large t'Hooft coupling $\lambda \equiv N\,g_{YM}^2$ the scale $\Lambda$ and $E$ are of the same order and we cannot distinguish between them. Therefore we cannot expect to see the moduli space of the non-conformal theory and the emerging baryonic symmetry in the dual supergravity. But in the far IR we can expect to see the Goldstone mode and global string that correspond to the broken global baryonic symmetry. In the next subsections we want to show that ineed these \emph{can} be observed in the dual supergravity background.

We want to start by describing the full moduli space of the $\C^3/\Z_3$ theory. It will prove useful to understand what are the moduli in the supergravity that correspond to the VEVs that lead to the non-conformal theory.
\subsubsection{$\C^3/\Z_3$ theory - Moduli space}
From \eqref{Z3-Qt}, the VMS equations are just
\bea \label{vmsequ}
\nn
&&|p_3|^2\,+|p_4|^2\,+|p_5|^2\,-\,3\,|p_6|^2\,=\,a \\ \nn
&&|p_2|^2\,-\,|p_6|^2\,=\,-b \\ 
&&|p_1|^2\,+\,|p_2|^2\,-\,|p_3|^2\,+|p_4|^2\,+|p_5|^2\,+\,|p_6|^2\,=\,0
\eea
where one needs to mod out by the corresponding $U(1)$ transformations. We now show that these equations can be written in three different forms, in which two different internal $p$-fields can be eliminated. These correspond to three different chambers in the FI parameter space of the field theory. 

\subsection*{Chamber 1}
We rearrange \eqref{vmsequ} 
\bea \label{vmsequ1}
\nn
&&|p_3|^2\,+|p_4|^2\,+|p_5|^2\,=3\,|p_2|^2\,+\,a\,+\,3\,b \\ \nn
&&|p_6|^2\,=|p_2|^2\,+\,b \\ 
&&|p_1|^2\,=\,|p_2|^2\,+\,a\,+2\,b
\eea
such that it is obvious that $p_1$ and $p_6$ can be eliminated with respect to the other p-fields and their corresponding phases can be fixed. We have to take $b \geq 0$ and $a+2b \geq 0$ to guarantee that solutions to these fields always exist. We are left with the first line that describes branes on the resolved $\C^3/\Z_3$ where the K\"ahler class is proportional to $a+3b$. In every case we must necessarily find this geometry since the only Calabi-Yau resolution of the $\C^3/\Z_3$ singularity is $\mathcal{O}(-3)\rightarrow\mathbb{CP}^2$. This can be easily seen from the pq-web in Figure~\ref{Z3-toric} that represents the resolution of the Calabi-Yau. 

To obtain the pq-web one first needs to choose a triangulation of the toric diagram. This corresponds to a specific resolution of the Calabi-Yau space. For $\C^3/\Z_3$ there is just one choice, as can be seen from Figure~\ref{Z3-toric}, which also describes the fully resolved space. The pq-web is obtained by replacing faces by vertices, lines by orthogonal lines, and vertices by faces in the triangulation of the toric diagram. This allows one to map the topology of the resolved space into a two-dimensional diagram \cite{Aharony:1997bh,Leung:1997tw}. Indeed, the triangle in the pq-web represents the compact divisor $\mathbb{CP}^2$ which is in agreement with the GLSM picture. The $\mathbb{CP}^2$ zero section is at $p_2=0$ in this chamber of the gauge theory FI space.

If we take $|p_2| > 0$ we see from \eqref{vmsequ1} that some of the external $p$-fields and all the internal $p$-fields obtain a non-vanishing VEV. This corresponds to putting the D3 branes away from the four-cycle and in the field theory to a VEV of a closed loop of fields in the quiver that corresponds to a mesonic operator. This Higgses the $SU(N)^3$ gauge group to $SU(N)$ and the theory flows to $\mathcal{N}=4$ $SU(N)$ SYM in the IR. Notice that in order to obtain the non-conformal phase we were discussing, we need to blow-up the four-cycle by taking $a+3b>0$, put the D3 branes on the blown-up four-cycle by setting $p_2=0$ and take $b=0$ or $a+2b=0$, the latter condition guarantees that just one internal point has a non-vanishing VEV. Setting $b>0$ and $a+2b>0$ results in RG flow that ends with the $\mathcal{N}=4$ $SU(N)$ SYM in the IR. 

The position of the branes on the four-cycle is determined by $p_3$, $p_4$ and $p_5$ which are constrained by the 1st row in \eqref{vmsequ}. This is just a point on $\mathbb{CP}^2$ and all such points are equivalent, since $\mathbb{CP}^2$ is homogeneous, thus these directions can be suppressed when discussing the moduli space of the field theory. Therefore, the interesting information on the moduli space of the field theory is just the FI parameter space.

\subsection*{Chamber 2}
A straightforward manipulation of \eqref{vmsequ} leads to
\bea
\nn
&&|p_3|^2\,+|p_4|^2\,+|p_5|^2\,=\,3\,|p_6|^2\,+\,a \\ \nn
&&|p_2|^2\,=\,|p_6|^2\,-b \\ 
&&|p_1|^2\,=\,|p_6|^2\,+a+b~.
\eea
One can see that $p_1$ and $p_2$ can be eliminated and their phases can be fixed. This can be done if we restrict $a \geq 0$, $b \leq 0$ and $a+b \geq 0$. The discussion is then along the same lines as in the previous paragraph. This time, to obtain the non-conformal phase, we take $a>0$, $p_6=0$ and $b=0$ or $a+b=0$. Again, since $p_6=0$ the branes sit on the blown-up cycle with the resolved geometry being the same as before. 

\subsection*{Chamber 3}
A straightforward manipulation of \eqref{vmsequ} leads to
\bea
\nn
&&|p_3|^2\,+|p_4|^2\,+|p_5|^2\,=\,3\,|p_1|^2\,-\,2a-3b \\ \nn
&&|p_2|^2\,=\,|p_1|^2\,-a-2b \\ 
&&|p_6|^2\,=\,|p_1|^2\,-a-b~.
\eea
One can see that $p_2$ and $p_6$ can be eliminated and their phases can be fixed. We need to take $a \leq - \frac{3}{2}\,b$, $a \leq -2b$ and $a \leq - b$. We see similar behaviour as for the other chambers that we just described. To obtain the non-conformal phase we have to take $-\,2a-3b>0$, $p_1=0$ and $a+2b=0$ or $a+b=0$ such that the branes again sit on the blown-up cycle of the resolved geometry. 
\begin{figure}[ht]
\begin{center}
\includegraphics[scale=0.6]{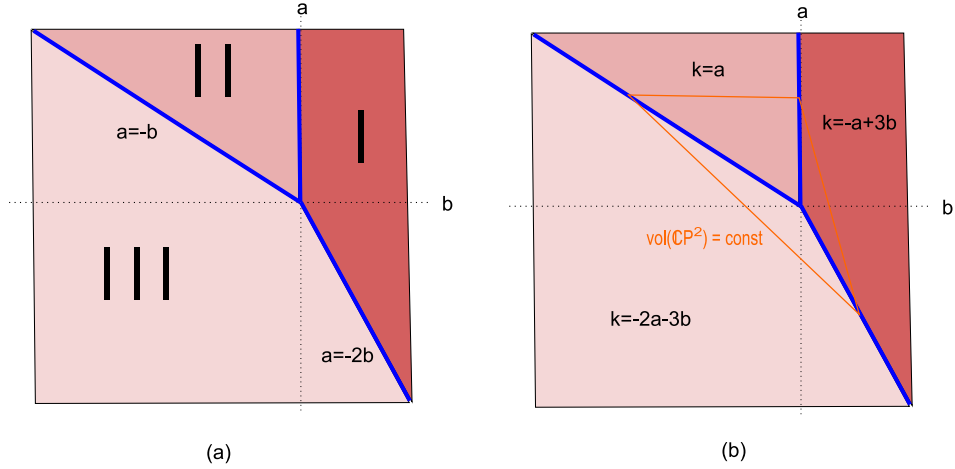}
\end{center}
\caption{The FI space of the $\C^3/\Z_3$ theory. (a) The three chambers of the moduli space. The blue lines correspond to the FI parameter values that result, after RG flow, in the non-conformal phase in Figure~\ref{HVZ-quiver}. (b) The FI path dual to the B-field period is plotted in orange line for the fixed FI $k$ which is the dual to the K\"ahler modulus in each chamber.}
\label{GKZ}
\end{figure}

To conclude, giving VEVs to fields such that no closed loop of fields in the quiver has a non-vanishing VEV results in a blown-up four-cycle and the D3 branes sit on the exceptional cycle $\mathbb{CP}^2$ in $\mathcal{O}(-3)\rightarrow \mathbb{CP}^2$. Figure~\ref{GKZ} (a) describes the part of the moduli space with vanishing VEVs to mesonic operators. The blue lines, which are the borderlines of the different regions, correspond to the values of the FI parameters $a$ and $b$ that brings us to the non-conformal theory after RG flow to the VEV scale. Any other point in the diagram corresponds to VEVs that induce RG flows that end with $\mathcal{N}=4$ $SU(N)$ SYM in the IR.
In Figure~\ref{GKZ} (b) we show the path that corresponds to a constant K\"ahler class. Following this path it is easy to see that the FI parameter that is "orthogonal" to the one that represents the size of the cycle is \emph{periodic}. This will be consistent with interpreting this on the gravity side with the period of the $B$-field through the $\R^2$ fibre of $\mathcal{O}(-3)\rightarrow \mathbb{CP}^2$. We will also see that the non-conformal phase corresponds in the supergravity to a resolved backgrounds with one of the three critical values of the B-field period as the diagram suggests.
%
\subsection{Gravity description}
The gravity background corresponding to the resolved orbifold $\C^3/\Z_3$ is
\begin{equation}
ds^2_X=\frac{d\rho^2}{\Big(1-\frac{a^6}{\rho^6}\Big)}+\frac{\rho^2}{9}\, \Big(1-\frac{a^6}{\rho^6}\Big)\, (d\psi-\mathcal{A})^2+\rho^2\, ds^2_{\mathbb{CP}^2}
\end{equation}
where
\bea
\mathcal{A}&=&\frac{3}{2}\,\sin^2\sigma\, (d\beta+\cos\theta\, d\phi),\nn\\
 ds^2_{\mathbb{CP}^2}&=&d\sigma^2+\frac{1}{4}\,\sin^2\sigma\,\Big(d\theta^2+\sin^2\theta\, d\phi^2+\cos^2\sigma\, (d\beta+\cos\theta\, d\phi)^2 \Big)~.
\eea
This is the usual Calabi ansatz, which works for the canonical line bundle over any K\"ahler-Einstein manifold. It is natural to introduce the vielbein
\bea
\nn
&&g_5=d\psi-\mathcal{A} \ , \quad e_{\beta}=\frac{1}{2}\, \sin\sigma\, \cos\sigma\, (d\beta+\cos\theta\, d\phi)\ , \\
&&e_{\theta}=\frac{1}{2}\,\sin\sigma\,d\theta\ , \quad e_{\phi}=\frac{1}{2}\,\sin\sigma\,\sin\theta\, d\phi~.
\eea
The metric then becomes
\begin{equation}
ds^2_X=\frac{d\rho^2}{\Big(1-\frac{a^6}{\rho^6}\Big)}+\frac{\rho^2}{9}\, \Big(1-\frac{a^6}{\rho^6}\Big)\, g_5^2+\rho^2\, d\sigma^2+\rho^2\, e_{\beta}^2+\rho^2\,e_{\theta}^2+\rho^2\, e_{\phi}^2~.
\end{equation}
Plugging $ds^2_X$ in \eqref{resolved-metric} one obtains the metric of the ten-dimensional Type IIB background. The warp factor $H$ is given by \eqref{green}. As usual, we can solve \eqref{green} by writing
\begin{equation}
\label{hI}
H=\sum\, h_I(\rho,\,\xi;\rho_p,\, \xi_p)\, Y_I^{\star}(\xi_p)\, Y_I(\xi)
\end{equation}
where the $Y_I$ are the relevant harmonics, $\xi$ denots collectively the angular coordinates in the internal manifold and $(\rho_p,\xi_p)$ stands for the particular point where the stack of D3 branes sits. We refer the reader to \cite{Krishnan:2008kv} for the explicit solution for this warp factor.

\subsubsection{Gravity moduli: B-field and $C_2$} \label{sec:2-form-moduli}

As discussed in Section~\ref{section3}, the classical gauge theory has a large VMS. One thus expects to find massless scalar fields associated to these flat directions in field space. We are now interested in discussing the form field moduli in the resolved background. As we reviewed, it is possible to turn on flat form fields in the resolved background that restrict to trivial classes on the UV boundary. In this subsection we want to discuss the B-field and $C_2$ RR two-form. The B-field moduli will become an important factor in the following discussion. The $C_4$ RR four-form will not be discussed here, however fluctuations in this direction will become important later on in this section. 

The B-field and $C_2$ moduli correspond to harmonic 2-forms that are $L^2$ normalizable with respect to the unwarped metric. For the resolved $\C^3/\Z_3$, $b_4=1$, thus we are looking for one such 2-form. Let us define $J=d\mathcal{A}$
\begin{eqnarray}
J&=&3\sin\sigma\, \cos\sigma\, d\sigma\wedge \, (d\beta+\cos\theta\, d\phi)-\frac{3}{2}\,\sin^2\sigma\,\sin\theta\, d\theta\wedge d\phi\nonumber \\&=&6\, d\sigma\wedge e_{\beta}-6\, e_{\theta}\wedge e_{\phi}~.
\end{eqnarray}
We can consider the following closed 4-form
\begin{equation}
\omega_4=J\wedge J + d\Big(f(\rho)\, g_5\wedge J\Big)~.
\end{equation}
Expanding it
\begin{equation}
\omega_4=(1-f)\, J\wedge J+f'\, d\rho\wedge g_5\wedge J
\end{equation}
one can easily show that
\begin{equation}
\star\, J\wedge J=-\frac{24}{\rho^3}\, d\rho\wedge g_5,\qquad \star\, d\rho\wedge g_5\wedge J=-\frac{3}{\rho}\, J~,
\end{equation}
so
\begin{equation}
\star\omega_4=-\frac{24}{\rho^3}\,(1-f)\, d\rho\wedge g_5-\frac{3}{\rho}\,f'\,\, J~.
\end{equation}
Thus, the condition $d\star\omega_4=0$ boils down to
\begin{equation}
\partial_{\rho}\Big(\rho^{-1}\,\partial_{\rho}f\Big)+\frac{8}{\rho^3}=0~.
\end{equation}
The solution to this equation which vanishes at infinity is
\begin{equation}
f=1+\frac{A}{\rho^2}~,
\end{equation}
$A$ being an integration constant.

We can now compute the square $L^2$ norm of the form, which is finite
\begin{equation}
\int\omega_4\wedge \star\omega_4=864\,a^{-6}\,A^2\,\pi\,{\rm Vol}(\mathbb{CP}^2)~.
\end{equation}
In turn, we have that $\omega_2=\star\omega_4$ can be written, as expected, as
\begin{equation} \label{2f-unwarped}
\omega_2=-6A\,d\Big(\rho^{-4}\,g_5\Big)~.
\end{equation}
We can now consider the following ansatz for the supergravity zero modes
\begin{equation}
B_2=b(x)\, \omega_2\, ,\qquad C_2=c(x)\, \omega_2~.
\end{equation}
From the supergravity equations of motion one can see that the scalar fields $b(x)$ and $c(x)$ are decoupled \cite{Martelli:2008cm}. Given the properties of $\omega_2$, it is straightforward to see that the $b$, $c$ fields satisfy free field equations of motion in the Minkowski space directions.

It is convenient to choose the normalization factor $A$ to be
\bea
A = \frac{a^4}{6\cdot 2\pi}~.
\eea
With this choice the above calculations show that 
\bea
\int_{\R^2_{\mathrm{fibre}}} \omega_2 = 1~.
\eea
Since $H^2_{\mathrm{cpt}}(X,\Z)\cong \Z$, this shows that $\omega_2$ is then an $L^2$ harmonic form 
representing the generator of this group. Here $X$ is the total space of $\mathcal{O}(-3)\rightarrow \mathbb{CP}^2$. 
An exact sequence shows that $\omega_2$ maps to $3$ times the generator of $H^2(\mathbb{CP}^2,\Z)\cong\Z$, where 
the $3$ is the same as in $\mathcal{O}(-3)$. Indeed, a simple calculation shows that
\bea \label{mult3} 
\int_{\mathbb{CP}^1} \omega_2 = 3~.
\eea

\subsubsection{Euclidean string condensates and the periodicity of $b$ and $c$}

Consider now a Euclidean fundamental string running along $\{\rho,\, \psi\}$. Its (Euclidean) action is
\begin{equation}
S_E=\frac{T_F}{3}\int_{a}^{\rho_c} d\rho\, \int_0^{2\pi} d\psi\, \, H\, \rho+\ii\, T_F\, b
\end{equation}
where $\rho_c$ is some $UV$ cut-off and $T_F$ is the string tension. Actually, this cut-off is not required, notice, since the integral
is convergent.

Under a shift
\begin{equation} \label{bperiod}
b\rightarrow b+\frac{2\pi n}{T_F}~, \qquad n\in\Z~,
\end{equation}
the partition function will not change. Thus, we have that $b$ is a periodic variable of period $2\pi/T_F$ (and so is $c$, had we considered a D1 string, upon changing $T_F\leftrightarrow T_1$).

We identify this B-field period, up to a constant factor, with the FI parameter in the field theory that corresponds to the close path in Figure~\ref{GKZ}. Recall that supersymmetry pairs the B-field with $C_2$ and the latter corresponds to a fiber over the FI parameter space in the field theory.

\subsubsection{D3 branes on the resolved cone and global strings}
A spontaneously broken global symmetry in a field theory generally leads to Goldstone bosons and global strings. The first corresponds to the flat directions given by acting with a broken symmetry generator. The Goldstones are fluctuations of RR fields, and are hence pseudo-scalars. Their scalar supersymmetric partners come from metric and B-field fluctuations. 

In \cite{Martelli:2008cm} the authors have shown the existence of massless scalar fields on $\R^4$ associated with linearized deformations of the B-field and RR field moduli and argued that such modes can be obtained also from fluctuations of the metric. In general $b_4(X)$ such modes are coming from B-field and $C_2$ fluctuations and $b_2(X)$ from $C_4$ and metric fluctuations. Thus it was argued that the dual field theory then includes corresponding massless particles. 

The linearized fluctuations just discussed may be associated with the Goldstone bosons and their supersymmetric partners. However as discussed in \cite{Martelli:2008cm}, these modes cannot be interpreted as Goldstone bosons when the corresponding broken symmetries are anomalous since these symmetries do not survive in the quantum theory. 

Coming back to our example, in the $\C^3/\Z_3$ field theory all the baryonic global symmetries are anomalous. The supergravity backgrounds constructed from resolved cones are dual to field theories in which these anomalous baryonic symmetries are "broken"\footnote{They are really already broken by quantum effects.}. However, for three values of the B-field there is an emerging non-anomalous baryonic symmetry during the RG flow as we argued in the last section. We claimed that the supergravity solution is dual to a field theory in which this symmetry is broken. We want to argue that the corresponding Goldstone boson and its supersymmetric partner originate from specific fluctuations in the RR 4-form and metric respectively. Since the fluctuations in the metric are somewhat involved, as explained in \cite{Martelli:2008cm}, we will discuss just the fluctuations for the RR four-form in detail in the next subsection. 

As discussed in \cite{Vilenkin:1982ni} the breaking of global $U(1)$ symmetry results in global strings around which the Goldstone boson has a monodromy. Global strings in the Minkowski space associated with the broken global symmetry were discussed in the AdS/CFT context for the conifold case in \cite{Klebanov:2007cx}. It was shown that a D3 brane wrapping the two-cycle in the bottom of the resolved cone sources fluctuations that contain the Goldstone boson and that this boson has a monodromy around the string-like wrapped D3 brane in the Minkowski space. 

Coming back to the $\C^3/\Z_3$ theory, we will show that a global string, obtained by wrapping a D3 brane on two-cycle, appears in supergravity \emph{just} for three critical values of the B-field as expected from the field theory analysis. For other values of the B-field the massless fields are still there but the wrapped D3 brane is not SUSY, this suggests that indeed these fields cannot be interpreted as Goldstone bosons in these cases. This is expected since for non-critical values of the B-field, the only "broken" baryonic symmetries are anomalous. 

We want now to discuss the wrapped D3 brane in more detail. Notice that the blown-up four-cycle, being a $\mathbb{CP}^2$, contains a topologically non-trivial $\mathbb{CP}^1$. So we can consider a D3 brane wrapping this two-cycle by taking the worldvolume of the brane to be $\{t,\, x,\,\mathbb{CP}^1\}$. This brane, having half of its worldvolume on the Minkowski space and half in the internal space, does not feel the warp factor. Thus, its energy at $\rho=a$ is a constant proportional to $a^2$. From the Minkowski space point of view, it looks like a string. 

Notice that if we were considering a D5 brane wrapping the four-cycle, similar arguments show that the tension of this brane blows up in the IR. This shows that in the IR of the field theory the corresponding string is infinitely massive and therefore completely decoupled in the low energy limit. It is interesting that such D5 branes source $C_2$ fluctuations. Such a fluctuation that solves the supergravity equations and corresponds to a massless field in $\R^4$ should have an interpretation in the field theory. One might consider this as the fluctuation that contain the Goldstone boson coming from the broken emerging baryonic symmetry. However, since in this case the global string, namely a D5 brane wrapped on four-cycle, is not part of the field theory spectrum, we do not want to interpret it as such. Similarly, if we consider the RR four-form fluctuation for non-critical B-field period discussed above, since in such background the D3 brane wrapping the two-cycle is not SUSY the global string is absent and the interpretation of the corresponding massless field is not clear to us.

To check when the D3 brane wrapping the two-cycle in the resolved cone is SUSY we follow \cite{Martucci:2005ht}. In their notations, the two supersymmetry parameters $\epsilon_{1,2}$ are Majorana-Weyl real spinors of positive ten-dimensional chirality and can be written as
\bea
\epsilon_a(y)=\xi_+ \otimes \eta_+^{(a)}(y)+\xi_- \otimes \eta_-^{(a)}(y)\ , \quad a=1,2
\eea
where $\xi$ and $\eta$ are the spinors in the internal four-dimensional and external six-dimensional space respectively.
 
We can start from the conditions (3.6) and (3.7) of \cite{Martucci:2005ht}
\bea \label{con1}
\gamma_{\underline{0..q}} \xi_+\,=\,\alpha^{-1}\,\xi_{(-)^{q+1}}
\eea
and
\bea \label{con2}
\hat{\gamma}^{'}_{(p-q)} \eta^{(2)}_{(-)^{p+1}}\,=\,\alpha\,\eta^{(1)}_{(-)^{q+1}}
\eea
where
\bea
\hat{\gamma}^{'}_{(r)}=\frac{1}{\sqrt{\det(g+\mathcal{F})}}\sum_{2l+s=r}\frac{\epsilon^{\alpha_1...\alpha_{2l}\beta_1...\beta_s}}{l!s!2^l}\mathcal{F}_{\alpha_1\alpha_2}\cdots\,\mathcal{F}_{\alpha_{2l-1}\alpha_{2l}}\hat{\gamma}_{\beta_1...\beta_s}~.
\eea
The four-dimensional gamma matrices $\gamma$ are real, the six-dimensional ones $\hat{\gamma}$ are antisymmetric and purely imaginary and underlined indices represent flat space. $\mathcal{F} \equiv 2\pi\ell_s^2\,F-B$ here stands for the gauge-invariant two-form that lives on the D3 brane. The D3 brane wrapped over a two-cycle corresponds to $q = 1$ and $p = 3$. In addition, from the discussion after (3.7) in \cite{Martucci:2005ht}, $\alpha = \pm 1$ in our case. 
       
We want to rephrase the conditions in terms of the geometrical objects $\Psi^+$ and $\Psi^-$ introduced in \cite{Martucci:2005ht}. For the branes which are strings in the 4d Minkowski space the equations \eqref{con1} and \eqref{con2} reduce to 
\bea \label{con1b}
\lbrace P[Re(i\,\Psi^+)] \wedge e^{\mathcal{F}} \rbrace_{(2)}=0
\eea
and
\bea \label{con2b}
\lbrace P[(\imath_m + g_{mn}\,\diff\,x^n\,\wedge)\Psi^-]\wedge\,e^{\mathcal{F}} \rbrace_{(2)}=0
\eea
respectively. The $\lbrace ... \rbrace_{(2)}$ denote that just the two-forms inside the brackets should be considered. 

Our warped background is a special case with $SU(3)$ structure\footnote{read section 5 in \cite{Martucci:2005ht} for more details} in which, from (5.2) in \cite{Martucci:2005ht}, $\Psi^{+}$ and $\Psi^{-}$ reads
\bea
\Psi^+\,=\,\frac{a\,\bar{b}}{8}\,e^{-iJ} \qquad , \, \Psi^-\,=\,-\,\frac{i\,a\,b}{8}\,\Omega \qquad \, \frac{a}{b}=e^{i\,\phi}~.
\eea
Following for example \cite{Gomis:2005wc}, it is easy to see that in the presence of a non-trivial warp factor $H$ the phase $\phi$ takes a fixed value such that $e^{i\,\phi}$ is purely imaginary\footnote{Note that in \cite{Gomis:2005wc} $e^{i\,\phi}=-1$, however, there the authors are working in euclidean signature in which $\gamma_{\underline{0123}}^2=1$ where here we have $\gamma_{\underline{0123}}^2=-1$.}. Therefore the equations \eqref{con1b} and \eqref{con2b} reduce to
\bea\label{con1c}
\lbrace P[Re(e^{-i\,J})] \wedge e^{\mathcal{F}} \rbrace_{(2)}=0
\eea
and
\bea\label{con2c}
\lbrace P[(\imath_m + g_{mn}\,\diff\,x^n\,\wedge)\Omega]\wedge\,e^{\mathcal{F}} \rbrace_{(2)}=0
\eea
respectively. Here $J$ is the almost complex structure with respect to which the six-dimensional metric
$g_{mn}$ is Hermitian and $\Omega$ is a $(3,0)$-form constructed from the spinor as explained in \cite{Martucci:2005ht}. The two equations \eqref{con1c} and \eqref{con2c} are easily translated to the SUSY conditions that read 
\bea 
\mathcal{F}\,=\,0\ ,
\eea 
and 
\bea \label{conD3}
\imath_m \Omega\,=\,0 \ .
\eea
The second condition is equivalent to demanding that the two-cycle is holomorphically embedded. In addition we see that we have to set $\mathcal{F}=0$ to have a brane wrapped over the two-cycle. Going back to (\ref{con1}), this condition is satisfied for half of the background spinors, so our strings are one-half BPS.

The $\mathcal{F}=0$ condition reduces to  
\bea \label{Mequ}
\int_{\mathbb{CP}^1}\,2\pi\ell_s^2\,F-B=4\pi^2\ell_s^2\,n-3b=0\ , \quad  n\in \Z
\eea
where we used \eqref{mult3} and denoted the quantized period of $F$ as $n$. From \eqref{bperiod} and using $T_F=1/2\pi\ell^2$ the fundamental string tension we see that $b \in [0,4\pi^2\ell_s^2)$. Thus we get that $\mathcal{F}=0$ for $(b=0,n=0)$, $(b=\frac{4\pi^2\ell_s^2}{3},n=1)$ and $(b=\frac{8\pi^2\ell_s^2}{3},n=2)$. We would like to identify these three special $B$-field values with the three values of the $B$-field implied by the gauge theory analysis. In the next subsection we will study the RR four-form fluctuations that are sourced by this wrapped brane.

Another interesting brane that one might consider is the E4 brane wrapping the blown-up four-cycle. We discuss in Appendix~\ref{app-E4} the SUSY conditions for such a brane. The various types of strings stretching between these branes can give rise to non-perturbative contributions to the superpotential of the field theory at the Calabi-Yau singularity. In
order for this non-perturbative superpotential to be generated at all the right number of zero modes must be present. An important remark here is that the Ep-Ep sector sees the full $\mathcal{N} = 2$ Calabi-Yau three-fold background, thus leading to four zero modes which are too many to saturate the $\mathcal{N} = 1$ superspace measure. However this brane might contribute to higher F-terms. As discussed in \cite{Beasley:2004ys,Beasley:2005iu}, instantons with four zero modes generate in general quantum deformations to the moduli space that correspond to the $N_f=N_c$ corrections in the corresponding gauge field theories. In our case we indeed observe such corrections in the field theory, however, these corrections are relevant in the middle of the RG flow at the scale of the given VEV and not in the far IR which is what we argued our supergravity solution corresponds to. This seems consistent as the volume of the E4 brane blows-up due to the warp factor. Since the contribution of the E4 brane to the F-term is proportional to $e^{-V}$, where $V$ is the volume of the cycle the branes wrap, one expects such contributions to indeed vanish. 

\subsubsection{The fluctuation containing the Goldstone mode} \label{sec:fluctuation}

We want to further examine the global string we have found. Specifically, we want to study the linearized backreaction
in the background due to the presence of this probe D3 brane. To linearized order this probe sources fluctuations in the RR four-form potential containing the term $a_2(x) \wedge W$ where $a_2$ is a two-form in $\R^{3,1}$ and $W$ is a closed two-form in the resolved cone. The latter is proportional to the volume form of the two-cycle in the bottom of the resolved cone wrapped by the D3 brane. The linearized equations of motion that should be satisfied are
\bea
\diff \delta G_5 = 0 \ , \quad \delta G_5 = \star\,\delta\,G_5~. 
\eea
The self-duality condition is satisfied by taking
\bea \label{flac}
\delta\,G_5=(1+\star)\diff(a_2(x)\,\wedge\,W)~.
\eea
Then the equations of motion reduce to
\bea
\diff\,\star_4\,\diff\,a_2=0 
\eea
and
\bea
\diff\,(H\,\star_6\,W)=0~,
\eea
where $\star_4$, $\star_6$ are the Hodge duals with respect to the unwarped Minkowski and resolved orbifold metrics
respectively. 

\subsubsection*{Solving for $W$}
From \cite{Krishnan:2008kv} we learn that the warp factor is a function of $\rho$ and $\sigma$ when the $N$ D3 branes sit on the bottom of the resolved cone. We start with the ansatz
\bea
W=\diff(f_1(\rho,\sigma)\,g_5+f_2(\rho,\sigma)e_\beta)~.
\eea
The $\diff(H(\rho,\sigma)\star_6\,W)=0$ condition reduces to
\bea \label{harm1}
\nn
&&\frac{4}{3}(3\,f_1-\cot\sigma\,f_2)H\,\cot\sigma+\frac{2}{3}(-6\,f_1+2\cot(2\sigma)f_2+\frac{\partial\,f_2}{\partial\sigma})H\,\cot\sigma+\\ 
&&+\frac{\partial}{\partial\sigma}(\frac{1}{3}\,\rho\,H(-6\,f_1+2\cot(2\sigma)f_2+\frac{\partial}{\partial\sigma}f_2))+\frac{\partial}{\partial\rho}(\frac{(-a^6+\rho^6)H\,\frac{\partial}{\partial\rho}f_2}{3\rho^3})=0
\eea
and
\bea \label{harm2}
\nn
&&4\rho(\cot\sigma\,f_2-3\,f_1)H+\frac{3\rho^7\,H(-3\cot\sigma+\tan\sigma)\frac{\partial}{\partial\sigma}f_1}{a^6-\rho^6}-\frac{\partial}{\partial\sigma}(\frac{3\rho^7\,H\,\frac{\partial}{\partial\sigma}f_1}{a^6-\rho^6})+\\
&&+\frac{\partial}{\partial\rho}(3\,\rho^3\,H\,\frac{\partial}{\partial\rho}f_1)+2\rho(-6\,f_1+2\,\cot(2\sigma)f_2+\frac{\partial}{\partial\sigma}f_2)H=0~.
\eea
We have two equations and two functions and together with the right boundary conditions these equations are expected to have a solution. We take the boundary conditions to be $f_1(\rho,\sigma) \sim f_1(\sigma)\rho^m$ and $f_2(\rho,\sigma) \sim f_2(\sigma)\rho^n$ with $m,n<0$ when $\rho$ goes to infinity. Since the warp factor behaves as $H(\rho,\sigma) \sim 1/\rho^4$ when $\rho$ goes to infinity, the equations \eqref{harm1} and \eqref{harm2} reduce to
\bea
\nn
&&f_2''(\sigma)+f_2'(\sigma) (3 \cot\sigma-\tan\sigma)-6\rho^{m-n} f_1'(\sigma)+\\
&&+f_2(\sigma) \left((n-2) (n-3) \csc^2\sigma-\sec ^2\sigma\right)=0
\eea
and
\bea
\nn
&&f_1''(\sigma)+f_1'(\sigma)\,3\cot\sigma-\tan\sigma+(m-4)(m+2)f_1(\sigma)+\\
&&+\frac{2}{3}\rho^{n-m}
   \left(f_2'(\sigma)+f_2(\sigma) (3\cot\sigma-\tan\sigma)\right)=0~.
\eea
These equations have decaying solutions when $m=-2>n$ and $f_1(\sigma)$ is a constant. Therefore for large $\rho$
\bea
W \sim \frac{3\rho(\diff\sigma\wedge e_{\beta}-e_{\theta}\wedge e_{\phi})+\diff\rho\wedge g_5}{\rho ^3}~.
\eea

On the other hand, when taking $\rho=a$ it is straight forward to see that the equations are solved for constant $f_1(a,\sigma)$ and $f_2(a,\sigma)=0$, thus the corresponding two-form proportional to the one in \eqref{2f-unwarped}. 

This is indeed the boundary conditions that one expects from the discussion in \cite{Martelli:2008cm}. From the discussion there we learn that $H^2_{L^2}(X,Hg_X)\cong H^2(X,\R)$. The following exact sequence
\bea
\nn
&&0\cong H^1(Y;\R)\rightarrow H^2(X,Y;\R)\rightarrow H^2(X;\R)\rightarrow \\
&&\rightarrow H^2(Y;\R) \rightarrow H^3(X,Y;\R)\cong 0
\eea
together with the fact that $b_4(X)=b_2(X)-b_3(Y)=b_2(X)=1$ show that the two-forms in $H^2(X;\R)$ restrict to trivial two-form class in $H^2(Y;\R)$. Recall that the pullback of the Thom class, which is represented by forms in $H^2_{cpt}(X;\R)$ that give 1 when integrated over the fiber, to the zero section $\mathbb{CP}^2$ is the Euler class, which in our case is just $[\text{vol}_{\mathbb{CP}^1}]$, where $\mathbb{CP}^1$ is the two-cycle inside $\mathbb{CP}^2$. Thus the pullback of the $L^2$ normalizable two-form with respect to the warped metric on $\mathbb{CP}^2$ is also two-form in this class. This then shows that the $\alpha_2 \wedge W$ part in this fluctuation is sourced electrically by D3 brane wrapping the $\mathbb{CP}^1$.

Following \cite{Klebanov:2007cx} we want to show that the fluctuations just discussed contain the Goldstone boson. We start by introducing the field $p(x)$ by dualizing the two-form $a_2$ 
\bea
 \star_4\,\diff\,a_2 = \diff\,p~.
\eea
Plugging this back into \eqref{flac}, the fluctuation in the five-form field strength reads
\bea
\delta\,G_5=\diff\,a_2\,\wedge\,W\,+\,\diff\,p\,\wedge\,h\,\star_6\,W~ 
\eea
and thus the fluctuation of the four-form potential is
\bea \label{dc4}
\delta\,C_4=a_2(x)\,\wedge\,W\,+\,p\,h\,\star_6\,W~.
\eea
For the conifold \cite{Klebanov:2007cx} it was shown that the $p(x)$ in \eqref{dc4} is the Goldstone boson for the broken baryonic $U(1)$. To show that, the authors demonstrated that the fluctuation couples through the Wess-Zumino term to the E4 condensate that corresponds to the VEV of the baryonic operator. Moreover, it was shown that the asymptotic behaviour of \eqref{dc4} in the UV corresponds to a VEV for the baryonic current in the field theory. 

For the broken emerging baryonic symmetry these checks cannot be repeated due to the fact that the gravity solution does not capture the intermediate non-conformal phase. Nevertheless, we interpret the $p(x)$ massless field as the Goldstone boson coming from the broken emerging baryonic symmetry. This indeed has a monodromy around the global string that electrically sources $a_2$. For a more elaborate discussion we refer the reader to \cite{Klebanov:2007cx}.

\section{The cone over $\mathbb{F}_0$}
As we mentioned in the introduction, intermediate non-conformal phases and emerging non-anomalous baryonic symmetries are expected to appear whenever the resolved geometry contains four-cycles. Thus, after completing the analysis for the simplest of such geometries, it will be interesting to check how our discussion applies to other more complicated examples. In this section we repeat the analysis for the cone over $\mathbb{F}_0$, which is just a $\Z_2$ freely acting orbifold of the confiold, and its dual field theory. The blowing-up of the four-cycle in such background and the corresponding Higgsing were studied in \cite{Benvenuti:2005qb}.
\subsection{Field theory description}
The cone over $\mathbb{F}_0$ is described through two Seiberg dual phases encoded in the following table
{\small
\begin{displaymath}
\begin{array}{||c | c||}
\hline\hline
Phase\, I & Phase\, II \\ \hline
\includegraphics[scale=.6]{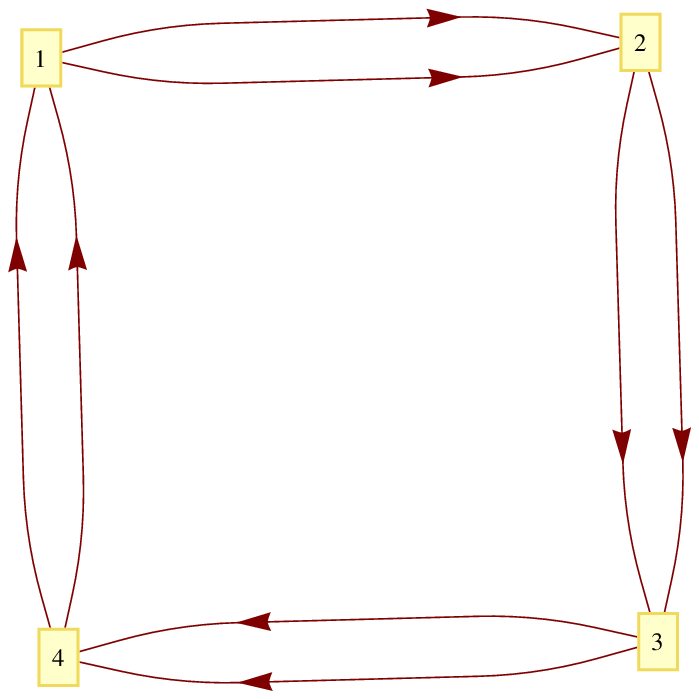}
& 
\includegraphics[scale=.7]{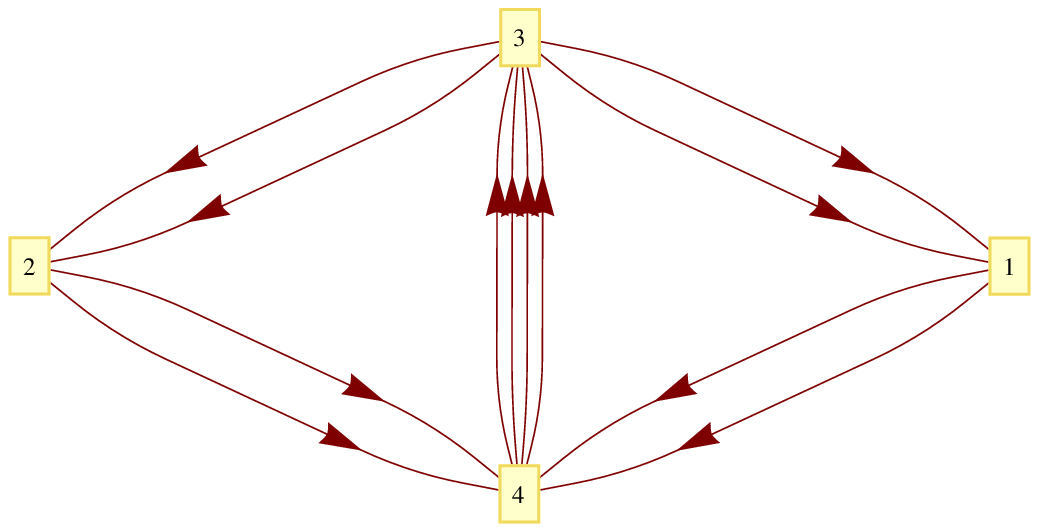}\\
W_I\,=\ \epsilon_{ij}\, \epsilon_{mn}\, X_{1,2}^i\, X_{2,3}^m\, X_{3,4}^j\, X_{4,1}^n & W_{II}\, =\, \epsilon_{ij}\, \epsilon_{mn}\, X_{3,2}^i\, X_{2,4}^m\, X_{4,3}^{jn}-\epsilon_{ij}\, \epsilon_{mn}\, X_{3,1}^m\, X_{1,4}^i\, X_{4,3}^{jn}\\
\hline\hline
\end{array}
\end{displaymath}
}

Here, and in the rest of the paper, we leave traces implicit. In the UV the gauge group is just $U(N)^4$ and there are two combinations of central $U(1)$ factors which are anomalous and one which is non-anomalous as can be seen from \eqref{anom}. These become global symmetries in the IR. Concentrating on Phase I, the anomalous $U(1)$s are generated by
\bea
\mathcal{A}_{a_1}=\mathcal{A}_1-\mathcal{A}_2-\mathcal{A}_3+\mathcal{A}_4  \ , \quad \mathcal{A}_{a_2}=\mathcal{A}_1+\mathcal{A}_2-\mathcal{A}_3-\mathcal{A}_4
\eea
and the non-anomalous 
\bea
\mathcal{A}_B=\mathcal{A}_1-\mathcal{A}_2+\mathcal{A}_3-\mathcal{A}_4~.
\eea
where $\mathcal{A}_i$ are the generators of the corresponding $U(1)$s in the quiver. The symmetries of Phase II can be analysed similarly.

In the following subsection we want to compute the chiral ring of the mesonic operators of the two phases, and using that show that the mesonic moduli space of these two phases is Sym$^N(\mathcal{C}_{\C}(\mathbb{F}_0))$\footnote{$\mathcal{C}_{\C}(\mathbb{F}_0)$ stands for the complex cone over $\mathbb{F}_0$.} as desired. 
\subsubsection{Chiral ring in Phase I}
Mesonic operators correspond to closed loops in the quiver. We can immediately write down 16 quartic objects corresponding to all possible length four loops around the quiver:
\begin{equation}
X^{ij,mn}_I=X^i_{12} X^m_{23} X^j_{34} X^n_{41}~.
\end{equation}
The R-charge 2 chiral operators are ${\rm Tr} X^{ij,mn}_I$, but there are only 9 of them. Applying the superpotential F-term relations to them,
\begin{equation}
X^1_{12} X^m_{23} X^2_{34} = X^2_{12} X^m_{23} X^1_{34}\ , \qquad X^1_{23} X^j_{34} X^2_{41}= X^2_{23} X^j_{34} X^1_{41}\ , \quad
etc.
\end{equation}
we find that the $SU(2)_1$ and $SU(2)_2$ indices are symmetrised. Therefore, these operators have $R=2$ and spins $J_1=J_2=1$. In general, the $R=2n$ chiral operators take the form
\begin{equation}
{\rm Tr} \prod_{a=1}^n X_I^{i_aj_a,m_a,n_a} \, ,
\end{equation}
with $SU(2)_1$ and $SU(2)_2$ indices symmetrised due to the F-term relations. These operators thus have spins $J_1=J_2=n$, matching the expectations from the algebraic geometry of $\mathbb{F}_0$.

\subsubsection{Chiral ring in Phase II}

Since in four dimensions this theory is a Seiberg dual of phase I, we expect to find the same spectrum of chiral operators. Let us work it out explicitly. As a warm-up, we write down the 9 spin $(1,1)$, $R=2$, gauge-invariant chiral operators
\begin{equation}
{\rm Tr} X^i_{14} X^{jm}_{43} X^n_{31} \, ,
\end{equation}
where $SU(2)_1$ and $SU(2)_2$ indices are symmetrised due to the F-term equations. These operators have $R=2$ due to marginality of the superpotential. There is an additional set of operators of the same form, where we change the gauge group index $1 \to 2$. They are equal to the operators above via the F-term relation
\begin{equation}
X^i_{32} X^m_{24} = X^m_{31} X^i_{14} \,.
\end{equation}
In general, the $R=2n$ chiral operators are given by
\begin{equation}
{\rm Tr} \prod_{a=1}^n X_{II}^{i_aj_a,m_a,n_a} \, ,
\end{equation}
where $X_{II}^{ij,mn} = X^i_{14} X^{jm}_{43} X^n_{31}$. Symmetrization over $SU(2)_1$ and $SU(2)_2$ indices follows from the superpotential F-term conditions, leading to spin $J_1=J_2=n$ and again matches the gravity result.
\subsubsection{Moduli space in Phase I and the space of FI parameters}
The master spaces for the two Seiberg dual phases are different. In general the master space is not Seiberg invariant, but we expect the physics to be. The master space is simply the space of VEVs one can give to the conformal theory, so there must be some physical explanation of the non-Seiberg invariance. Beasely and Plesser conjectured in their paper \cite{Beasley:2001zp} that there can be additional quantum relations in the chiral ring that then account for the differences. 

Actually we are interested here just in diagonal VEVs and therefore one can replace the baryon relation simply with the relation between bifundamental fields. Then this is just another F-term to impose i.e. one should add this new F-term to the set of classical ones, and then recompute the master space. So, if one introduce the new quantum relations for Phase II, one can prove that the master space of Phase II is now isomorphic to that of Phase I. To see that we refer to the results of \cite{Forcella:2008ng}, in particular section 4.1 where they discuss the $F_0$ theory, and its two phases. The irreducible component of the Phase I master space is the direct product of two copies of the conifold. Phase II has a completely different, much more complicated, master space.

However, in the notations of \cite{Forcella:2008ng}, the quantum relations are 
\bea
X_{3,1}^{ij} = X_{1,4}^j X_{4,3}^i = X_{1,2}^i X _{2,3}^j~.
\eea
This allows one to algebraically eliminate the $X_{3,1}$ fields completely. In the master space equations for Phase II in (4.22) in their paper, the first two F-term equations define again the direct product of two conifolds. However, the remaining 13 equations are completely redundant with the above relations. So, this proves with the additional relations above, the master spaces become equal. It is completely trivial that also the lower dimensional components work too. 

\begin{figure}
\begin{center}
\includegraphics[scale=1]{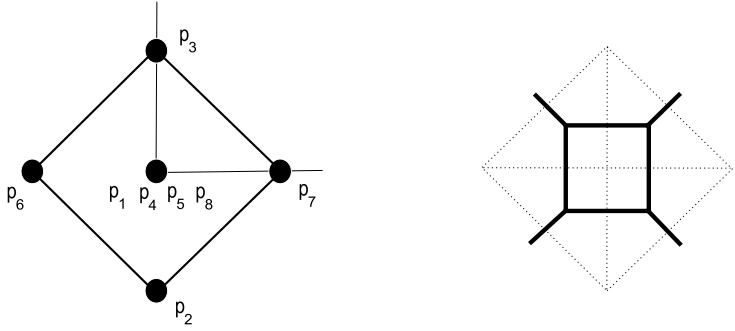}
\end{center}
\caption{The $\mathcal{C}_{\C}(\mathbb{F}_0)$ geometry. 1. The toric $\mathcal{C}_{\C}(\mathbb{F}_0)$ diagram. 2. The pq-web of the fully resolved $\mathcal{C}_{\C}(\mathbb{F}_0)$.}
\label{toric-F0}
\end{figure}

We have learnt that we can concentrate on Phase I. The relations between the $X_{i,j}^k$ fields to the $p$-fields is
\bea
\nn
&&X_{1,2}^1=p_1\,p_2 \ , X_{1,2}^2=p_1\,p_3 \ , X_{2,3}^1=p_5\,p_6 \ , X_{2,3}^2=p_5\,p_7 \\
&&X_{3,4}^1=p_2\,p_4 \ , X_{3,4}^2=p_3\,p_4 \ , X_{4,1}^1=p_6\,p_8 \ , X_{4,1}^2=p_7\,p_8 \ .
\eea
Using the forward algorithm one can show that 
\bea
Q_t=
\left(
\begin{array}{cccccccc|c}
p_1 & p_2 & p_3 & p_4 & p_5 & p_6 & p_7 & p_8 & FI \\ \hline
 0 & 1 & 1 & -1 & 0 & 1 & 1 & -3 & a \\
 0 & 1 & 1 & 0 & 0 & -1 & -1 & 0 & b \\
 0 & 2 & 2 & -3 & 0 & 0 & 0 & -1 & c \\
 0 & 0 & 0 & 0 & 1 & -1 & -1 & 1 & 0 \\
 1 & -1 & -1 & 1 & 0 & 0 & 0 & 0 & 0
\end{array}
\right)~.
\eea
By taking the null-space of this matrix one can verify that the toric diagram is the one we present in Figure~\ref{toric-F0}. This is indeed the toric diagram of the the $\mathcal{C}_{\C}(\mathbb{F}_0)$ singularity.

We will be interested mostly in the full resolution of the $\mathcal{C}_{\C}(\mathbb{F}_0)$ singularity that corresponds to the $\mathcal{O}(-2,-2)\rightarrow \mathbb{CP}^1 \times \mathbb{CP}^1$. The compact divisor in such resolved geometry is just $\mathbb{CP}^1 \times \mathbb{CP}^1$, this corresponds to the rectangle in the pq-web in Figure~\ref{toric-F0}.

We want to start with an analysis of the moduli space of the field theory. In a similar manner to the orbifold studied before the $Q_t$ matrix can be re-arranged this time in four different ways. These correspond to four different chambers in the master space

\begin{tiny}
\bea
\nn
Q_1=
\left(
\begin{array}{ccccccccc}
 1 & 0 & 0 & 0 & 0 & 0 & 0 & -1 & a+b+c \\ 
 0 & 1 & 1 & 0 & 0 & 0 & 0 & -2 & 3 a+3 b-c \\ 
 0 & 0 & 0 & 1 & 0 & 0 & 0 & -1 & 2 (a+b-c) \\ 
 0 & 0 & 0 & 0 & 1 & 0 & 0 & -1 & 3 a-b-c \\ 
 0 & 0 & 0 & 0 & 0 & 1 & 1 & -2 & 3 a-b-c 
\end{array} 
\right),\,
Q_2=
\left(
\begin{array}{ccccccccc}
 -1 & 0 & 0 & 0 & 0 & 0 & 0 & 1 & -a-b-c \\
 -2 & 1 & 1 & 0 & 0 & 0 & 0 & 0 & a+b-3 c \\
 -1 & 0 & 0 & 1 & 0 & 0 & 0 & 0 & a+b-3 c \\
 -1 & 0 & 0 & 0 & 1 & 0 & 0 & 0 & 2 (a-b-c) \\
 -2 & 0 & 0 & 0 & 0 & 1 & 1 & 0 & a-3 (b+c)
\end{array}
\right) 
\eea
\bea
\nn
Q_3=
\left(
\begin{array}{ccccccccc}
 1 & 0 & 0 & -1 & 0 & 0 & 0 & 0 & 3c-a-b \\
 0 & 1 & 1 & -2 & 0 & 0 & 0 & 0 & 3c-a-b \\
 0 & 0 & 0 & -1 & 0 & 0 & 0 & 1 & 2(c-a-b) \\
 0 & 0 & 0 & -1 & 1 & 0 & 0 & 0 & a-3 b+c \\
 0 & 0 & 0 & -2 & 0 & 1 & 1 & 0 & 3c-a-5 b
\end{array}
\right),\,
Q_4=
\left(
\begin{array}{ccccccccc}
 1 & 0 & 0 & 0 & -1 & 0 & 0 & 0 & 2 (c-a+b) \\
 0 & 1 & 1 & 0 & -2 & 0 & 0 & 0 & c-3 a+5 b \\
 0 & 0 & 0 & 1 & -1 & 0 & 0 & 0 & 3 b-a-c \\
 0 & 0 & 0 & 0 & -1 & 0 & 0 & 1 & c-3 a+b \\
 0 & 0 & 0 & 0 & -2 & 1 & 1 & 0 & c-3 a+b
\end{array}
\right)~.
\eea
\end{tiny}

In each case, using the 1st, 3rd and 4th rows in the $Q$ matrix, we can eliminate three of the internal points. This can be done for some region in the $\{a,b,c\}$ space for which the last column in the $Q$ matrix is non-negative. These regions are four different chambers in the FI parameter space spanned by $\{a,b,c\}$. After eliminating these internal points we need to impose the 2nd and 5th rows. The FI parameter in each such row corresponds to the size of one of the two $\mathbb{CP}^1$'s. Setting the p-field with charge $-2$ in these rows to zero corresponds to the $\mathbb{CP}^1 \times \mathbb{CP}^1$ zero section. Non-vanishing VEV to this p-field corresponds to moving the branes from the bottom of the (resolved) cone. 

In each chamber we have one combination of FI parameters which is orthogonal to the two FI that control the size of the two-cycles and this combination corresponds to the B-field period in supergravity. Indeed since $b_4=1$ for the resolved cone over $\mathbb{F}_0$, the B-field moduli is one-dimensional, i.e a circle.

We will be interested in keeping the branes on the blown-up four-cycle $\mathbb{CP}^1 \times \mathbb{CP}^1$ as this leads to the interesting intermediate non-conformal phases. We see that moving the branes from the four-cycle by giving VEVs to $p_8$, $p_1$, $p_4$ or $p_5$ in $Q_1$, $Q_2$, $Q_3$ or $Q_4$ respectively, leads to a non-vanishing VEV to the rest of the internal points. This then corresponds to a VEV of closed loop of fields in the quiver and thus to $\mathcal{N}=4$ $SU(N)$ SYM theory in the new IR.

\begin{figure}[ht] 
\begin{center} 
\includegraphics[scale=1.2]{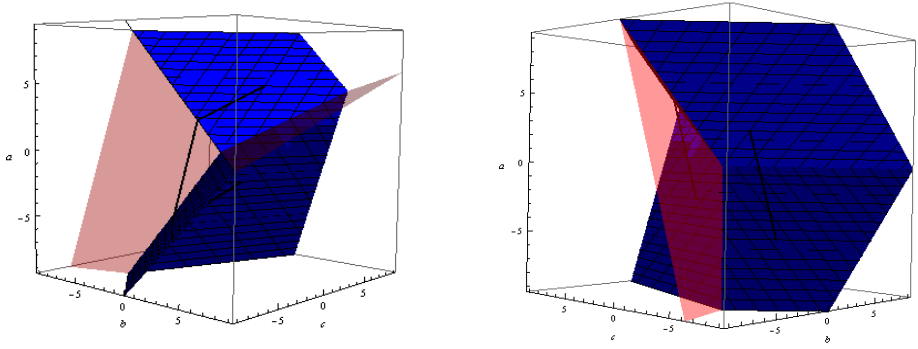}
\end{center}
\caption{The FI parameter space diagram for the $\mathcal{C}_{\C}(\mathbb{F}_0)$ theory from two different angles. a rotatable 3d diagram can be found in \cite{files}}
\label{GKZ-F0}
\end{figure}

\begin{figure}[ht]
\begin{center} 
\includegraphics[scale=0.7]{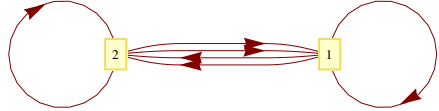}
\end{center}
\caption{The quiver diagram for the $\C \times \C^2/\Z_2$ conformal phase.}
\label{2node-quiver}
\end{figure}

The FI parameter space is presented in Figure~\ref{GKZ-F0}. We present two diagrams, each one rotated with respect to the other. For a rotatable 3d diagram, presented as Mathematica file, in which the details of the FI space can be seen more easily we refer the reader to \cite{files}. We have drawn the path of constant K\"ahler class. In the first chamber, for example, we just drew $3 a + 3 b - c = 10$ and $ 3 a - b - c = 10$. Indeed, this path is closed as expected from the periodicity of the B-field.
Points inside any of the chambers in the FI space correspond to the VEVs that initiate RG flows that end in $\mathcal{N}=4$ $SU(N)$ SYM theory in the IR. Red walls correspond to the $\C \times \C^2/\Z_2$ conformal phase. The quiver of this phase is presented in Figure~\ref{2node-quiver} 
and its superpotential reads
\bea
W=X_{1,1} X_{1,2}^1 X_{2,1}^2-X_{1,1} X_{1,2}^2 X_{2,1}^1+X_{1,2}^2 X_{2,1}^1 X_{2,2}-X_{1,2}^1 X_{2,1}2 X_{2,2}~.
\eea
One obtains this phase after blowing up just one of the $\mathbb{CP}^1$'s and putting the stack of D3 branes on the singular point in the bottom of the partly resolved Calabi-Yau three-fold. As we see, there are two such walls as there are two $\mathbb{CP}^1$'s that can remain blown-down. These walls meet at the origin of course, which is the singular geometry.

The blue meshed walls in the FI space correspond to VEVs that lead after RG flow to the non-conformal theory that was studied earlier in this paper and whose quiver is presented in Figure~\ref{HVZ-quiver}. The corresponding VEVs Higgs three adjacent $SU(N)$ factors in the quiver to one $SU(N)$. Notice that this Higgsing breaks one anomalous and one non-anomalous baryonic symmetry in the $\mathcal{C}_{\C}(\mathbb{F}_0)$ theory. As we claimed in the $\C^3/\Z_3$ section, the surviving anomalous baryonic symmetry becomes non-anomalous during the RG flow. The same discussion, as in Section~\ref{sec:noncon}, on the confining of the $N_c=N_f$ node and the breaking of the emerging non-anomalous symmetry applies also here. Each blue wall corresponds to a constant value of the FI parameter that is orthogonal to the two FI parameters that are dual to the K\"ahler classes. Thus, four critical values of the B-field period are expected to be observed in the supergravity. 

\begin{figure}[ht]
\begin{center} 
\includegraphics[scale=0.7]{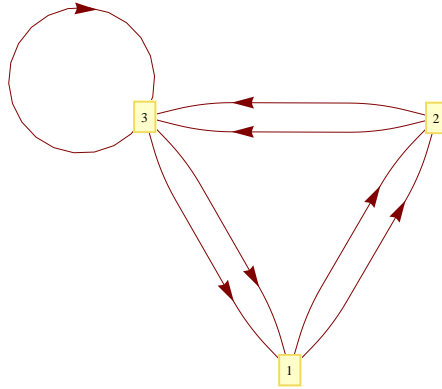}
\end{center}
\caption{The quiver diagram for the 3-noded non-conformal phase.}
\label{3node-quiver}
\end{figure}

Interestingly, the VEVs that correspond to the intersections between blue and red walls correspond to yet another non-conformal phase. The quiver of this phase is presented in Figure~\ref{3node-quiver},
and the superpotential is
\bea
W=X_{1,2}^1 X_{2,3}^1 X_{3,1}^2-X_{1,2}^1 X_{2,3}^2 X_{3,1}^1+X_{1,2}^2 X_{2,3}^2 X_{3,1}^1 X_{3,3}-X_{1,2}^2 X_{2,3}^1 X_{3,1}^2 X_{3,3}~.
\eea
The fact that this theory sits on a red wall in the FI space tells us that the stack of D3 branes probe a $\C \times \C^2/\Z_2$ singularity. Thus, the corresponding supergravity background is not smooth and contains $\Z_2$ singularities in the horizon. Thus, the supergravity solution itself cannot be trusted and additional twisted sectors should be added on top of it. We therefore leave this more complicated case to future study and concentrate on the non-conformal theory that corresponds to a smooth supergravity background. 

\subsection{Gravity description}

The most convenient form for the metric on the resolved $\mathcal{C}_{\C}(\mathbb{F}_0)$ for our purposes can be borrowed from Appendix \ref{sec:app-F0}
\begin{equation}
ds^2=U^{-1}\, dr^2+U\, g_5^2+\sum (r+\ell_i^2)\, \Big[d\theta_i^2+\sin^2\theta_id\phi_i^2\Big]
\end{equation}
where
\begin{equation}
U=\frac{r\,\Big\{ 3\, \ell_1^2\, (2\, \ell_2^2+r)+r\, (3\, \ell_2^2+2\,r)\Big\}}{3\, (\ell_1^2+r)\, (\ell_2^2+r)}
\end{equation}
and the $\ell_i$ are proportional to the K\"ahler classes. We will use the obvious vielbein in which $d\theta_i=e_{\theta_i}$ and $\sin\theta_i\, d\phi_i=e_{\phi_i}$.

\subsubsection{Gravity moduli: B-field and $C_2$}

We repeat the analysis that was done in Section~\ref{sec:2-form-moduli} for the resolved $\mathcal{C}_{\C}(\mathbb{F}_0)$ space. The B-field and $C_2$ moduli correspond to harmonic two-forms that are $L^2$ normalizable with respect to the unwarped metric. $b_4=1$ for the resolved $\mathcal{C}_{\C}(\mathbb{F}_0)$, thus we are looking for one such two-form.

Let us consider the following closed four-form
\begin{equation}
\omega_4=e_{\theta_1}\wedge e_{\phi_1}\wedge e_{\theta_2}\wedge e_{\phi_2}+d\Big(\, \sum\, f_i\, g_5\wedge e_{\theta_i}\wedge e_{\phi_i}\Big)
\end{equation}
and we will assume $f_i=f_i(r)$. Let us impose that $d\star\omega_4=0$. This translates into
\begin{equation} \label{form-eq}
\partial_r\Big( \frac{f_1'\, (r+\ell_2^2)}{(r+\ell_1^2)}\Big)=\frac{1+f_1+f_2}{(r+\ell_1^2)\, (r+\ell_2^2)}\, ,\quad \partial_r\Big( \frac{f_2'\, (r+\ell_1^2)}{(r+\ell_2^2)}\Big)=\frac{1+f_1+f_2}{(r+\ell_1^2)\, (r+\ell_2^2)}~.
\end{equation}
We can now consider the following ansatz
\begin{equation}
C_2=c\, \omega_2
\end{equation}
where $\omega_2=\star \omega_4$ and $c=c(x)$. Upon assuming that the field $c$ satisfies the free field equation in Minkowski space, the above fluctuation is a supergravity zero mode. We can of course consider $B_2=b(x)\, \omega_2$ and the same conclusions would go through. One can easily see that the solution to \eqref{form-eq} is
\bea
\nn
f_1(r)\,&=&1 - \frac{C_1}{6 (\ell_2^2 + r)} - 
 C_2 - \frac{( \ell_1^4 - \ell_1^2 \ell_2^2 + r (\ell_1^2 + r)) C_3}{\ell_2^2+r} +\\ \nn 
 &+& \frac{(2 \ell_1^4 \ell_2^2 + 2 \ell_1^2 r (3 \ell_2^2 + r) + r^2 (3 \ell_2^2 + r)) C_4]}{\ell_2^2 + r}\ , \\
f_2(r)\,&=&\,-\frac{C_1}{6 (\ell_1^2 + r)} + C_2 + r C_3 + r^2 C_4~.
\eea
Thus,
\bea
\label{generalform}
\nn
w_2\,&=&\,\big(\frac{(-\ell_1^2 - \ell_2^2 - 2 r) C_1}{6 (\ell_1^2 + r)^2 (\ell_2^2 + r)^2} - \frac{(\ell_1^2 - \ell_2^2)  C_3}{(\ell_2^2 + r)^2} +\frac{2(\ell_1^2 \ell_2^2 + r (2 \ell_2^2 + r)) C_4}{(\ell_2^2 + r)^2}\big) e_r \wedge g_5 - \\ \nn
&-& \frac{C_1 + 
       6 (\ell_1^2 + r) ((\ell_1^2 - 2 \ell_2^2 - r) C_3 + 
          2 (-\ell_1^2 \ell_2^2 + 3 \ell_2^4 + 3 \ell_2^2 r + r^2) C_4)}{6 (\ell_1^2 + r) (\ell_2^2 + 
     r)} e_{\theta_2} \wedge e_{\phi_2} -\\ 
 &-& \frac{C_1 + 6 (\ell_1^2 + r)^2 (C_3 + 2 r C_4)}{6 (\ell_1^2 + r) (\ell_2^2 + 
     r)} e_{\theta_1} \wedge e_{\phi_1}~.
\eea
The normalizable mode corresponds to taking $C_3=C_4=0$. We normalize this mode by integrating over the fiber
\bea
\int_{r=0}^{\infty}\,\int_{\psi=0}^{2\,\pi} - \frac{C_1 (\ell_1^2 + \ell_2^2 + 2 r)}{6 (\ell_1^2 + r)^2 (\ell_2^2 + r)^2} \diff r \, \diff \psi =\,1~.
\eea
Thus
\bea
C_1=-\frac{6 \ell_1^2 \ell_2^2}{2\pi}
\eea
and we get that the normalizable mode with respect to the unwarped metric is
\bea \label{two-form-trivial}
w^{N}_2 \equiv \frac{\ell_1^2\,\ell_2^2\, (\ell_1^2 + \ell_2^2 + 2 r) e_r \wedge g_5} {2\,\pi\,(\ell_1^2 + r)^2 (\ell_2^2 + 
   r)^2} + \frac{\ell_1^2\,\ell_2^2\, (e_{\theta_1} \wedge e_{\phi_1} + e_{\theta_2} \wedge e_{\phi_2})}{2\,\pi\,(\ell_1^2 + r) (\ell_2^2 + 
   r)}~.
\eea
It is easy to see that the pull back of this form to the four-cycle at $r=0$ is
\bea \label{pullbackF0}
w^{N}_2\mid_{r=0} = \frac{(e_{\theta_1} \wedge e_{\phi_1} + e_{\theta_2} \wedge e_{\phi_2})}{2\,\pi}
\eea
and when $r\rightarrow \infty$ 
\bea
w^{N}_2 \simeq \frac{\ell_1^2\,\ell_2^2}{2\,\pi\,r^2}(\frac{2 e_r \wedge g_5}{r} + e_{\theta_1} \wedge e_{\phi_1} + e_{\theta_2} \wedge e_{\phi_2})~.
\eea

One can extract an additional two-form from \eqref{generalform} which is not normalizable with respect to the unwarped metric but normalizable with respect to the warped one. Turning on a B-field and $C_2$ RR two-form, propotional to this two-form, in supergravity induces marginal deformations in the gauge couplings of the field theory that lives on the UV boundary. We will consider these modes to be turned off in the rest of the paper.

\subsubsection{D3 branes on the resolved cone and global strings}
The $\mathbb{F}_0$ field theory has one non-anomalous baryonic symmetry and two anomalous ones. The analysis now is more involved than that for the $\C^3/\Z_3$ theory since a distinction should be made between the Goldstone bosons coming from the non-anomalous symmetries breaking in the UV and the emerging one. 

Four-form fluctuations that solve the supergravity equations can be written as follows
\bea \label{flac2}
\delta\,C^4=a_2(x)\,\wedge\,W\,+\,p\,h\,\star_6\,W 
\eea
where $W$ is a two-form in the resolved cone. As explained in \cite{Martelli:2008cm}, using their notations, there are $b_3(Y)$ four-form fluctuations of type I where $Y$ is the Sasaki-Einstein base of the Calabi-Yau cone. For this type of fluctuation $W$ goes to a harmonic two-from on the $(Y,g_{Y})$ boundary at infinite $r$. These fluctuations were interpreted there, generalizing the conifold case \cite{Klebanov:2007cx}, as containing the Goldstone bosons that correspond to the broken non-anomalous $U(1)^{b_3(Y)}$ baryonic symmetry in the field theory. For the example discussed in this section $b_3(Y)=1$, thus we expect to find one such Goldstone boson.

The second type of four-form fluctuations is the one we encountered in the discussion for the $\C^3/\Z_3$ theory. These $b_4(X)$ four-forms are constructed with $W$'s of type III$^+$ in \cite{Martelli:2008cm} notations that vanish on $Y$. As for the $\C^3/\Z_3$ theory, we want to identify these fluctuations as the ones containing the Goldstone boson of the broken emerging baryonic symmetry. We expect to find one such Goldstone boson.

In the resolved cone over $\mathbb{F}_0$ the blown-up four-cycle, being a $\mathbb{CP}^1 \times \mathbb{CP}^1$, contains two topologically non-trivial $\mathbb{CP}^1$'s. In the next subsection we will motivate singling out the diagonal and anti-diagonal two-cycles. We will show that the D3-brane that wraps the anti-diagonal two-cycle sources just the Goldstone boson that comes from the baryonic symmetry in the UV. The D3 brane that wraps the diagonal two-cycle, however, sources just the Goldstone boson that comes from the emerging baryonic symmetry. 

\subsubsection{The fluctuation containing the Goldstone mode}
Following subsection \ref{sec:fluctuation} we want to study the fluctuations containing the Goldstone modes. From Section~\ref{sec:warp-F0} we see that the warp factor depends on $r$, $\theta_1$ and $\theta_2$ whenever the $N$ D3 branes sit in the bottom of the resolved cone. For the $W$ in \eqref{flac2} we can consider the following ansatz
\bea
W = \diff (f_2(r,\theta_1,\theta_2) g_5 + f_5(r,\theta_1,\theta_2) e_{\phi_1}+ f_6(r,\theta_1,\theta_2) e_{\phi_2})~.
\eea
For simplicity we will discuss the equations for $W$ just for $\ell_1=\ell_2=\ell$. The condition $\diff(H\,\star_6\,W)=0$ reduces to
\bea
\nn
&&\frac{\partial}{\partial_{\theta_2}}\Big( H(r,\theta_1,\theta_2) (f_2(r,\theta_1,\theta_2)-\frac{1}{\sin\theta_2}\frac{\partial}{\partial_{\theta_2}}(\sin\theta_2\,f_6(r,\theta_1,\theta_2))) \Big)-\\ \nn
&&-\frac{1}{\sin\theta_1}\frac{\partial}{\partial_{\theta_1}}\Big( \sin\theta_1\,H(r,\theta_1,\theta_2) \frac{\partial}{\partial_{\theta_1}}\,f_6(r,\theta_1,\theta_2) \Big)-\\
&&-\frac{\partial}{\partial_{r}}\Big( H(r,\theta_1,\theta_2) (\ell^2+r)\,U(r)\,\frac{\partial}{\partial_r}\,f_6(r,\theta_1,\theta_2) \Big)=0~,
\eea
\bea
\nn
&&\frac{\partial}{\partial_{\theta_1}}\Big( H(r,\theta_1,\theta_2) (f_2(r,\theta_1,\theta_2)-\frac{1}{\sin\theta_1}\frac{\partial}{\partial_{\theta_1}}(\sin\theta_1\,f_6(r,\theta_1,\theta_2))) \Big)-\\ \nn
&&-\frac{1}{\sin\theta_2}\frac{\partial}{\partial_{\theta_2}}\Big( \sin\theta_2\,H(r,\theta_1,\theta_2) \frac{\partial}{\partial_{\theta_2}}\,f_5(r,\theta_1,\theta_2) \Big)-\\
&&-\frac{\partial}{\partial_{r}}\Big( H(r,\theta_1,\theta_2) (\ell^2+r)\,U(r)\,\frac{\partial}{\partial_r}\,f_5(r,\theta_1,\theta_2) \Big)=0
\eea
and
\bea
\nn
&&\frac{1}{\sin\theta_2}\frac{\partial}{\partial_{\theta_2}}\Big( \sin\theta_2\,H(r,\theta_1,\theta_2)\frac{\ell^2+r}{U(r)} \frac{\partial}{\partial_{\theta_2}}\,f_2(r,\theta_1,\theta_2) \Big)-\\ \nn
&&-H(r,\theta_1,\theta_2) (f_2(r,\theta_1,\theta_2)-\frac{1}{\sin\theta_2}\frac{\partial}{\partial_{\theta_2}}(\sin\theta_2\,f_6(r,\theta_1,\theta_2))) + \\ \nn
&&\frac{1}{\sin\theta_1}\frac{\partial}{\partial_{\theta_1}}\Big( \sin\theta_1\,H(r,\theta_1,\theta_2)\frac{\ell^2+r}{U(r)} \frac{\partial}{\partial_{\theta_1}}\,f_2(r,\theta_1,\theta_2) \Big)-\\ \nn
&&-H(r,\theta_1,\theta_2) (f_2(r,\theta_1,\theta_2)-\frac{1}{\sin\theta_1}\frac{\partial}{\partial_{\theta_1}}(\sin\theta_1\,f_5(r,\theta_1,\theta_2))) + \\ 
&&+\frac{\partial}{\partial_{r}}\Big((\ell^2+r)^2\,H(r,\theta_1,\theta_2)\,\frac{\partial}{\partial_r}\,f_2(r,\theta_1,\theta_2) \Big)=0~.
\eea
Thus we obtain three equations for three unknown functions. These equations are too complicated to be solved analytically. However, it is reasonable that with the correct boundary conditions these solutions exist. 

Let us concentrate now on the boundary conditions. In the large $r$ limit, using the results in Appendix~\ref{sec:warp-F0}, we see that $H(r,\theta_1,\theta_2)\sim\frac{1}{r^2}$. Thus after writing the leading terms in the ansatz as
\bea
\tilde{\omega}_2 \simeq \diff \Big((C_1\,r^{m_1}+C_2\,r^{m_2}) g_5 + r^{n_1}\,f_{5}(\theta_1) e_{\phi_1}+ r^{n_1}\,f_6(\theta_2) e_{\phi_2}\Big)
\eea
the equations that should be satisfied have two solutions. The first corresponds to $m_1=-1$ and $m_2=-2$ and $n_1<-2$. In this case
\bea \label{bound1}
\tilde{\omega}_2 \sim \frac{e_r\,\wedge\,g_5+r(e_{\theta_1}\wedge\,e_{\phi_1}+e_{\theta_2}\wedge\,e_{\phi_2})}{r^2}~.
\eea
The second solution corresponds to $m_1=0,n_1=0$
\bea
\tilde{\omega}_2 \sim e_{\theta_1}\wedge\,e_{\phi_1}-e_{\theta_2}\wedge\,e_{\phi_2}~.
\eea

Obviously the boundary condition at $r=0$ of the $W$ that does not vanish at infinite $r$ is not unique. This is the case because one can always add to it the mode that vanishes at infinity. However for the $W$ that vanishes at infinite $r$ one can argue that $f_2(0,\theta_1,\theta_2)$ is constant and $f_5(0,\theta_1,\theta_2)=f_6(0,\theta_1,\theta_2)=0$ and this results in
\bea \label{bound2}
\tilde{\omega}_2(r=0) \sim e_{\theta_1} \wedge e_{\phi_1} + e_{\theta_2} \wedge e_{\phi_2}~.
\eea
This mode then can be related to the boundary condition at $r \rightarrow \infty$ using the shooting method. 
However, from the discussion in \cite{Martelli:2008cm} we learn that 
\bea \label{iso}
H^2_{L^2}(X,Hg_X)\cong H^2(X,\R)~. 
\eea
The following exact sequence
\bea
\nn
&&0\cong H^1(Y;\R)\rightarrow H^2(X,Y;\R)\rightarrow H^2(X;\R)\rightarrow \\
&&\rightarrow H^2(Y;\R) \rightarrow H^3(X,Y;\R)\cong 0
\eea
implies that one can choose a basis for $H^2(X;\R)$ such that $b_3(Y)=1$ and $b_4(X)=b_2(X)-b_3(Y)=1$ two-forms restrict to non-trivial and trivial two-form classes in $H^2(Y;\R)$ respectively. Recall that the pullback of the Thom class, which is represented by forms in $H^2_{cpt}(X;\R)$ that give 1 when integrated over the fiber, on the zero section $\mathbb{CP}^1_1 \times \mathbb{CP}^1_2$ is the Euler class, which in our case is just $[\text{vol}_{\mathbb{CP}^1_1}+\text{vol}_{\mathbb{CP}^1_2}]$. Thus from \eqref{iso} we see that the $L^2$ normalizable two-form with respect to the warped metric that vanishes at infinite $r$ is pulled-back in $r=0$ to a two-form in this class. This then shows that the $\alpha_2 \wedge W$ part in this fluctuation is sourced electrically by D3 brane wrapping the diagonal two-cycle. Note that this mode is not coupled to the D3 brane wrapping the anti-diagonal two-cycle.

As we already mentioned the boundary condition at $r=0$ of the $W$ that does not vanish at $r\rightarrow \infty$ is not unique. To fix the boundary condition we thus need to use the fact that the D3 brane that wraps the anti-diagonal two-cycle is special since the background compactly-supported B-field vanishes when pulled back to it, as can be seen from \eqref{pullbackF0}. This means that for a trivial gauge field on this brane $\mathcal{F}$ always vanishes and therefore this brane satisfies the SUSY conditions without any dependence on the VEV of the compactly-supported B-field. This is expected from the field theory analysis as the UV non-anomalous baryonic symmetry, in field theory that is dual to background with blown-up four-cycle, is always broken. 

Thus, the D3 brane that wraps the anti-diagonal two-cycle sources the RR four-form fluctuation that contains the Goldstone mode coming from the breaking of the baryonic symmetry in the UV. From the $\alpha_2 \wedge W$ part in this fluctuation we see that $W$ is expected to sit in the same class as the volume form of the anti-diagonal two-cycle. Thus we see that this fluctuation is not going to be coupled to the D3 brane wrapping the diagonal two-cycle. So we obtain two D3 branes that uniquely source the two distinct four-form fluctuations as desired.

The field theory analysis contains four walls that correspond to the intermediate non-conformal quiver described in Figure~\ref{HVZ-quiver}. During the RG flow there is an emerging non-anomalous baryonic symmetry in such a theory, as we argued. We claimed that the supergravity background correspond to the breaking of this symmetry by the VEV of the baryonic fields. We want to check now for which B-field periods the D3 brane can be wrapped on the diagonal two-cycle. Using \eqref{pullbackF0} we can write $\mathcal{F}$ on the D3 branes as follows
\bea 
\mathcal{F}\,=\,2\pi\ell_s^2\,n\,\frac{(e_{\theta_1} \wedge e_{\phi_1} + e_{\theta_2} \wedge e_{\phi_2})}{4}-b\,\frac{(e_{\theta_1} \wedge e_{\phi_1} + e_{\theta_2} \wedge e_{\phi_2})}{2\,\pi}\ , \quad n \in \Z
\eea
where $b \in [0,4\pi^2\ell_s^2)$. So $\mathcal{F}=0$ when $\{b=0,n=0\}$,$\{b=\pi^2\ell_s^2,n=1\}$,$\{b=2\pi^2\ell_s^2,n=2\}$ and $\{b=3\pi^2\ell_s^2,n=3\}$. These are exactly four critical values of the B-field, as one expected to see from the field theory analysis.

As a remark, for partial resolved cones in which just one of the $\mathbb{CP}^1$'s is blown-up, the same discussion for D3 branes wrapping one of the blown-up $\mathbb{CP}^1$'s results in two critical values for the B-field that satisfy the SUSY conditions. This is the same as the number of intersections of one of the red walls with blue walls in the FI parameter space. This suggests that in the RG flow that leads to the 3-noded non-conformal quiver the wrapped D3 branes also play an important role. Again, such singular backgrounds are more involved and we leave this subject for future work.
\section{The $\mathbb{C}^3/\mathbb{Z}_5$ orbifold}
So far we concentrated on examples with one four-cycle in the resolved geometry. It will be interesting to study more involved geometries in which there are more four-cycles in order to see how our proposal is generalizes. It is natural to start with the $\mathbb{C}^3/\mathbb{Z}_5$ orbifold which has two four-cycles. Since $b_3(S^5/\Z_5)=0$ the dual theory does not have non-anomalous baryonic symmetries. This simplifies the discussion since we avoid the need to distinguish between such symmetries and emerging ones. 

The space $\C^3/\Z_5$ is defined as the three-dimensional complex space $\C^3$ under the identification
\bea \label{Z5-weights}
\{x_1,x_2,x_2\}\sim\{w^2\,x_1,w^2\,x_2,w\,x_3\} \quad , \, w^5=1~.
\eea
The fixed point under this identification is just the origin, therefore the near horizon geometry close to the point-like $N$ D3 branes is smooth.

In this section we will carry out the first steps of such a study by analyzing the moduli space of the field theory. As the explicit metric of the resolved cone is not known, the analysis of the supergravity seems to be more involved. We hope to come back to this problem in the future. 
\subsection{Field theory description}
 
\begin{figure}[ht]
\begin{center} 
\includegraphics[scale=.5]{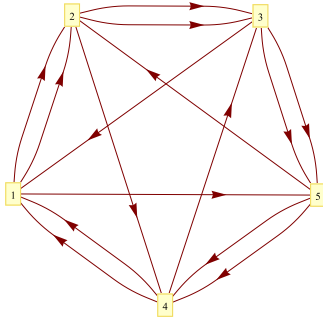} 
\end{center}
\caption{The quiver diagram for the $\mathbb{C}^3/\mathbb{Z}_5$ theory.}
\label{Z5-quiver}
\end{figure}

A stack of $N$ D3 branes propagating on the $\mathbb{C}^3/\mathbb{Z}_5$ orbifold is dual to a $(3+1)$d SCFT described by the quiver in Figure~\ref{Z5-quiver}.

\noindent
The field theory has a superpotential that reads
\bea
\nn
W=\,X_{1,2}^1 X_{2,3}^2 X_{3,1} -X_{1,2}^2 X_{2,3}^1 X_{3,1} -
  X_{2,3}^2 X_{3,5}^1 X_{5,2} + 
 X_{2,3}^1 X_{3,5}^2 X_{5,2} + X_{1,2}^2 X_{2,4} X_{4,1}^1 - \\
  X_{1,5} X_{5,4}^2 X_{4,1}^1 - 
 X_{1,2}^1 X_{2,4} X_{4,1}^2 + X_{1,5} X_{5,4}^1 X_{4,1}^2 -
  X_{3,5}^2 X_{5,4}^1 X_{4,3} + X_{3,5}^1 X_{5,4}^2 X_{4,3}~.
\eea
In the UV the gauge group is just $U(N)^5$. All four combinations of central $U(1)$ factors except the diagonal one,  which is decoupled, are anomalous as can be seen from \eqref{anom}. These become global symmetries in the IR. The anomalous $U(1)$s are generated by
\bea
\nn
&&\mathcal{A}_{a_1}=\mathcal{A}_1-\mathcal{A}_2 \ , \quad \mathcal{A}_{a_2}=\mathcal{A}_1+\mathcal{A}_2+\mathcal{A}_3+\mathcal{A}_4-4\mathcal{A}_5 \\
&&\quad \mathcal{A}_{a_3}=\mathcal{A}_3-\mathcal{A}_4 \ , \quad \mathcal{A}_{a_4}=\mathcal{A}_1+\mathcal{A}_2-\mathcal{A}_3-\mathcal{A}_4
\eea
where $\mathcal{A}_i$ are the generators of the corresponding $U(1)$s in the quiver.

\subsubsection{The GLSM description}

We can compute then the moduli space in the usual way. By making use of gauge rotations, we can set all the fields to be diagonal. Then, the effective theory reduces to $N$ copies of the $U(1)$ theory. As is standard, we can easily obtain the perfect matchings. The relations between the $X_{i,j}^k$ fields to the $p$-fields is encoded by
\bea
P=
\left(
\begin{array}{c | c c c c c c c c c c c c c} 
 & p_1 & p_2 & p_3 & p_4 & p_5 & p_6 & p_7 & p_8 & p_9 & p_{10} & p_{11} & p_{12} & p_{13} \\ \hline
 X_{1,2}^1 & 0 & 1 & 0 & 1 & 0 & 0 & 0 & 0 & 1 & 0 & 0 & 1 & 0 \\ 
 X_{1,2}^2 & 0 & 0 & 1 & 1 & 0 & 0 & 0 & 0 & 1 & 0 & 0 & 1 & 0 \\
 X_{1,5} & 1 & 0 & 0 & 1 & 1 & 0 & 1 & 0 & 1 & 0 & 0 & 0 & 0 \\
 X_{2,3}^1 & 0 & 1 & 0 & 0 & 0 & 0 & 1 & 1 & 0 & 0 & 1 & 0 & 0 \\ 
 X_{2,3}^2 & 0 & 0 & 1 & 0 & 0 & 0 & 1 & 1 & 0 & 0 & 1 & 0 & 0 \\ 
 X_{2,4} & 1 & 0 & 0 & 0 & 1 & 0 & 1 & 0 & 0 & 0 & 1 & 0 & 1 \\ 
 X_{3,1} & 1 & 0 & 0 & 0 & 1 & 1 & 0 & 0 & 0 & 1 & 0 & 0 & 1 \\ 
 X_{3,5}^1 & 0 & 1 & 0 & 1 & 1 & 1 & 0 & 0 & 0 & 0 & 0 & 0 & 0 \\ 
 X_{3,5}^2 & 0 & 0 & 1 & 1 & 1 & 1 & 0 & 0 & 0 & 0 & 0 & 0 & 0 \\ 
 X_{4,1}^1 & 0 & 1 & 0 & 0 & 0 & 1 & 0 & 1 & 0 & 1 & 0 & 0 & 0 \\ 
 X_{4,1}^2 & 0 & 0 & 1 & 0 & 0 & 1 & 0 & 1 & 0 & 1 & 0 & 0 & 0 \\ 
 X_{4,3} & 1 & 0 & 0 & 0 & 0 & 0 & 1 & 1 & 1 & 1 & 0 & 0 & 0 \\ 
 X_{5,2} & 1 & 0 & 0 & 0 & 0 & 0 & 0 & 0 & 1 & 1 & 0 & 1 & 1 \\ 
 X_{5,4}^1 & 0 & 1 & 0 & 0 & 0 & 0 & 0 & 0 & 0 & 0 & 1 & 1 & 1 \\ 
 X_{5,4}^2 & 0 & 0 & 1 & 0 & 0 & 0 & 0 & 0 & 0 & 0 & 1 & 1 & 1 \\ 
\end{array}~,
\right)
\eea
such that $X_{i,j}^k=\prod_{\ell} p_{\ell}^{P_{m,\ell}}$, where $m$ stands for the row that corresponds to the field $X_{i,j}^k$.

\begin{figure}[ht]
\begin{center}
\includegraphics[scale=2]{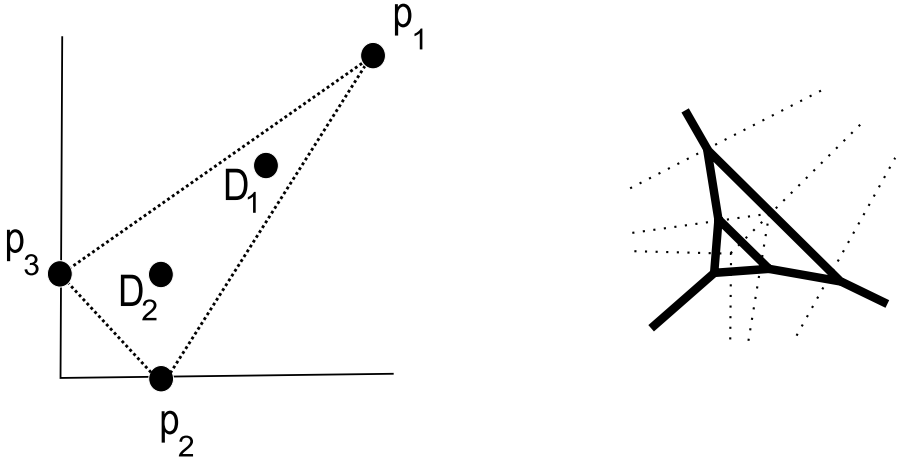} 
\end{center}
\caption{The $\mathbb{C}^3/\mathbb{Z}_5$ geometry. 1. The toric diagram. 2. The pq-web of the fully resolved $\mathbb{C}^3/\mathbb{Z}_5$ orbifold}.
\label{Z5-toric}
\end{figure}

Using the forward algorithm one can show that 
\bea
Q_t=
\left(
\begin{array}{cccccccccccccc}
p_1 & p_2 & p_3 & p_4 & p_5 & p_6 & p_7 & p_8 & p_9 & p_{10} & p_{11} & p_{12} & p_{13} & FI \\ \hline
 2 & 1 & 1 & 0 & 0 & 0 & 0 & 0 & 0 & -2 & -1 & 0 & -1 & a\\
 1 & 0 & 0 & 0 & 0 & 0 & 0 & 0 & 0 & -2 & 0 & 1 & 0 & b\\
 1 & -1 & -1 & 0 & 0 & 0 & 0 & 0 & 0 & 0 & 2 & 2 & -3 & c\\
 1 & 0 & 0 & 0 & 0 & 0 & 0 & 0 & 0 & -1 & -1 & 2 & -1 & d\\
 0 & -1 & -1 & 2 & -2 & 1 & 1 & 0 & -1 & 0 & 0 & 0 & 1 &\\
 0 & -1 & -1 & 1 & -1 & 1 & 1 & 0 & -1 & 0 & 0 & 1 & 0 &\\
 0 & -1 & -1 & 1 & -1 & 1 & 0 & 0 & 0 & 0 & 1 & 0 & 0 &\\
 0 & 0 & 0 & 1 & 0 & -1 & 0 & 0 & -1 & 1 & 0 & 0 & 0 &\\
 1 & 0 & 0 & 1 & -1 & 0 & 0 & 0 & -1 & 0 & 0 & 0 & 0 &\\
 0 & 0 & 0 & 0 & 1 & -1 & -1 & 1 & 0 & 0 & 0 & 0 & 0 &
\end{array}
\right) 
\label{Q-Z5}
\eea
and taking the null-space of this matrix we get
\bea
G_t=
\left(
\begin{array}{ccccccccccccc}
p_1 & p_2 & p_3 & p_4 & p_5 & p_6 & p_7 & p_8 & p_9 & p_{10} & p_{11} & p_{12} & p_{13} \\ \hline
 3 & 1 & 0 & 1 & 2 & 1 & 2 & 1 & 2 & 2 & 1 & 1 & 2 \\
 3 & 0 & 1 & 1 & 2 & 1 & 2 & 1 & 2 & 2 & 1 & 1 & 2 \\
 1 & 1 & 1 & 1 & 1 & 1 & 1 & 1 & 1 & 1 & 1 & 1 & 1 \\
\end{array}
\right)~.
\eea
The columns of the $G_t$ matrix encodes the vertices in the toric diagram which is presented in Figure~\ref{Z5-toric}.
Indeed, using the Delzant construction, we see that this is the toric diagram of the $\mathbb{C}^3/\mathbb{Z}_5$ orbifold described by \eqref{Z5-weights}. The faces in the pq-web correspond to the exceptional divisors in the fully resolved geometry. The triangle is the projective plane $\mathbb{CP}^2$ and the rectangle is $\mathbb{F}_2$, which is a $\mathbb{CP}^1$-bundle over $\mathbb{CP}^1 \subset \mathbb{CP}^2$. Notice that the intersection of the two four-cycles is the $\mathbb{CP}^1 \subset \mathbb{CP}^2$. This can be seen from the pq-web as the common edge that the two faces share. We refer the reader to \cite{Mukhopadhyay:2001sr} for more details on this resolved geometry.

\subsubsection{$\C^3/\Z_5$ theory - Moduli space}
The first four rows in \eqref{Q-Z5} correspond to D-terms. Resolving the singularity and turning on compactly-supported B-field in the background corresponds to adding FI parameters to these terms. After simple redefinitions of the linear sigma model gauge groups\footnote{i.e redefining rows in the charge matrix as independent linear combinations of rows.} and the following change of variables $a\rightarrow x + y, c \rightarrow x - y$ one obtains the following charge matrix
\bea
Q_t=
\left(
\begin{array}{cccccccccccccc}
 0 & 1 & 1 & 0 & 0 & 0 & 0 & 0 & 0 & 0 & 0 & -3 & 1 & \frac{1}{5} (-b-6 d+3
   x+5 y) \\
 1 & 0 & 0 & 0 & 0 & 0 & 0 & 0 & 0 & 0 & 0 & 1 & -2 & \frac{1}{5} (-3 b+2
   d+4 x) \\
 0 & 0 & 0 & 1 & 0 & 0 & 0 & 0 & 0 & 0 & 0 & -1 & 0 & \frac{1}{5} (-b-d+3
   x+5 y) \\
 0 & 0 & 0 & 0 & 1 & 0 & 0 & 0 & 0 & 0 & 0 & 0 & -1 & \frac{1}{5} (-b-d+3
   x+5 y) \\
 0 & 0 & 0 & 0 & 0 & 1 & 0 & 0 & 0 & 0 & 0 & -1 & 0 & \frac{1}{5} (x-2
   (b+d))+y \\
 0 & 0 & 0 & 0 & 0 & 0 & 1 & 0 & 0 & 0 & 0 & 0 & -1 & -\frac{2}{5} (b+d-3 x)
   \\
 0 & 0 & 0 & 0 & 0 & 0 & 0 & 1 & 0 & 0 & 0 & -1 & 0 & \frac{4 x}{5}-\frac{3
   (b+d)}{5} \\
 0 & 0 & 0 & 0 & 0 & 0 & 0 & 0 & 1 & 0 & 0 & 0 & -1 & \frac{1}{5} (-3 b+2
   d+4 x) \\
 0 & 0 & 0 & 0 & 0 & 0 & 0 & 0 & 0 & 1 & 0 & 0 & -1 & \frac{1}{5} (-4 b+d+2
   x) \\
 0 & 0 & 0 & 0 & 0 & 0 & 0 & 0 & 0 & 0 & 1 & -1 & 0 & \frac{1}{5} (b-4 d+2
   x)
\end{array}
\right)~.
\label{Q-Z5-2}
\eea
We can use the 3rd to last rows in \eqref{Q-Z5-2} to eliminate the internal points $p_4$-$p_{11}$ in terms of $p_{12}$ and $p_{13}$. This can be done when the last entry in these rows is non-negative. By redefining $Q_t$ , similar to what we did for $\C^3/\Z_3$ and $\mathbb{F}_0$, we can define $24$ additional $Q_t$ matrices for which these rows in the charge matrix can be used to eliminate different sets of internal p-fields. We present these matrices in Appendix~\ref{sec:App-Z5}. 

\begin{figure}[ht]
\begin{center} 
\includegraphics[scale=.5]{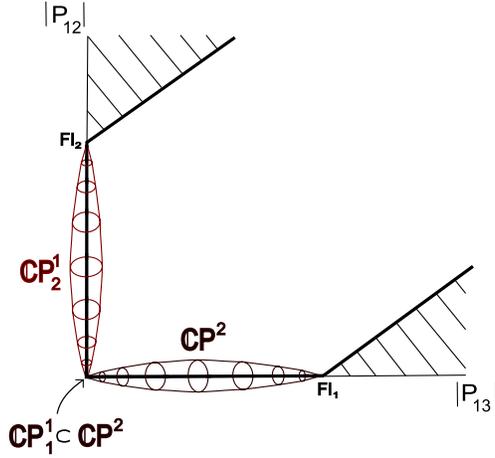} 
\end{center}
\caption{The mesonic moduli space of the $\mathbb{C}^3/\mathbb{Z}_5$ theory. FI$_1$ and FI$_2$ correspond to the FI parameters in the 1st and 2nd rows of \eqref{Q-Z5-2} respectively.}
\label{Z5-resolved}
\end{figure}

Coming back to \eqref{Q-Z5-2}, after eliminating the internal points mentioned, we are left with the first two rows that encode the geometry. The first row in \eqref{Q-Z5-2} corresponds to the $\mathbb{CP}^2$ while the second row corresponds to $\mathbb{CP}_2^1$ fibered over the $\mathbb{CP}_1^1 \subset \mathbb{CP}^2$, where the $\mathbb{CP}^2$ corresponds to $p_{13}=0$. Taking non-vanishing FI parameters in these rows corresponds to the case in which the geometry is completely resolved. The mesonic moduli space in this case is presented in Figure~\ref{Z5-resolved}. Note that the circles around the $|p_{13}|$ axes correspond to directions that parametrize the $\mathbb{CP}^2$ that were suppressed. The circles around the $|p_{12}|$ axes correspond to the phase of $p_{12}$ that are fibered over the line to form $\mathbb{CP}^1$.
Indeed, the two two-cycles, $\mathbb{CP}_1^1$ and $\mathbb{CP}_2^1$, are generators for the $H_2(X,\Z)$ for the resolved orbifold. 

Putting the stack of D3 branes in the corner of the resolved geometry is somehow a special case. This corresponds in the field theory to setting $p_{12}=p_{13}=0$ as can be seen from Figure~\ref{Z5-resolved}. In this point the stack of D3 branes intersect the two two-cycles at a point and thus the D3 branes wrapped over these cycles form strings in the space in which the IR field theory lives. Thus in this section we will explore this part of the moduli space. We leave the study of other scenarios to future work.

\begin{figure}[ht]
\begin{center}
\includegraphics[scale=2]{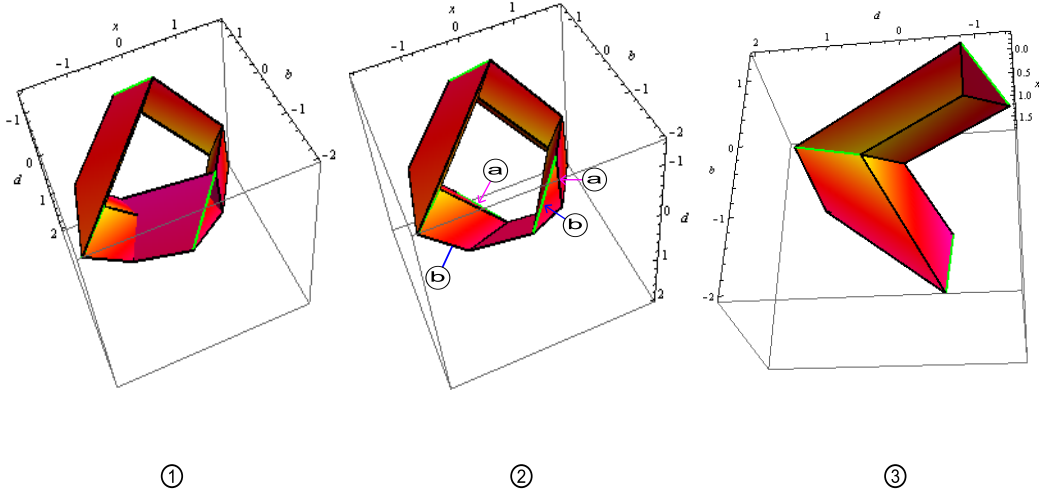} 
\end{center}
\caption{1. The $\C^3/\Z_5$ FI parameter space. 2. The $\C^3/\Z_5$ FI parameter space not including two chambers. The border lines of the surface space(each represents a chamber) that were deleted are labeled by (a) and (b) for the 1st and 2nd chamber respectively. Notice that for (a) the chopped arrow corresponds to a line that cannot be seen in this figure. From the color of the surfaces that were deleted one sees that in the four-dimensional space the 1st and 2nd deleted chambers intersect with the other chambers solely in the lines that are labeled by (a) and (b) respectively. 3. A partial section in the moduli space for the convenience of the reader. A rotatable Mathematica diagrams are available in \cite{files}.}
\label{GKZ-Z5}
\end{figure}

Recall that in addition to two FI parameters corresponding to the K\"ahler moduli there are $b_2(X)=2$ FI parameters that correspond to B-field moduli. We want to fix the K\"ahler class, so the $25$ chambers describe the B-field moduli which is 2-dimensional embedded in 4-dimensional space. We describe this moduli space as a two-dimensional surface in Figure~\ref{GKZ-Z5} where we encode the 4th coordinate as the color of each point in the 3-dimensional space. Rotatable Mathematica diagrams are available in \cite{files}.
We see that there are two cycles in the FI parameter space - the shape of one is a triangle and the other is pentagon. Indeed one expects to find a torus-like FI parameter space as the two compactly-supported B-fields modes are periodic. However the structure of the FI space encodes more information than just the periodicity of the B-field. This structure is due to the critical values of FI parameters that lead to intermediate non-conformal phases. In the following subsection we discuss in more detail the appearance of such phases in this moduli space.

\subsection{Non-conformal phases in $\C^3/\Z_5$}
\begin{figure}[ht]
\begin{center}
\includegraphics[scale=1]{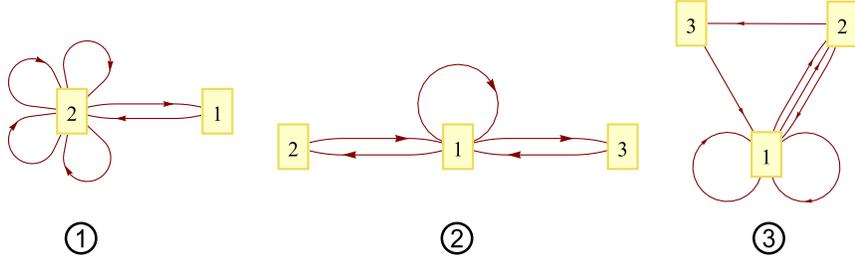} 
\end{center}
\caption{The three non-conformal quiver diagrams obtained from the Higgsing of the $\C^3/\Z_5$ theory.}
\label{Z5-noncon}
\end{figure}

A general point on the two-dimensional surface in Figure~\ref{GKZ-Z5} corresopnds to a VEV that initiate RG flow that ends in the $\mathcal{N}=4$ $SU(N)$ SYM theory in the IR. In addition, there are four non-conformal phases that appear in the part of the moduli space that we just plotted. The first is described by the quiver in Figure~\ref{HVZ-quiver} and the related superpotential \eqref{HVZ-W}. This is the same theory that we found when Higgsing the $\C^3/\Z_3$ and $\mathbb{F}_0$ theories. This theory corresponds to the black lines in Figure~\ref{GKZ-Z5}. The quivers of the other phases are described in Figure~\ref{Z5-noncon}.

\noindent
with the following superpotentials
\bea
\nn
&&W_1=X_{1,2}\,X_{2,1}\,X_{2,2}^1\,-\,X_{2,2}^2\,X_{2,2}^3\,X_{2,2}^1\,-\,X_{1,2}\,X_{2,1}\,X_{2,2}^4\,+\,X_{2,2}^2\,X_{2,2}^4\,X_{2,2}^3\ .
\\ \nn
&&W_2=X_{3,1}\,X_{1,3}\,X_{3,2}\,X_{2,3}\,X_{3,3}-\,X_{3,2}\,X_{2,3}\,X_{3,1}\,X_{1,3}\,X_{3,3}\ .
\\
&&W_3=X_{1,1}^1\,X_{1,2}^1\,X_{2,1}\,-\,X_{1,1}^1\,X_{1,1}^2\,X_{1,2}^2\,X_{2,1}\,- X_{1,2}^1\,X_{2,3}\,X_{3,1}\,+\,X_{1,1}^2\,X_{1,2}^2\,X_{2,3}\,X_{3,1}~.
\eea
The 1st theory, corresponding in Figure~\ref{GKZ-Z5} to the green lines, is similar to the theory in Figure~\ref{HVZ-quiver} with two additional adjoint fields. Following the same discussion we see that this theory also has just one non-anomalous baryonic symmetry coming from one non-broken anomalous baryonic symmetry in the UV. After node 1 confines the superpotential reads
\bea
\nn
W_1&=&\mathcal{M}\,(X_{2,2}^1-X_{2,2}^4)\,-\,X_{2,2}^2\,X_{2,2}^3\,X_{2,2}^1\,+\,X_{2,2}^2\,X_{2,2}^4\,X_{2,2}^3+ \\
&+&\lambda\, \Big({\rm det}\,\mathcal{M}-\mathcal{B}\, \tilde{\mathcal{B}}-\Lambda^{2N}\Big)~. 
\eea
where 
\bea
\nn
&&\mathcal{M}^i_j=(X_{1,2})^j_{\alpha}\,(X_{2,1})^{\alpha}_i\, ,\qquad \mathcal{B}^{i_1\cdots i_N}=\epsilon^{\alpha_1\cdots \alpha_N}\, (X_{1,2})^{i_1}_{\alpha_1}\cdots (X_{1,2})_{\alpha_N}^{i_N}\, , \\ &&\tilde{\mathcal{B}}_{i_1\cdots i_N}=\epsilon_{\alpha_1\cdots \alpha_N}\, (X_{2,1})_{i_1}^{\alpha_1}\cdots (X_{2,1})^{\alpha_N}_{i_N}
\eea
where greek indices stand for $SU(N)_1$ while latin ones stand for $SU(N)_2$.

We see that $\mathcal{M}$ is massive and therefore this time we are left just with the baryonic branch. After setting $\mathcal{B}\, \tilde{\mathcal{B}}=-\Lambda^{2N}$ the effective superpotential in the IR reads
\bea
&&W=\,X_{2,2}^2\,X_{2,2}^+\,X_{2,2}^3-\,X_{2,2}^2\,X_{2,2}^3\,X_{2,2}^+
\eea
where $X_{2,2}^+\equiv \frac{1}{2}(X_{2,2}^1+X_{2,2}^4)$. Thus we obtain the $\mathcal{N}=4$ $SU(N)$ SYM theory as before. In the supergravity side one expects to find a Goldstone boson and the related global string coming from the breaking of the emerging non-anomalous symmetry.

The 2nd theory, corresponding in Figure~\ref{GKZ-Z5} to the vertices that connect four black lines in each of the five triangular junctions, and the 3rd one sits in the vertices that connect green and black lines in each of the five triangular junctions. The Higgsings in the UV that lead to the 2nd and 3rd theories correspond to two non-broken anomalous baryonic symmetries in the $\C^3/\Z_5$ theory. These, we argue, become non-anomalous during the RG flow. We see that the 2nd and 3rd theories sits in the moduli space in points that are expected to correspond to supergravity backgrounds in which both B-field modes obtain critical values. This is expected as, for critical B-field values, one can now wrap D3 branes on the two blown-up two-cycles. These then correspond to two global strings that comes from the breaking of the two emerging non-anomalous baryonic symmetries. However, a more detailed study of the supergravity should be done to see if indeed the number of critical B-field values matches with the field theory FI space. 

\section{Final comments and summary}
In this paper we studied the moduli spaces of several field theories that are dual to D3 branes probing toric Calabi-Yau three-folds that contain four-cycles in their resolution. We have shown that there is a remarkable agreement between the moduli space in the field theory and supergravity. One new result in this paper is the fact that the directions in the moduli space of the field theory which are dual to the B-field moduli are indeed periodic. This was demonstrated for the three examples discussed in this paper. For a general resolved Calabi-Yau space $X$, one expects $b_4(X)$ such directions.

The main aim of this paper was to show that there is extra information in the field theory moduli space that singles out critical values of the compactly-supported background B-field. In the field theory these critical values correspond to RG flows that result in intermediate non-conformal phases. We claimed that part of the surviving anomalous baryonic symmetries in the UV become non-anomalous during the RG flow with dependence on the B-field in the background. We focused on the fully resolved geometries. However there are non-conformal phases that appear when the geometry is just partly resolved. It will be interesting to study this in more detail. 

One can consider giving non-vanishing VEVs to mesonic operators in the confined theories. This was done in \cite{Krishnan:2008kv} for the $\C^3/\Z_3$ orbifold. This seems to correspond to moving the branes from the bottom of the resolved cone. More interesting scenarios seem to appear when one keeps the branes in the bottom of the cone by choosing vanishing VEVs for mesonic operators and non-vanishing ones for baryonic operators in the confined theory. In this case the D3 branes that form global strings intersect the worldvolume of the stack of D3 branes. 

We made a distinction between two different types of D3 branes that wrap two-cycles and form global strings in the Minkowski space. The first type corresponds to D3 branes that source fluctuations that contain the Goldstone bosons coming from the broken non-anomalous baryonic symmetries in the UV. There are $b_3(Y)$ such branes corresponding with the $U(1)^{b_3(Y)}$ baryonic symmetry in the field theory \cite{Martelli:2008cm}. The second type of D3 branes, we suggest, source fluctuations that contain the Goldstone bosons coming from broken emerging non-anomalous baryonic symmetries. In general, one expects to see $b_4(X)=b_2(X)-b_3(Y)$ such D3 branes. These branes can be wrapped just for critical values of the $b_4(X)$ B-field moduli, as explained. 

\subsection*{Acknowledgments}
\noindent 
I thank Amihay Hanany for discussions and Dario Martelli, Diego Rodriguez-Gomez and James Sparks for collaborations at the initial stages of this project. I am very grateful to James Sparks for many helpful discussions and for reading and commenting on the manuscript.
I would also like to thank ISEF for their support; this work was completed with the support of a University of Oxford Clarendon Fund Scholarship.

\begin{appendix}
\section{E4 branes wrapping four-cycles} \label{app-E4}
Let us consider a Euclidean brane wrapping the blown-up four-cycle. Being $\mathbb{CP}^2$, which is not a spin manifold, this will suffer from the Freed-Witten anomaly \cite{Freed:1999vc}. In particular, to cancel this global anomaly requires a half-integer worldvolume field through $\mathbb{CP}^1\subset \mathbb{CP}^2$. Recall that the gauge-invariant quantity on the worldvolume of the brane is $\mathcal{F}=2\pi\,\ell_s^2\, F-B$. 
The SUSY conditions for this worldvolume field were discussed in \cite{Marino:1999af}. Here we have a Euclidean D3 brane 
wrapping a divisor in a CY 3-fold, and in this case the SUSY condition and EOM on $\mathcal{F}$ is that it should be 
primitive, so that $J\wedge\mathcal{F}=0$, and also Hodge type $(1,1)$. Of course, it is also closed (in the absence of 
strings ending on the E3-brane). These conditions imply in particular that $\mathcal{F}$ is anti-self-dual on the 
four-cycle and harmonic. However, for $\mathbb{CP}^2$ there are no such harmonic forms, so the worldvolume SUSY conditions 
imply that $\mathcal{F}=0$. We must then turn on a background $B$-field to cancel the half-integer flux of $F$ that 
cancels the FW anomaly:
\begin{equation}
B_2=b_0\,\omega_2\, ,\qquad b_0\sim b_0+\frac{2\pi n}{T_F}
\end{equation}
where now $b_0$ is a constant. In order to cancel the anomalous term to obtain $\mathcal{F}=0$, we have to tune this constant such that
\begin{equation}
\frac{T_F}{2\pi}\int_{\mathbb{CP}^1}\, B_2=\frac{1}{2}+m~, \qquad  m\in\mathbb{Z}~.
\end{equation}
Here of course $T_F=1/2\pi \ell_s^2$ is the fundamental string tension.
It is straightforward to perform the integral, getting that
\begin{equation}
T_F\int_{\mathbb{CP}^1}B_2=3\,T_F\,b_0
\end{equation}
which is, as it should be, 3 times the integral of $B_2$ over $\{\rho,\, \psi\}$. From here, we have
\begin{equation}
b_0=\left(\frac{1}{6}+\frac{m}{3}\right)\frac{2\pi}{T_F}~.
\end{equation}
From here we can read that there are actually three inequivalent values of $b_0$ in its  periodicity range (corresponding to $m=0,\,1,\,2$)
\begin{equation}
b_0^{(0)}=\frac{1}{6}\cdot \frac{2\pi}{T_F}\,  ,\qquad b_0^{(1)}=\frac{3}{6}\cdot \frac{2\pi}{T_F}\,  ,\qquad b_0^{(2)}=\frac{5}{6}\cdot \frac{2\pi}{T_F}~.
\end{equation}
In particular, the integral of $B_2$ on the $\mathbb{CP}^1\subset\mathbb{CP}^2$ becomes

\begin{equation}
T_F\int_{\mathbb{CP}^1}B_2=k\, \pi\, , \qquad k=1,\,3,\,5~.
\end{equation}
On the other hand, the volume of the cycle is $\pi\, a^2$. Thus, the complexified K\"ahler parameter is, for each choice of $k$

\begin{equation}
\xi=\pi\, a^2+i\,\frac{k\, \pi}{T_F}\, , \qquad k=1,\,3,\,5~.
\end{equation}

\section{Geometry of $\mathcal{C}_{\C}(\mathbb{F}_0)$ and its resolutions} 
\subsection{Metric for the resolved $\mathcal{C}_{\C}(\mathbb{F}_0)$} \label{sec:app-F0}

The $\mathcal{C}_{\C}(\mathbb{F}_0)$ can be seen as a certain $\mathbb{Z}_2$ orbifold of the conifold. Thus, we consider
\begin{equation}
\{z_i\,\in \mathbb{C}\,\,; i=1\cdots4\, |\, z_1\, z_2-z_3\, z_4=0\}~.
\end{equation}
The above relation can be solved by taking
\bea
\label{zs}
\nn
&&z_1=r\, e^{\frac{i}{2}\, (\psi-\phi_1-\phi_2)}\, \sin\frac{\theta_1}{2}\, \sin\frac{\theta_2}{2} \ , \quad z_3=r\, e^{\frac{i}{2}\, (\psi+\phi_1-\phi_2)}\, \cos\frac{\theta_1}{2}\, \sin\frac{\theta_2}{2}\\ \nn
&&z_2=r\, e^{\frac{i}{2}\, (\psi+\phi_1+\phi_2)}\, \cos\frac{\theta_1}{2}\, \cos\frac{\theta_2}{2} \ , \quad z_4=r\, e^{\frac{i}{2}\, (\psi-\phi_1+\phi_2)}\, \sin\frac{\theta_1}{2}\, \cos\frac{\theta_2}{2}~.
\eea
Assuming $\theta_i\in[0,\pi]$, $\phi_i=[0,2\pi]$; if we consider $\psi\in[0,4\pi]$ we have $\mathcal{C}(T^{1,1})$, while if we consider $\psi\in [0,2\pi]$ we have $\mathcal{C}_{\C}(\mathbb{F}_0)$. Since we are interested in a K\"ahler manifold, the most generic K\"ahler potential we can write is
\begin{equation}
K=F(r^2)+4\, a^2\, \log(1+|\lambda|^2)~.
\end{equation}
Here $\lambda$ is a local coordinate on the $\mathbb{P}^1$ which we can choose to blow-up by setting a non-zero $a$. Imposing Ricci-flatness leads to
\begin{equation}
F'\, (F'+r^2\, F'')\, (4a^2+r^2\, F')=\frac{2}{3}
\end{equation}
where $'\equiv \frac{d}{d(r^2)}$. Following now the usual steps, it is convenient to introduce
\begin{equation}
\gamma=r^2\, F'
\end{equation}
such that the Ricci-flatness condition leads to
\begin{equation}
\gamma\, \gamma'\, (4a^2+\gamma)=\frac{2}{3}\, r^2~.
\end{equation}
This can be integrated into
\begin{equation}
\gamma^3+6a`2\, \gamma^2-r^4-\Big(\frac{2\, b^2}{3})^3=0
\end{equation}
where $b$ is an integration constant. It is further convenient to introduce
\begin{equation}
\rho^2=\frac{3}{2}\gamma
\end{equation}
since then the metric becomes
\begin{equation}
ds^2=\kappa^{-1}\, d\rho^2+\frac{\kappa}{9}\,\rho^2\, g_5^2+\Big(\frac{\rho^2}{6}+a^2\Big)\, \Big[d\theta_1^2+\sin^2\theta_1\,d\phi_1^2\Big]+\frac{\rho^2}{6}\, \Big[d\theta_2^2+\sin^2\theta_2\,d\phi_2^2\Big]
\end{equation}
where $g_5=d\psi+\sum\cos\theta_i\, d\phi_i$ and
\begin{equation}
\kappa=\frac{\rho^6+9a^2\, \rho^4-b^6}{\rho^4\, (\rho^2+6a^2)}~.
\end{equation}
Since the metric has to be positive definite, we have to impose $\rho^6+9a^2\, \rho^4-b^6\ge 0$. Thus the range of $\rho$ is limited as
\begin{equation}
\rho\in [\rho_{\star},\infty)\, ,\quad \rho_{\star}^6+9a^2\, \rho_{\star}^4-b^6=0~.
\end{equation}
The equation defining $\rho_{\star}$, being cubic in $\rho_{\star}^2$, has a cumbersome explicit solution. However, we stress that $\rho_{\star}=\rho_{\star}(a,\, b)$. Furthermore, $\rho_{\star}(a,0)=0$. 

Finally, we note that as long as $b\ne 0$, for $\rho\sim \rho_{\star}$ the metric develops an orbifold singularity which is cured iff $\psi\in [0,2\pi]$; that is, if we consider $\mathcal{C}_{\C}(\mathbb{F}_0)$.

\subsection{Warp factor for the resolved $\mathcal{C}_{\C}(\mathbb{F}_0)$} \label{sec:warp-F0}

We are interested in considering D3 branes moving in the resolved $\mathcal{C}_{\C}(\mathbb{F}_0)$ geometry we have just discussed. Since we will consider a large amount of them, they will back-react on the space, giving rise to a near-horizon $AdS_5$ geometry. This geometry is obtained from \eqref{resolved-metric} where the warped factor is the solution of the following equation
\begin{equation}
\Box_{\mathcal{C}_{\C}(\mathbb{F}_0)_{Resolved}}H=\frac{\mathcal{N}}{\sqrt{\det g_{\mathcal{C}_{\C}(\mathbb{F}_0)_{Resolved}}}}\, \delta(r-r_p,\, \xi-\xi_p)
\end{equation}
where $\xi$ denotes collectively the angular coordinates in the internal manifold, $(r_p,\, \xi_p)$ stands for the particular point where the stack of D3 sits and we collected the numerical factors in $\mathcal{N}$.

\subsubsection{A more useful coordinate system}

It turns out that in the coordinates we have for the metric, the Laplace equation above becomes more complicated to solve. It is therefore useful to define a new radial coordinate $r$, not to be confused with the $r$ of the $z's$ in (\ref{zs}), as
\begin{equation}
\rho^2=6\, r+\rho_{\star}^2~.
\end{equation}
After a bit of algebra, the metric reduces to
\begin{equation}
ds^2=U^{-1}\, dr^2+U\, g_5^2+\sum (r+\ell_i^2)\, \Big[d\theta_i^2+\sin^2\theta_id\phi_i^2\Big]
\end{equation}
where
\begin{equation}
U=\frac{r\,\Big\{ 3\, \ell_1^2\, (2\, \ell_2^2+r)+r\, (3\, \ell_2^2+2\,r)\Big\}}{3\, (\ell_1^2+r)\, (\ell_2^2+r)}~.
\end{equation}
The parameters $\ell_i$ are defined as
\begin{equation}
\ell_1^2=\frac{\rho_{\star}^2}{6}+a^2\, ,\qquad \ell_2^2=\frac{\rho_{\star}^2}{6}~,
\end{equation}
so that $\ell_i^2=\ell_i^2(a,\, b)$. 

Going back to the Laplace equation we are interested in, in this coordinate system it reads
\begin{equation}
\frac{1}{(r+\ell_1^2)\, (r+\ell_2^2)}\, \partial_r\Big( (r+\ell_1^2)\, (r+\ell_2^2)\, U\, \partial_r H\Big)+\hat{A}H=\frac{\mathcal{N}\, \delta(r-r_p)}{(r+\ell_1^2)\, (r+\ell_2^2)}\, \frac{\delta(\xi-\xi_p)}{\sqrt{g_{\Omega}}}
\end{equation}
where we collectively denote by $g_{\Omega}$ the angular metric (which is like that of the base of the singular cone). Also, as is standard
\begin{equation}
\hat{A}H=U^{-1}\partial_{\psi}^2H+\sum \frac{1}{(r+\ell_i^2)}\, \Delta_iH
\end{equation}
where
\begin{equation}
\Delta_i=\frac{1}{\sin\theta_i}\, \partial_{\theta_i}(\sin\theta_i\partial_{\theta_i}\, )+\frac{1}{\sin^2\theta_i}(\partial_{\phi_i}-\cos\theta_i\partial_{\psi}\, )^2~.
\end{equation}
As usual, the eigenfunctions of $\hat{A}$ are those of the singular cone (we denote them by $Y_I$). Thus, we can use these, in particular their completeness relation, to write the angular $\delta$ function in terms of them. Then, we may write
\bea
H=\sum \psi_I(r)\,Y_I(\xi_p)^{*}\,Y_I(\xi) \ .
\eea
Let us recall that the multi-index $I$ stands for the angular quantum numbers $R,\, (l_i,\, m_i)$. Here $(l_i,\, m_i)$ are the angular quantum numbers linked with each of the two-spheres parametrized by $(\theta_i,\, \phi_i)$ and $R$ the R-charge.

After a bit of algebra, we have that $\psi_I$ satisfies the following equation (we drop the subscript $I$)
\begin{equation}
\frac{\partial_r\Big( (r+\ell_1^2)\, (r+\ell_2^2)\, U\, \partial_r\psi\Big)}{(r+\ell_1^2)\, (r+\ell_2^2)}\,-\Big\{ U^{-1}\, R^2+\sum\frac{l_i(l_i+1)-R^2}{(r+\ell_i^2)}\Big\}\, \psi=\frac{\mathcal{N}\, \delta(r-r_p)}{(r+\ell_1^2)\, (r+\ell_2^2)}~.
\end{equation}

\subsubsection{Resolving the four-cycle}

Let us at this point assume $a=0$ and $b\ne 0$, that is, $\ell_1=\ell_2=\ell$. Under that assumption, there is a permutation symmetry between $(\theta_1,\,\phi_1)\leftrightarrow (\theta_2,\, \phi_2)$ which suggests $l_1=l_2=l$. Furthermore, we can take $R=0$. Under those assumptions the equation for the radial eigenfunction is 
\begin{equation} \label{eq-wa}
\frac{2}{3\,(r+\ell^2)^2}\, \partial_r\Big( r\, (3\, \ell^4+3\,\ell^2\, r+r^2)\, \partial_r\psi\Big)-2\frac{l(l+1)}{(r+\ell^2)}\, \psi=\frac{\mathcal{N}\, \delta(r-r_p)}{(r+\ell^2)^2}~.
\end{equation}
The solutions of this equation are
\begin{equation}
\psi_1=_2F_1\Big(\frac{1}{3}-\frac{1}{3}\, \beta,\,\frac{1}{3}+\frac{1}{3}\, \beta,\, \frac{2}{3},\, -\frac{(3\, \ell^4+6\ell^2\, r+3\, r^2)^{\frac{3}{2}}}{3\, \sqrt{3}\, \ell^6} \Big)
\end{equation}
and
\begin{equation}
\psi_2=\sqrt{\ell^4+r^2+2\ell^2\, r}\,_2F_1\Big(\frac{2}{3}-\frac{1}{3}\, \beta,\,\frac{2}{3}+\frac{1}{3}\, \beta,\, \frac{4}{3},\, -\frac{(3\, \ell^4+6\ell^2\, r+3\, r^2)^{\frac{3}{2}}}{3\, \sqrt{3}\, \ell^6} \Big)
\end{equation}
where $\beta \equiv \sqrt{1+3\, l(l+1)}$. Using these solutions one can construct the well-behaved radial functions for $r>r_p$ and $r<r_p$ that should match at $r=r_p$. 

In the main text we are discussing the case in which the stack of D3 branes sits in the bottom of the resolved cone. This corresponds to $r_p=0$. It is easy to see from \eqref{eq-wa} that for large $r$ the leading decaying solution in this case corresponds to $l=0$ for which 
\bea
\psi \sim \frac{1}{r^2}
\eea
\section{$\C^3/\Z_5$ : details on FI space} \label{sec:App-Z5}
Here we present the charge matrices corresponding to the 25 chambers discussed in the text
{\tiny
\bea
Q_t^1=
\left(
\begin{array}{cccccccccccccc}
 0 & 1 & 1 & 0 & 0 & 0 & 0 & 0 & 0 & 0 & 0 & -3 & 1 & 2 \\
 1 & 0 & 0 & 0 & 0 & 0 & 0 & 0 & 0 & 0 & 0 & 1 & -2 & 3 \\
 0 & 0 & 0 & 1 & 0 & 0 & 0 & 0 & 0 & 0 & 0 & -1 & 0 & d+2 \\
 0 & 0 & 0 & 0 & 1 & 0 & 0 & 0 & 0 & 0 & 0 & 0 & -1 & d+2 \\
 0 & 0 & 0 & 0 & 0 & 1 & 0 & 0 & 0 & 0 & 0 & -1 & 0 & \frac{1}{3} (2 d-2 x+9) \\
 0 & 0 & 0 & 0 & 0 & 0 & 1 & 0 & 0 & 0 & 0 & 0 & -1 & \frac{2}{3} (-d+x+3) \\
 0 & 0 & 0 & 0 & 0 & 0 & 0 & 1 & 0 & 0 & 0 & -1 & 0 & 3-d \\
 0 & 0 & 0 & 0 & 0 & 0 & 0 & 0 & 1 & 0 & 0 & 0 & -1 & 3 \\
 0 & 0 & 0 & 0 & 0 & 0 & 0 & 0 & 0 & 1 & 0 & 0 & -1 & \frac{1}{3} (-d-2 x+12) \\
 0 & 0 & 0 & 0 & 0 & 0 & 0 & 0 & 0 & 0 & 1 & -1 & 0 & \frac{1}{3} (-2 d+2 x-3)
\end{array}
\right)~,
\eea
\bea
Q_t^2=
\left(
\begin{array}{cccccccccccccc}
 0 & 1 & 1 & 0 & 1 & 0 & 0 & 0 & 0 & 0 & 0 & -3 & 0 & 2 \\
 1 & 0 & 0 & 0 & -2 & 0 & 0 & 0 & 0 & 0 & 0 & 1 & 0 & 3 \\
 0 & 0 & 0 & 1 & 0 & 0 & 0 & 0 & 0 & 0 & 0 & -1 & 0 & \frac{d+2}{2} \\
 0 & 0 & 0 & 0 & -1 & 0 & 0 & 0 & 0 & 0 & 0 & 0 & 1 & \frac{1}{2} (-d-2) \\
 0 & 0 & 0 & 0 & 0 & 1 & 0 & 0 & 0 & 0 & 0 & -1 & 0 & \frac{1}{6} (3 d-4 x+16) \\
 0 & 0 & 0 & 0 & -1 & 0 & 1 & 0 & 0 & 0 & 0 & 0 & 0 & \frac{1}{6} (-3 d+4 x+14) \\
 0 & 0 & 0 & 0 & 0 & 0 & 0 & 1 & 0 & 0 & 0 & -1 & 0 & 5 \\
 0 & 0 & 0 & 0 & -1 & 0 & 0 & 0 & 1 & 0 & 0 & 0 & 0 & \frac{d+8}{2} \\
 0 & 0 & 0 & 0 & -1 & 0 & 0 & 0 & 0 & 1 & 0 & 0 & 0 & \frac{1}{6} (3 d-4 x+34) \\
 0 & 0 & 0 & 0 & 0 & 0 & 0 & 0 & 0 & 0 & 1 & -1 & 0 & \frac{1}{3} (-3 d+2 x-5)
\end{array}
\right)~,
\eea
\bea
Q_t^3=
\left(
\begin{array}{cccccccccccccc}
 0 & 1 & 1 & 0 & 0 & 0 & 1 & 0 & 0 & 0 & 0 & -3 & 0 & 2 \\
 1 & 0 & 0 & 0 & 0 & 0 & -2 & 0 & 0 & 0 & 0 & 1 & 0 & 3 \\
 0 & 0 & 0 & 1 & 0 & 0 & 0 & 0 & 0 & 0 & 0 & -1 & 0 & -d+2 x+8 \\
 0 & 0 & 0 & 0 & 1 & 0 & -1 & 0 & 0 & 0 & 0 & 0 & 0 & -3 d+4 x+14 \\
 0 & 0 & 0 & 0 & 0 & 1 & 0 & 0 & 0 & 0 & 0 & -1 & 0 & 5 \\
 0 & 0 & 0 & 0 & 0 & 0 & -1 & 0 & 0 & 0 & 0 & 0 & 1 & -2 d+2 x+6 \\
 0 & 0 & 0 & 0 & 0 & 0 & 0 & 1 & 0 & 0 & 0 & -1 & 0 & 3 d-4 x-9 \\
 0 & 0 & 0 & 0 & 0 & 0 & -1 & 0 & 1 & 0 & 0 & 0 & 0 & 2 d-2 x-3 \\
 0 & 0 & 0 & 0 & 0 & 0 & -1 & 0 & 0 & 1 & 0 & 0 & 0 & 3 d-4 x-6 \\
 0 & 0 & 0 & 0 & 0 & 0 & 0 & 0 & 0 & 0 & 1 & -1 & 0 & -2 d+2 x+3
\end{array}
\right)~,
\eea
\bea
Q_t^4=
\left(
\begin{array}{cccccccccccccc}
 0 & 1 & 1 & 0 & 0 & 0 & 0 & 0 & 1 & 0 & 0 & -3 & 0 & 2 \\
 1 & 0 & 0 & 0 & 0 & 0 & 0 & 0 & -2 & 0 & 0 & 1 & 0 & 3 \\
 0 & 0 & 0 & 1 & 0 & 0 & 0 & 0 & 0 & 0 & 0 & -1 & 0 & d+5 \\
 0 & 0 & 0 & 0 & 1 & 0 & 0 & 0 & -1 & 0 & 0 & 0 & 0 & d+8 \\
 0 & 0 & 0 & 0 & 0 & 1 & 0 & 0 & 0 & 0 & 0 & -1 & 0 & \frac{2}{3} (d-x+6) \\
 0 & 0 & 0 & 0 & 0 & 0 & 1 & 0 & -1 & 0 & 0 & 0 & 0 & \frac{1}{3} (-2 d+2 x+3) \\
 0 & 0 & 0 & 0 & 0 & 0 & 0 & 1 & 0 & 0 & 0 & -1 & 0 & -d-3 \\
 0 & 0 & 0 & 0 & 0 & 0 & 0 & 0 & -1 & 0 & 0 & 0 & 1 & 3 \\
 0 & 0 & 0 & 0 & 0 & 0 & 0 & 0 & -1 & 1 & 0 & 0 & 0 & \frac{1}{3} (-d-2 x-3) \\
 0 & 0 & 0 & 0 & 0 & 0 & 0 & 0 & 0 & 0 & 1 & -1 & 0 & \frac{1}{3} (-2 d+2 x+3)
\end{array}
\right)~,
\eea
\bea
Q_t^5=
\left(
\begin{array}{cccccccccccccc}
 0 & 1 & 1 & 0 & 0 & 0 & 0 & 0 & 0 & 1 & 0 & -3 & 0 & 2 \\
 1 & 0 & 0 & 0 & 0 & 0 & 0 & 0 & 0 & -2 & 0 & 1 & 0 & 3 \\
 0 & 0 & 0 & 1 & 0 & 0 & 0 & 0 & 0 & 0 & 0 & -1 & 0 & \frac{2}{5} (2 d-x+11) \\
 0 & 0 & 0 & 0 & 1 & 0 & 0 & 0 & 0 & -1 & 0 & 0 & 0 & \frac{1}{5} (3 d-4 x+34) \\
 0 & 0 & 0 & 0 & 0 & 1 & 0 & 0 & 0 & 0 & 0 & -1 & 0 & \frac{1}{5} (3 d-4 x+19) \\
 0 & 0 & 0 & 0 & 0 & 0 & 1 & 0 & 0 & -1 & 0 & 0 & 0 & \frac{1}{5} (-3 d+4 x+6) \\
 0 & 0 & 0 & 0 & 0 & 0 & 0 & 1 & 0 & 0 & 0 & -1 & 0 & \frac{1}{5} (-3 d+4 x-9) \\
 0 & 0 & 0 & 0 & 0 & 0 & 0 & 0 & 1 & -1 & 0 & 0 & 0 & \frac{1}{5} (d+2 x+3) \\
 0 & 0 & 0 & 0 & 0 & 0 & 0 & 0 & 0 & -1 & 0 & 0 & 1 & \frac{1}{5} (-d-2 x+12) \\
 0 & 0 & 0 & 0 & 0 & 0 & 0 & 0 & 0 & 0 & 1 & -1 & 0 & \frac{1}{5} (-4 d+2 x+3)
\end{array}
\right)~,
\eea
\bea
Q_t^6=
\left(
\begin{array}{cccccccccccccc}
 0 & 1 & 1 & -3 & 0 & 0 & 0 & 0 & 0 & 0 & 0 & 0 & 1 & 2 \\
 1 & 0 & 0 & 1 & 0 & 0 & 0 & 0 & 0 & 0 & 0 & 0 & -2 & 3 \\
 0 & 0 & 0 & -1 & 0 & 0 & 0 & 0 & 0 & 0 & 0 & 1 & 0 & \frac{d+2}{2} \\
 0 & 0 & 0 & 0 & 1 & 0 & 0 & 0 & 0 & 0 & 0 & 0 & -1 & \frac{1}{2} (-d-2) \\
 0 & 0 & 0 & -1 & 0 & 1 & 0 & 0 & 0 & 0 & 0 & 0 & 0 & \frac{1}{6} (-d-4 x+8) \\
 0 & 0 & 0 & 0 & 0 & 0 & 1 & 0 & 0 & 0 & 0 & 0 & -1 & \frac{1}{3} (-d+2 x+8) \\
 0 & 0 & 0 & -1 & 0 & 0 & 0 & 1 & 0 & 0 & 0 & 0 & 0 & 5 \\
 0 & 0 & 0 & 0 & 0 & 0 & 0 & 0 & 1 & 0 & 0 & 0 & -1 & \frac{d+8}{2} \\
 0 & 0 & 0 & 0 & 0 & 0 & 0 & 0 & 0 & 1 & 0 & 0 & -1 & \frac{1}{3} (d-2 x+16) \\
 0 & 0 & 0 & -1 & 0 & 0 & 0 & 0 & 0 & 0 & 1 & 0 & 0 & \frac{1}{3} (-d+2 x-1)
\end{array}
\right)~,
\eea
\bea
Q_t^7=
\left(
\begin{array}{cccccccccccccc}
 0 & 1 & 1 & -3 & 1 & 0 & 0 & 0 & 0 & 0 & 0 & 0 & 0 & 2 \\
 1 & 0 & 0 & 1 & -2 & 0 & 0 & 0 & 0 & 0 & 0 & 0 & 0 & 3 \\
 0 & 0 & 0 & -1 & 0 & 0 & 0 & 0 & 0 & 0 & 0 & 1 & 0 & d+2 \\
 0 & 0 & 0 & 0 & -1 & 0 & 0 & 0 & 0 & 0 & 0 & 0 & 1 & d+2 \\
 0 & 0 & 0 & -1 & 0 & 1 & 0 & 0 & 0 & 0 & 0 & 0 & 0 & \frac{1}{3} (-2 d-2 x+1) \\
 0 & 0 & 0 & 0 & -1 & 0 & 1 & 0 & 0 & 0 & 0 & 0 & 0 & \frac{1}{3} (-d+2 x+8) \\
 0 & 0 & 0 & -1 & 0 & 0 & 0 & 1 & 0 & 0 & 0 & 0 & 0 & 3-d \\
 0 & 0 & 0 & 0 & -1 & 0 & 0 & 0 & 1 & 0 & 0 & 0 & 0 & 3 \\
 0 & 0 & 0 & 0 & -1 & 0 & 0 & 0 & 0 & 1 & 0 & 0 & 0 & -\frac{2}{3} (d+x-5) \\
 0 & 0 & 0 & -1 & 0 & 0 & 0 & 0 & 0 & 0 & 1 & 0 & 0 & \frac{1}{3} (2 d+2 x+5)
\end{array}
\right)~,
\eea
\bea
Q_t^8=
\left(
\begin{array}{cccccccccccccc}
 0 & 1 & 1 & -3 & 0 & 0 & 1 & 0 & 0 & 0 & 0 & 0 & 0 & 2 \\
 1 & 0 & 0 & 1 & 0 & 0 & -2 & 0 & 0 & 0 & 0 & 0 & 0 & 3 \\
 0 & 0 & 0 & -1 & 0 & 0 & 0 & 0 & 0 & 0 & 0 & 1 & 0 & \frac{1}{5} (b+4 x+33) \\
 0 & 0 & 0 & 0 & 1 & 0 & -1 & 0 & 0 & 0 & 0 & 0 & 0 & \frac{1}{5} (b-6 x-17) \\
 0 & 0 & 0 & -1 & 0 & 1 & 0 & 0 & 0 & 0 & 0 & 0 & 0 & \frac{1}{5} (-b-4 x-8) \\
 0 & 0 & 0 & 0 & 0 & 0 & -1 & 0 & 0 & 0 & 0 & 0 & 1 & \frac{2}{5} (b-x+8) \\
 0 & 0 & 0 & -1 & 0 & 0 & 0 & 1 & 0 & 0 & 0 & 0 & 0 & \frac{1}{5} (-2 b+2 x+9) \\
 0 & 0 & 0 & 0 & 0 & 0 & -1 & 0 & 1 & 0 & 0 & 0 & 0 & \frac{1}{5} (-b+6 x+32) \\
 0 & 0 & 0 & 0 & 0 & 0 & -1 & 0 & 0 & 1 & 0 & 0 & 0 & -\frac{2}{5} (b-x-12) \\
 0 & 0 & 0 & -1 & 0 & 0 & 0 & 0 & 0 & 0 & 1 & 0 & 0 & \frac{1}{5} (2 b-2 x+1)
\end{array}
\right)~,
\eea
\bea
Q_t^9=
\left(
\begin{array}{cccccccccccccc}
 0 & 1 & 1 & -3 & 0 & 0 & 0 & 0 & 1 & 0 & 0 & 0 & 0 & 2 \\
 1 & 0 & 0 & 1 & 0 & 0 & 0 & 0 & -2 & 0 & 0 & 0 & 0 & 3 \\
 0 & 0 & 0 & -1 & 0 & 0 & 0 & 0 & 0 & 0 & 0 & 1 & 0 & d+5 \\
 0 & 0 & 0 & 0 & 1 & 0 & 0 & 0 & -1 & 0 & 0 & 0 & 0 & 3 \\
 0 & 0 & 0 & -1 & 0 & 1 & 0 & 0 & 0 & 0 & 0 & 0 & 0 & -\frac{2}{3} (d+x+4) \\
 0 & 0 & 0 & 0 & 0 & 0 & 1 & 0 & -1 & 0 & 0 & 0 & 0 & \frac{1}{3} (-d+2 x+8) \\
 0 & 0 & 0 & -1 & 0 & 0 & 0 & 1 & 0 & 0 & 0 & 0 & 0 & -d-3 \\
 0 & 0 & 0 & 0 & 0 & 0 & 0 & 0 & -1 & 0 & 0 & 0 & 1 & d+8 \\
 0 & 0 & 0 & 0 & 0 & 0 & 0 & 0 & -1 & 1 & 0 & 0 & 0 & -\frac{2}{3} (d+x+4) \\
 0 & 0 & 0 & -1 & 0 & 0 & 0 & 0 & 0 & 0 & 1 & 0 & 0 & \frac{1}{3} (2 d+2 x+23)
\end{array}
\right)~,
\eea
\bea
Q_t^{10}=
\left(
\begin{array}{cccccccccccccc}
 0 & 1 & 1 & -3 & 0 & 0 & 0 & 0 & 0 & 1 & 0 & 0 & 0 & 2 \\
 1 & 0 & 0 & 1 & 0 & 0 & 0 & 0 & 0 & -2 & 0 & 0 & 0 & 3 \\
 0 & 0 & 0 & -1 & 0 & 0 & 0 & 0 & 0 & 0 & 0 & 1 & 0 & \frac{1}{3} (2 d-x+11) \\
 0 & 0 & 0 & 0 & 1 & 0 & 0 & 0 & 0 & -1 & 0 & 0 & 0 & \frac{1}{3} (-d-x+5) \\
 0 & 0 & 0 & -1 & 0 & 1 & 0 & 0 & 0 & 0 & 0 & 0 & 0 & \frac{1}{3} (-d-x-4) \\
 0 & 0 & 0 & 0 & 0 & 0 & 1 & 0 & 0 & -1 & 0 & 0 & 0 & \frac{1}{3} (-d+2 x+8) \\
 0 & 0 & 0 & -1 & 0 & 0 & 0 & 1 & 0 & 0 & 0 & 0 & 0 & \frac{1}{3} (-d+2 x-1) \\
 0 & 0 & 0 & 0 & 0 & 0 & 0 & 0 & 1 & -1 & 0 & 0 & 0 & \frac{1}{3} (d+x+4) \\
 0 & 0 & 0 & 0 & 0 & 0 & 0 & 0 & 0 & -1 & 0 & 0 & 1 & \frac{1}{3} (d-2 x+16) \\
 0 & 0 & 0 & -1 & 0 & 0 & 0 & 0 & 0 & 0 & 1 & 0 & 0 & 5
\end{array}
\right)~,
\eea
\bea
Q_t^{11}=
\left(
\begin{array}{cccccccccccccc}
 0 & 1 & 1 & 0 & 0 & -3 & 0 & 0 & 0 & 0 & 0 & 0 & 1 & 2 \\
 1 & 0 & 0 & 0 & 0 & 1 & 0 & 0 & 0 & 0 & 0 & 0 & -2 & 3 \\
 0 & 0 & 0 & 1 & 0 & -1 & 0 & 0 & 0 & 0 & 0 & 0 & 0 & \frac{1}{5} (d+4 x-8) \\
 0 & 0 & 0 & 0 & 1 & 0 & 0 & 0 & 0 & 0 & 0 & 0 & -1 & \frac{1}{5} (-d+6 x-17) \\
 0 & 0 & 0 & 0 & 0 & -1 & 0 & 0 & 0 & 0 & 0 & 1 & 0 & \frac{1}{5} (2 d-2 x+9) \\
 0 & 0 & 0 & 0 & 0 & 0 & 1 & 0 & 0 & 0 & 0 & 0 & -1 & -\frac{2}{5} (d-x-8) \\
 0 & 0 & 0 & 0 & 0 & -1 & 0 & 1 & 0 & 0 & 0 & 0 & 0 & \frac{1}{5} (-d-4 x+33) \\
 0 & 0 & 0 & 0 & 0 & 0 & 0 & 0 & 1 & 0 & 0 & 0 & -1 & \frac{2}{5} (d-x+12) \\
 0 & 0 & 0 & 0 & 0 & 0 & 0 & 0 & 0 & 1 & 0 & 0 & -1 & \frac{1}{5} (d-6 x+32) \\
 0 & 0 & 0 & 0 & 0 & -1 & 0 & 0 & 0 & 0 & 1 & 0 & 0 & \frac{1}{5} (-2 d+2 x+1)
\end{array}
\right)~,
\eea
\bea
Q_t^{12}=
\left(
\begin{array}{cccccccccccccc}
 0 & 1 & 1 & 0 & 1 & -3 & 0 & 0 & 0 & 0 & 0 & 0 & 0 & 2 \\
 1 & 0 & 0 & 0 & -2 & 1 & 0 & 0 & 0 & 0 & 0 & 0 & 0 & 3 \\
 0 & 0 & 0 & 1 & 0 & -1 & 0 & 0 & 0 & 0 & 0 & 0 & 0 & \frac{1}{7} (2 d+2 x-1) \\
 0 & 0 & 0 & 0 & -1 & 0 & 0 & 0 & 0 & 0 & 0 & 0 & 1 & \frac{1}{7} (d-6 x+17) \\
 0 & 0 & 0 & 0 & 0 & -1 & 0 & 0 & 0 & 0 & 0 & 1 & 0 & \frac{1}{7} (3 d-4 x+16) \\
 0 & 0 & 0 & 0 & -1 & 0 & 1 & 0 & 0 & 0 & 0 & 0 & 0 & \frac{1}{7} (-3 d+4 x+19) \\
 0 & 0 & 0 & 0 & 0 & -1 & 0 & 1 & 0 & 0 & 0 & 0 & 0 & \frac{1}{7} (-3 d+4 x+19) \\
 0 & 0 & 0 & 0 & -1 & 0 & 0 & 0 & 1 & 0 & 0 & 0 & 0 & \frac{2}{7} (d+x+10) \\
 0 & 0 & 0 & 0 & -1 & 0 & 0 & 0 & 0 & 1 & 0 & 0 & 0 & 3 \\
 0 & 0 & 0 & 0 & 0 & -1 & 0 & 0 & 0 & 0 & 1 & 0 & 0 & \frac{1}{7} (-2 d-2 x+15)
\end{array}
\right)~,
\eea
\bea
Q_t^{13}=
\left(
\begin{array}{cccccccccccccc}
 0 & 1 & 1 & 0 & 0 & -3 & 1 & 0 & 0 & 0 & 0 & 0 & 0 & 2 \\
 1 & 0 & 0 & 0 & 0 & 1 & -2 & 0 & 0 & 0 & 0 & 0 & 0 & 3 \\
 0 & 0 & 0 & 1 & 0 & -1 & 0 & 0 & 0 & 0 & 0 & 0 & 0 & -d+2 x+8 \\
 0 & 0 & 0 & 0 & 1 & 0 & -1 & 0 & 0 & 0 & 0 & 0 & 0 & -3 d+4 x+19 \\
 0 & 0 & 0 & 0 & 0 & -1 & 0 & 0 & 0 & 0 & 0 & 1 & 0 & 5 \\
 0 & 0 & 0 & 0 & 0 & 0 & -1 & 0 & 0 & 0 & 0 & 0 & 1 & 2 (-d+x+8) \\
 0 & 0 & 0 & 0 & 0 & -1 & 0 & 1 & 0 & 0 & 0 & 0 & 0 & 3 d-4 x-19 \\
 0 & 0 & 0 & 0 & 0 & 0 & -1 & 0 & 1 & 0 & 0 & 0 & 0 & 2 (d-x-4) \\
 0 & 0 & 0 & 0 & 0 & 0 & -1 & 0 & 0 & 1 & 0 & 0 & 0 & 3 d-4 (x+4) \\
 0 & 0 & 0 & 0 & 0 & -1 & 0 & 0 & 0 & 0 & 1 & 0 & 0 & -2 d+2 x+13
\end{array}
\right)~,
\eea
\bea
Q_t^{14}=
\left(
\begin{array}{cccccccccccccc}
 0 & 1 & 1 & 0 & 0 & -3 & 0 & 0 & 1 & 0 & 0 & 0 & 0 & 2 \\
 1 & 0 & 0 & 0 & 0 & 1 & 0 & 0 & -2 & 0 & 0 & 0 & 0 & 3 \\
 0 & 0 & 0 & 1 & 0 & -1 & 0 & 0 & 0 & 0 & 0 & 0 & 0 & \frac{1}{2} (d+x+4) \\
 0 & 0 & 0 & 0 & 1 & 0 & 0 & 0 & -1 & 0 & 0 & 0 & 0 & \frac{1}{2} (d+x+10) \\
 0 & 0 & 0 & 0 & 0 & -1 & 0 & 0 & 0 & 0 & 0 & 1 & 0 & \frac{1}{2} (d-x+6) \\
 0 & 0 & 0 & 0 & 0 & 0 & 1 & 0 & -1 & 0 & 0 & 0 & 0 & \frac{1}{2} (-d+x+4) \\
 0 & 0 & 0 & 0 & 0 & -1 & 0 & 1 & 0 & 0 & 0 & 0 & 0 & -d-3 \\
 0 & 0 & 0 & 0 & 0 & 0 & 0 & 0 & -1 & 0 & 0 & 0 & 1 & \frac{1}{2} (d-x+12) \\
 0 & 0 & 0 & 0 & 0 & 0 & 0 & 0 & -1 & 1 & 0 & 0 & 0 & \frac{1}{2} (-d-x-4) \\
 0 & 0 & 0 & 0 & 0 & -1 & 0 & 0 & 0 & 0 & 1 & 0 & 0 & 5
\end{array}
\right)~,
\eea
\bea
Q_t^{15}=
\left(
\begin{array}{cccccccccccccc}
 0 & 1 & 1 & 0 & 0 & -3 & 0 & 0 & 0 & 1 & 0 & 0 & 0 & 2 \\
 1 & 0 & 0 & 0 & 0 & 1 & 0 & 0 & 0 & -2 & 0 & 0 & 0 & 3 \\
 0 & 0 & 0 & 1 & 0 & -1 & 0 & 0 & 0 & 0 & 0 & 0 & 0 & \frac{2}{7} (d+x+4) \\
 0 & 0 & 0 & 0 & 1 & 0 & 0 & 0 & 0 & -1 & 0 & 0 & 0 & 3 \\
 0 & 0 & 0 & 0 & 0 & -1 & 0 & 0 & 0 & 0 & 0 & 1 & 0 & \frac{1}{7} (3 d-4 x+19) \\
 0 & 0 & 0 & 0 & 0 & 0 & 1 & 0 & 0 & -1 & 0 & 0 & 0 & \frac{1}{7} (4 (x+4)-3 d) \\
 0 & 0 & 0 & 0 & 0 & -1 & 0 & 1 & 0 & 0 & 0 & 0 & 0 & \frac{1}{7} (-3 d+4 x-5) \\
 0 & 0 & 0 & 0 & 0 & 0 & 0 & 0 & 1 & -1 & 0 & 0 & 0 & \frac{2}{7} (d+x+4) \\
 0 & 0 & 0 & 0 & 0 & 0 & 0 & 0 & 0 & -1 & 0 & 0 & 1 & \frac{1}{7} (d-6 x+32) \\
 0 & 0 & 0 & 0 & 0 & -1 & 0 & 0 & 0 & 0 & 1 & 0 & 0 & \frac{1}{7} (-2 d-2 x+27)
\end{array}
\right)~,
\eea
\bea
Q_t^{16}=
\left(
\begin{array}{cccccccccccccc}
 0 & 1 & 1 & 0 & 0 & 0 & 0 & -3 & 0 & 0 & 0 & 0 & 1 & 2 \\
 1 & 0 & 0 & 0 & 0 & 0 & 0 & 1 & 0 & 0 & 0 & 0 & -2 & 3 \\
 0 & 0 & 0 & 1 & 0 & 0 & 0 & -1 & 0 & 0 & 0 & 0 & 0 & 5 \\
 0 & 0 & 0 & 0 & 1 & 0 & 0 & 0 & 0 & 0 & 0 & 0 & -1 & \frac{13-d}{2} \\
 0 & 0 & 0 & 0 & 0 & 1 & 0 & -1 & 0 & 0 & 0 & 0 & 0 & \frac{1}{6} (-d-4 x+33) \\
 0 & 0 & 0 & 0 & 0 & 0 & 1 & 0 & 0 & 0 & 0 & 0 & -1 & \frac{1}{3} (-d+2 x+3) \\
 0 & 0 & 0 & 0 & 0 & 0 & 0 & -1 & 0 & 0 & 0 & 1 & 0 & \frac{d-3}{2} \\
 0 & 0 & 0 & 0 & 0 & 0 & 0 & 0 & 1 & 0 & 0 & 0 & -1 & \frac{d+3}{2} \\
 0 & 0 & 0 & 0 & 0 & 0 & 0 & 0 & 0 & 1 & 0 & 0 & -1 & \frac{1}{3} (d-2 x+6) \\
 0 & 0 & 0 & 0 & 0 & 0 & 0 & -1 & 0 & 0 & 1 & 0 & 0 & \frac{1}{3} (-d+2 x-6)
\end{array}
\right)~,
\eea
\bea
Q_t^{17}=
\left(
\begin{array}{cccccccccccccc}
 0 & 1 & 1 & 0 & 1 & 0 & 0 & -3 & 0 & 0 & 0 & 0 & 0 & 2 \\
 1 & 0 & 0 & 0 & -2 & 0 & 0 & 1 & 0 & 0 & 0 & 0 & 0 & 3 \\
 0 & 0 & 0 & 1 & 0 & 0 & 0 & -1 & 0 & 0 & 0 & 0 & 0 & \frac{d-3}{2} \\
 0 & 0 & 0 & 0 & -1 & 0 & 0 & 0 & 0 & 0 & 0 & 0 & 1 & \frac{13-d}{2} \\
 0 & 0 & 0 & 0 & 0 & 1 & 0 & -1 & 0 & 0 & 0 & 0 & 0 & \frac{1}{6} (3 d-4 x-19) \\
 0 & 0 & 0 & 0 & -1 & 0 & 1 & 0 & 0 & 0 & 0 & 0 & 0 & \frac{1}{6} (-3 d+4 x+19) \\
 0 & 0 & 0 & 0 & 0 & 0 & 0 & -1 & 0 & 0 & 0 & 1 & 0 & 5 \\
 0 & 0 & 0 & 0 & -1 & 0 & 0 & 0 & 1 & 0 & 0 & 0 & 0 & \frac{d+3}{2} \\
 0 & 0 & 0 & 0 & -1 & 0 & 0 & 0 & 0 & 1 & 0 & 0 & 0 & \frac{1}{6} (3 d-4 x-1) \\
 0 & 0 & 0 & 0 & 0 & 0 & 0 & -1 & 0 & 0 & 1 & 0 & 0 & \frac{1}{3} (-3 d+2 x+20)
\end{array}
\right)~,
\eea
\bea
Q_t^{18}=
\left(
\begin{array}{cccccccccccccc}
 0 & 1 & 1 & 0 & 0 & 0 & 1 & -3 & 0 & 0 & 0 & 0 & 0 & 2 \\
 1 & 0 & 0 & 0 & 0 & 0 & -2 & 1 & 0 & 0 & 0 & 0 & 0 & 3 \\
 0 & 0 & 0 & 1 & 0 & 0 & 0 & -1 & 0 & 0 & 0 & 0 & 0 & -d+2 x+8 \\
 0 & 0 & 0 & 0 & 1 & 0 & -1 & 0 & 0 & 0 & 0 & 0 & 0 & \frac{1}{2} (-3 d+4 x+19) \\
 0 & 0 & 0 & 0 & 0 & 1 & 0 & -1 & 0 & 0 & 0 & 0 & 0 & \frac{1}{2} (-3 d+4 x+19) \\
 0 & 0 & 0 & 0 & 0 & 0 & -1 & 0 & 0 & 0 & 0 & 0 & 1 & d-2 x-3 \\
 0 & 0 & 0 & 0 & 0 & 0 & 0 & -1 & 0 & 0 & 0 & 1 & 0 & \frac{3 (d-3)}{2}-2 x \\
 0 & 0 & 0 & 0 & 0 & 0 & -1 & 0 & 1 & 0 & 0 & 0 & 0 & \frac{d+3}{2} \\
 0 & 0 & 0 & 0 & 0 & 0 & -1 & 0 & 0 & 1 & 0 & 0 & 0 & 3 \\
 0 & 0 & 0 & 0 & 0 & 0 & 0 & -1 & 0 & 0 & 1 & 0 & 0 & d-2 (x+3)
\end{array}
\right)~,
\eea
\bea
Q_t^{19}=
\left(
\begin{array}{cccccccccccccc}
 0 & 1 & 1 & 0 & 0 & 0 & 0 & -3 & 1 & 0 & 0 & 0 & 0 & 2 \\
 1 & 0 & 0 & 0 & 0 & 0 & 0 & 1 & -2 & 0 & 0 & 0 & 0 & 3 \\
 0 & 0 & 0 & 1 & 0 & 0 & 0 & -1 & 0 & 0 & 0 & 0 & 0 & \frac{1}{5} (-3 b+4 x+19) \\
 0 & 0 & 0 & 0 & 1 & 0 & 0 & 0 & -1 & 0 & 0 & 0 & 0 & \frac{1}{5} (-3 b+4 x+34) \\
 0 & 0 & 0 & 0 & 0 & 1 & 0 & -1 & 0 & 0 & 0 & 0 & 0 & \frac{2}{5} (-2 b+x+11) \\
 0 & 0 & 0 & 0 & 0 & 0 & 1 & 0 & -1 & 0 & 0 & 0 & 0 & \frac{1}{5} (b+2 (x+6)) \\
 0 & 0 & 0 & 0 & 0 & 0 & 0 & -1 & 0 & 0 & 0 & 1 & 0 & \frac{1}{5} (3 b-4 x-9) \\
 0 & 0 & 0 & 0 & 0 & 0 & 0 & 0 & -1 & 0 & 0 & 0 & 1 & \frac{1}{5} (3 b-4 x+6) \\
 0 & 0 & 0 & 0 & 0 & 0 & 0 & 0 & -1 & 1 & 0 & 0 & 0 & \frac{1}{5} (-b-2 x+3) \\
 0 & 0 & 0 & 0 & 0 & 0 & 0 & -1 & 0 & 0 & 1 & 0 & 0 & \frac{1}{5} (4 b-2 x+3)
\end{array}
\right)~,
\eea
\bea
Q_t^{20}=
\left(
\begin{array}{cccccccccccccc}
 0 & 1 & 1 & 0 & 0 & 0 & 0 & -3 & 0 & 1 & 0 & 0 & 0 & 2 \\
 1 & 0 & 0 & 0 & 0 & 0 & 0 & 1 & 0 & -2 & 0 & 0 & 0 & 3 \\
 0 & 0 & 0 & 1 & 0 & 0 & 0 & -1 & 0 & 0 & 0 & 0 & 0 & -d+2 x-1 \\
 0 & 0 & 0 & 0 & 1 & 0 & 0 & 0 & 0 & -1 & 0 & 0 & 0 & \frac{1}{2} (-3 d+4 x+1) \\
 0 & 0 & 0 & 0 & 0 & 1 & 0 & -1 & 0 & 0 & 0 & 0 & 0 & \frac{1}{2} (-3 d+4 x-5) \\
 0 & 0 & 0 & 0 & 0 & 0 & 1 & 0 & 0 & -1 & 0 & 0 & 0 & 3 \\
 0 & 0 & 0 & 0 & 0 & 0 & 0 & -1 & 0 & 0 & 0 & 1 & 0 & \frac{3 (d+3)}{2}-2 x \\
 0 & 0 & 0 & 0 & 0 & 0 & 0 & 0 & 1 & -1 & 0 & 0 & 0 & \frac{d+3}{2} \\
 0 & 0 & 0 & 0 & 0 & 0 & 0 & 0 & 0 & -1 & 0 & 0 & 1 & d-2 x+6 \\
 0 & 0 & 0 & 0 & 0 & 0 & 0 & -1 & 0 & 0 & 1 & 0 & 0 & d-2 x+6
\end{array}
\right)~,
\eea
\bea
Q_t^{21}=
\left(
\begin{array}{cccccccccccccc}
 0 & 1 & 1 & 0 & 0 & 0 & 0 & 0 & 0 & 0 & -3 & 0 & 1 & 2 \\
 1 & 0 & 0 & 0 & 0 & 0 & 0 & 0 & 0 & 0 & 1 & 0 & -2 & 3 \\
 0 & 0 & 0 & 1 & 0 & 0 & 0 & 0 & 0 & 0 & -1 & 0 & 0 & -d+2 x-1 \\
 0 & 0 & 0 & 0 & 1 & 0 & 0 & 0 & 0 & 0 & 0 & 0 & -1 & -2 d+3 x-\frac{5}{2} \\
 0 & 0 & 0 & 0 & 0 & 1 & 0 & 0 & 0 & 0 & -1 & 0 & 0 & -d+x+\frac{1}{2} \\
 0 & 0 & 0 & 0 & 0 & 0 & 1 & 0 & 0 & 0 & 0 & 0 & -1 & 3 \\
 0 & 0 & 0 & 0 & 0 & 0 & 0 & 1 & 0 & 0 & -1 & 0 & 0 & d-2 x+6 \\
 0 & 0 & 0 & 0 & 0 & 0 & 0 & 0 & 1 & 0 & 0 & 0 & -1 & d-x+\frac{9}{2} \\
 0 & 0 & 0 & 0 & 0 & 0 & 0 & 0 & 0 & 1 & 0 & 0 & -1 & d-2 x+6 \\
 0 & 0 & 0 & 0 & 0 & 0 & 0 & 0 & 0 & 0 & -1 & 1 & 0 & d-x+\frac{3}{2}
\end{array}
\right)~,
\eea
\bea
Q_t^{22}=
\left(
\begin{array}{cccccccccccccc}
 0 & 1 & 1 & 0 & 1 & 0 & 0 & 0 & 0 & 0 & -3 & 0 & 0 & 2 \\
 1 & 0 & 0 & 0 & -2 & 0 & 0 & 0 & 0 & 0 & 1 & 0 & 0 & 3 \\
 0 & 0 & 0 & 1 & 0 & 0 & 0 & 0 & 0 & 0 & -1 & 0 & 0 & \frac{1}{10} (2 d+2 x+5) \\
 0 & 0 & 0 & 0 & -1 & 0 & 0 & 0 & 0 & 0 & 0 & 0 & 1 & \frac{1}{10} (4 d-6 x+5) \\
 0 & 0 & 0 & 0 & 0 & 1 & 0 & 0 & 0 & 0 & -1 & 0 & 0 & \frac{1}{10} (-2 d-2 x+15) \\
 0 & 0 & 0 & 0 & -1 & 0 & 1 & 0 & 0 & 0 & 0 & 0 & 0 & \frac{1}{10} (-4 d+6 x+25) \\
 0 & 0 & 0 & 0 & 0 & 0 & 0 & 1 & 0 & 0 & -1 & 0 & 0 & -\frac{3 d}{5}+\frac{2 x}{5}+4 \\
 0 & 0 & 0 & 0 & -1 & 0 & 0 & 0 & 1 & 0 & 0 & 0 & 0 & \frac{1}{10} (2 d+2 x+35) \\
 0 & 0 & 0 & 0 & -1 & 0 & 0 & 0 & 0 & 1 & 0 & 0 & 0 & \frac{1}{10} (-2 d-2 x+45) \\
 0 & 0 & 0 & 0 & 0 & 0 & 0 & 0 & 0 & 0 & -1 & 1 & 0 & \frac{1}{5} (3 d-2 x+5)
\end{array}
\right)~,
\eea
\bea
Q_t^{23}=
\left(
\begin{array}{cccccccccccccc}
 0 & 1 & 1 & 0 & 0 & 0 & 1 & 0 & 0 & 0 & -3 & 0 & 0 & 2 \\
 1 & 0 & 0 & 0 & 0 & 0 & -2 & 0 & 0 & 0 & 1 & 0 & 0 & 3 \\
 0 & 0 & 0 & 1 & 0 & 0 & 0 & 0 & 0 & 0 & -1 & 0 & 0 & -d+2 x+8 \\
 0 & 0 & 0 & 0 & 1 & 0 & -1 & 0 & 0 & 0 & 0 & 0 & 0 & -2 d+3 x+\frac{25}{2} \\
 0 & 0 & 0 & 0 & 0 & 1 & 0 & 0 & 0 & 0 & -1 & 0 & 0 & -d+x+\frac{13}{2} \\
 0 & 0 & 0 & 0 & 0 & 0 & -1 & 0 & 0 & 0 & 0 & 0 & 1 & 3 \\
 0 & 0 & 0 & 0 & 0 & 0 & 0 & 1 & 0 & 0 & -1 & 0 & 0 & d-2 (x+3) \\
 0 & 0 & 0 & 0 & 0 & 0 & -1 & 0 & 1 & 0 & 0 & 0 & 0 & d-x-\frac{3}{2} \\
 0 & 0 & 0 & 0 & 0 & 0 & -1 & 0 & 0 & 1 & 0 & 0 & 0 & d-2 x-3 \\
 0 & 0 & 0 & 0 & 0 & 0 & 0 & 0 & 0 & 0 & -1 & 1 & 0 & d-x-\frac{3}{2}
\end{array}
\right)~,
\eea
\bea
Q_t^{24}=
\left(
\begin{array}{cccccccccccccc}
 0 & 1 & 1 & 0 & 0 & 0 & 0 & 0 & 1 & 0 & -3 & 0 & 0 & 2 \\
 1 & 0 & 0 & 0 & 0 & 0 & 0 & 0 & -2 & 0 & 1 & 0 & 0 & 3 \\
 0 & 0 & 0 & 1 & 0 & 0 & 0 & 0 & 0 & 0 & -1 & 0 & 0 & \frac{1}{4} (2 d+2 x+23) \\
 0 & 0 & 0 & 0 & 1 & 0 & 0 & 0 & -1 & 0 & 0 & 0 & 0 & \frac{1}{4} (2 d+2 x+35) \\
 0 & 0 & 0 & 0 & 0 & 1 & 0 & 0 & 0 & 0 & -1 & 0 & 0 & 5 \\
 0 & 0 & 0 & 0 & 0 & 0 & 1 & 0 & -1 & 0 & 0 & 0 & 0 & \frac{1}{4} (-2 d+2 x+3) \\
 0 & 0 & 0 & 0 & 0 & 0 & 0 & 1 & 0 & 0 & -1 & 0 & 0 & -d-3 \\
 0 & 0 & 0 & 0 & 0 & 0 & 0 & 0 & -1 & 0 & 0 & 0 & 1 & \frac{1}{4} (2 d-2 x+9) \\
 0 & 0 & 0 & 0 & 0 & 0 & 0 & 0 & -1 & 1 & 0 & 0 & 0 & \frac{1}{4} (-2 d-2 x-3) \\
 0 & 0 & 0 & 0 & 0 & 0 & 0 & 0 & 0 & 0 & -1 & 1 & 0 & \frac{1}{4} (2 d-2 x-3)
\end{array}
\right)~,
\eea
\bea
Q_t^{25}=
\left(
\begin{array}{cccccccccccccc}
 0 & 1 & 1 & 0 & 0 & 0 & 0 & 0 & 0 & 1 & -3 & 0 & 0 & 2 \\
 1 & 0 & 0 & 0 & 0 & 0 & 0 & 0 & 0 & -2 & 1 & 0 & 0 & 3 \\
 0 & 0 & 0 & 1 & 0 & 0 & 0 & 0 & 0 & 0 & -1 & 0 & 0 & 5 \\
 0 & 0 & 0 & 0 & 1 & 0 & 0 & 0 & 0 & -1 & 0 & 0 & 0 & \frac{1}{6} (-2 d-2 x+45) \\
 0 & 0 & 0 & 0 & 0 & 1 & 0 & 0 & 0 & 0 & -1 & 0 & 0 & \frac{1}{6} (-2 d-2 x+27) \\
 0 & 0 & 0 & 0 & 0 & 0 & 1 & 0 & 0 & -1 & 0 & 0 & 0 & \frac{1}{3} (-d+2 x+3) \\
 0 & 0 & 0 & 0 & 0 & 0 & 0 & 1 & 0 & 0 & -1 & 0 & 0 & \frac{1}{3} (-d+2 x-6) \\
 0 & 0 & 0 & 0 & 0 & 0 & 0 & 0 & 1 & -1 & 0 & 0 & 0 & \frac{1}{6} (2 d+2 x+3) \\
 0 & 0 & 0 & 0 & 0 & 0 & 0 & 0 & 0 & -1 & 0 & 0 & 1 & \frac{1}{3} (d-2 x+6) \\
 0 & 0 & 0 & 0 & 0 & 0 & 0 & 0 & 0 & 0 & -1 & 1 & 0 & \frac{1}{6} (4 d-2 x-3)
\end{array}
\right)~.
\eea
}
\end{appendix}


\begin{thebibliography}{99}

\bibitem{Maldacena:1997re}
  J.~M.~Maldacena,
  ``The large N limit of superconformal field theories and supergravity,''
  Adv.\ Theor.\ Math.\ Phys.\  {\bf 2}, 231 (1998)
  [Int.\ J.\ Theor.\ Phys.\  {\bf 38}, 1113 (1999)]
  [arXiv:hep-th/9711200].
  
\bibitem{Klebanov:1998hh}
  I.~R.~Klebanov and E.~Witten,
  ``Superconformal field theory on threebranes at a Calabi-Yau  singularity,''
  Nucl.\ Phys.\  B {\bf 536}, 199 (1998)
  [arXiv:hep-th/9807080].

\bibitem{Acharya:1998db}
  B.~S.~Acharya, J.~M.~Figueroa-O'Farrill, C.~M.~Hull and B.~J.~Spence,
  ``Branes at conical singularities and holography,''
  Adv.\ Theor.\ Math.\ Phys.\  {\bf 2}, 1249 (1999)
  [arXiv:hep-th/9808014].
  
\bibitem{Morrison:1998cs}
  D.~R.~Morrison and M.~R.~Plesser,
  ``Non-spherical horizons. I,''
  Adv.\ Theor.\ Math.\ Phys.\  {\bf 3}, 1 (1999)
  [arXiv:hep-th/9810201].

\bibitem{Gauntlett:2004zh}
  J.~P.~Gauntlett, D.~Martelli, J.~Sparks and D.~Waldram,
  ``Supersymmetric AdS(5) solutions of M-theory,''
  Class.\ Quant.\ Grav.\  {\bf 21}, 4335 (2004)
  [arXiv:hep-th/0402153].

\bibitem{Gauntlett:2004yd}
  J.~P.~Gauntlett, D.~Martelli, J.~Sparks and D.~Waldram,
  ``Sasaki-Einstein metrics on S(2) x S(3),''
  Adv.\ Theor.\ Math.\ Phys.\  {\bf 8}, 711 (2004)
  [arXiv:hep-th/0403002].

\bibitem{Cvetic:2005vk}
  M.~Cvetic, H.~Lu, D.~N.~Page and C.~N.~Pope,
  ``New Einstein-Sasaki and Einstein spaces from Kerr-de Sitter,''
  JHEP {\bf 0907}, 082 (2009)
  [arXiv:hep-th/0505223].
  
\bibitem{Martelli:2005wy}
  D.~Martelli and J.~Sparks,
  ``Toric Sasaki-Einstein metrics on S**2 x S**3,''
  Phys.\ Lett.\  B {\bf 621}, 208 (2005)
  [arXiv:hep-th/0505027].

\bibitem{Martelli:2004wu}
  D.~Martelli and J.~Sparks,
  ``Toric geometry, Sasaki-Einstein manifolds and a new infinite class of
  AdS/CFT duals,''
  Commun.\ Math.\ Phys.\  {\bf 262}, 51 (2006)
  [arXiv:hep-th/0411238].

\bibitem{Bertolini:2004xf}
  M.~Bertolini, F.~Bigazzi and A.~L.~Cotrone,
  ``New checks and subtleties for AdS/CFT and a-maximization,''
  JHEP {\bf 0412}, 024 (2004)
  [arXiv:hep-th/0411249].

\bibitem{Benvenuti:2004dy}
  S.~Benvenuti, S.~Franco, A.~Hanany, D.~Martelli and J.~Sparks,
  ``An infinite family of superconformal quiver gauge theories with
  Sasaki-Einstein duals,''
  JHEP {\bf 0506}, 064 (2005)
  [arXiv:hep-th/0411264].

\bibitem{Benvenuti:2005ja}
  S.~Benvenuti and M.~Kruczenski,
  ``From Sasaki-Einstein spaces to quivers via BPS geodesics: Lpqr,''
  JHEP {\bf 0604}, 033 (2006)
  [arXiv:hep-th/0505206].

\bibitem{Franco:2005sm}
  S.~Franco, A.~Hanany, D.~Martelli, J.~Sparks, D.~Vegh and B.~Wecht,
  ``Gauge theories from toric geometry and brane tilings,''
  JHEP {\bf 0601}, 128 (2006)
  [arXiv:hep-th/0505211].

\bibitem{Butti:2005sw}
  A.~Butti, D.~Forcella and A.~Zaffaroni,
  ``The dual superconformal theory for L(p,q,r) manifolds,''
  JHEP {\bf 0509}, 018 (2005)
  [arXiv:hep-th/0505220].

\bibitem{Hanany:2005ve}
  A.~Hanany and K.~D.~Kennaway,
  ``Dimer models and toric diagrams,''
  arXiv:hep-th/0503149.

\bibitem{Franco:2005rj}
  S.~Franco, A.~Hanany, K.~D.~Kennaway, D.~Vegh and B.~Wecht,
  ``Brane Dimers and Quiver Gauge Theories,''
  JHEP {\bf 0601}, 096 (2006)
  [arXiv:hep-th/0504110].

\bibitem{Hanany:2005ss}
  A.~Hanany and D.~Vegh,
  ``Quivers, tilings, branes and rhombi,''
  JHEP {\bf 0710}, 029 (2007)
  [arXiv:hep-th/0511063].

\bibitem{Feng:2005gw}
  B.~Feng, Y.~H.~He, K.~D.~Kennaway and C.~Vafa,
  ``Dimer models from mirror symmetry and quivering amoebae,''
  Adv.\ Theor.\ Math.\ Phys.\  {\bf 12}, 3 (2008)
  [arXiv:hep-th/0511287].

\bibitem{Kennaway:2007tq}
  K.~D.~Kennaway,
  ``Brane Tilings,''
  Int.\ J.\ Mod.\ Phys.\  A {\bf 22}, 2977 (2007)
  [arXiv:0706.1660 [hep-th]].
  
\bibitem{Klebanov:1999tb}
  I.~R.~Klebanov and E.~Witten,
  ``AdS/CFT correspondence and symmetry breaking,''
  Nucl.\ Phys.\  B {\bf 556}, 89 (1999)
  [arXiv:hep-th/9905104].
  
\bibitem{Fulton} W.~Fulton, ``Introduction to toric varieties,'' 
(AM-131) (Annals of Mathematics Studies),
Princeton University Press (1993), ISBN-10:0691000492.

\bibitem{craig1} C.~van Coevering, ``A Construction of Complete Ricci-flat K\"ahler Manifolds,'' 
arXiv:0803.0112 [math.DG].

\bibitem{craig2} C.~van Coevering, ``Ricci-flat K\"ahler metrics on crepant resolutions of K�hler cones,'' 
arXiv:0806.3728 [math.DG].

\bibitem{craig3} C.~van Coevering, ``Examples of asymptotically conical Ricci-flat K\"ahler manifolds,'' 
arXiv:0812.4745 [math.DG].

\bibitem{Benishti:2009ky}
  N.~Benishti, Y.~H.~He and J.~Sparks,
  ``(Un)Higgsing the M2-brane,''
  JHEP {\bf 1001}, 067 (2010)
  [arXiv:0909.4557 [hep-th]].

\bibitem{Benishti:2010jn}
  N.~Benishti, D.~Rodriguez-Gomez and J.~Sparks,
  ``Baryonic symmetries and M5 branes in the AdS$_4$/CFT$_3$ correspondence,''
  JHEP {\bf 1007}, 024 (2010)
  [arXiv:1004.2045 [hep-th]].

\bibitem{Martelli:2008cm}
  D.~Martelli and J.~Sparks,
  ``Symmetry-breaking vacua and baryon condensates in AdS/CFT,''
  Phys.\ Rev.\  D {\bf 79}, 065009 (2009)
  [arXiv:0804.3999 [hep-th]].
  
\bibitem{Beasley:2001zp}
  C.~E.~Beasley and M.~R.~Plesser,
  ``Toric duality is Seiberg duality,''
  JHEP {\bf 0112}, 001 (2001)
  [arXiv:hep-th/0109053].

\bibitem{Krishnan:2008kv}
  C.~Krishnan and S.~Kuperstein,
  ``Gauge Theory RG Flows from a Warped Resolved Orbifold,''
  JHEP {\bf 0804}, 009 (2008)
  [arXiv:0801.1053 [hep-th]].
  
\bibitem{Seiberg:1994bz}
  N.~Seiberg,
  ``Exact Results On The Space Of Vacua Of Four-Dimensional Susy Gauge
  Theories,''
  Phys.\ Rev.\  D {\bf 49}, 6857 (1994)
  [arXiv:hep-th/9402044].
  
\bibitem{Feng:2000mi}
  B.~Feng, A.~Hanany and Y.~H.~He,
  ``D-brane gauge theories from toric singularities and toric duality,''
  Nucl.\ Phys.\  B {\bf 595}, 165 (2001)
  [arXiv:hep-th/0003085].
  
\bibitem{Hanany:2008gx}
  A.~Hanany and Y.~H.~He,
  ``M2-Branes and Quiver Chern-Simons: A Taxonomic Study,''
  arXiv:0811.4044 [hep-th].

\bibitem{Martelli:2007mk}
  D.~Martelli and J.~Sparks,
  ``Baryonic branches and resolutions of Ricci-flat Kahler cones,''
  JHEP {\bf 0804}, 067 (2008)
  [arXiv:0709.2894 [hep-th]].

\bibitem{Gukov:1998kn}
  S.~Gukov, M.~Rangamani and E.~Witten,
  ``Dibaryons, strings, and branes in AdS orbifold models,''
  JHEP {\bf 9812}, 025 (1998)
  [arXiv:hep-th/9811048].
  
\bibitem{Imamura:2007dc}
  Y.~Imamura, H.~Isono, K.~Kimura and M.~Yamazaki,
  ``Exactly marginal deformations of quiver gauge theories as seen from   brane
  tilings,''
  Prog.\ Theor.\ Phys.\  {\bf 117}, 923 (2007)
  [arXiv:hep-th/0702049].
  
\bibitem{Vilenkin:1982ni}
  A.~Vilenkin and A.~E.~Everett,
  ``Cosmic Strings And Domain Walls In Models With Goldstone And
  Pseudogoldstone Bosons,'' Phys. Rev. Lett. 48, 1867-1870 (1982).
  
\bibitem{Martucci:2005ht}
  L.~Martucci and P.~Smyth,
  ``Supersymmetric D-branes and calibrations on general N = 1 backgrounds,''
  JHEP {\bf 0511}, 048 (2005)
  [arXiv:hep-th/0507099].
  
\bibitem{Gomis:2005wc}
  J.~Gomis, F.~Marchesano and D.~Mateos,
  ``An open string landscape,''
  JHEP {\bf 0511}, 021 (2005)
  [arXiv:hep-th/0506179].
  
\bibitem{Klebanov:2007cx}
  I.~R.~Klebanov, A.~Murugan, D.~Rodriguez-Gomez and J.~Ward,
  ``Goldstone Bosons and Global Strings in a Warped Resolved Conifold,''
  JHEP {\bf 0805}, 090 (2008)
  [arXiv:0712.2224 [hep-th]].
  
\bibitem{Beasley:2004ys}
  C.~Beasley and E.~Witten,
  ``New instanton effects in supersymmetric QCD,''
  JHEP {\bf 0501}, 056 (2005)
  [arXiv:hep-th/0409149].

\bibitem{Beasley:2005iu}
  C.~Beasley and E.~Witten,
  ``New instanton effects in string theory,''
  JHEP {\bf 0602}, 060 (2006)
  [arXiv:hep-th/0512039].
   
\bibitem{Forcella:2008ng}
  D.~Forcella, A.~Hanany and A.~Zaffaroni,
  ``Master Space, Hilbert Series and Seiberg Duality,''
  JHEP {\bf 0907}, 018 (2009)
  [arXiv:0810.4519 [hep-th]].

\bibitem{Forcella:2008bb}
  D.~Forcella, A.~Hanany, Y.~H.~He and A.~Zaffaroni,
  ``The Master Space of N=1 Gauge Theories,''
  JHEP {\bf 0808}, 012 (2008)
  [arXiv:0801.1585 [hep-th]].
  
\bibitem{Mukhopadhyay:2001sr}
  S.~Mukhopadhyay and K.~Ray,
  ``Fractional branes on a non-compact orbifold,''
  JHEP {\bf 0107}, 007 (2001)
  [arXiv:hep-th/0102146].
  
\bibitem{Marino:1999af}
  M.~Marino, R.~Minasian, G.~W.~Moore and A.~Strominger,
  ``Nonlinear instantons from supersymmetric p-branes,''
  JHEP {\bf 0001}, 005 (2000)
  [arXiv:hep-th/9911206].
  
\bibitem{Aharony:1997bh}
  O.~Aharony, A.~Hanany and B.~Kol,
  ``Webs of (p,q) 5-branes, five dimensional field theories and grid
  diagrams,''
  JHEP {\bf 9801}, 002 (1998)
  [arXiv:hep-th/9710116].

\bibitem{Leung:1997tw}
  N.~C.~Leung and C.~Vafa,
  ``Branes and toric geometry,''
  Adv.\ Theor.\ Math.\ Phys.\  {\bf 2}, 91 (1998)
  [arXiv:hep-th/9711013].

\bibitem{flip}  
I. V. Dolgachev, Y. Hu, 
"Variation of Geometric Invariant Theory Quotients,''
Publ. IHES 87, 5-56 
[arXiv:alg-geom/9402008v4].

\bibitem{Freed:1999vc}
  D.~S.~Freed and E.~Witten,
  ``Anomalies in string theory with D-branes,''
  arXiv:hep-th/9907189.
  
\bibitem{Benvenuti:2005qb}
  S.~Benvenuti, M.~Mahato, L.~A.~Pando Zayas and Y.~Tachikawa,
  ``The Gauge/gravity theory of blown up four cycles,''
  arXiv:hep-th/0512061.
  
\bibitem{files}
https://sites.google.com/site/nessibenishti/research-files

\end{thebibliography}
\end{document}